\documentclass[journal]{IEEEtran}

\usepackage[normalem]{ulem}
\usepackage[pagebackref=true,breaklinks=true,colorlinks,bookmarks=false]{hyperref}

\usepackage[pdftex]{graphicx}
\usepackage{array,animate}
\usepackage{xcolor}
\usepackage{multirow}

\usepackage{cite}
\usepackage{amsmath}
 \usepackage[caption=false,font=footnotesize]{subfig}
\usepackage{url}

\def\etal{\emph{et al.}}

\usepackage{pifont}
\newcommand{\cmark}{\ding{51}}%
\newcommand{\xmark}{\ding{55}}%

\usepackage{adjustbox}

\newlength\fsdurthree

\begin{document}
%
\title{Harnessing Multi-View Perspective of Light Fields for Low-Light Imaging}
%
%
%

\author{Mohit~Lamba*,
        Kranthi~Kumar~Rachavarapu*
        and~Kaushik~Mitra\\ \texttt{\url{https://mohitlamba94.github.io/L3Fnet/}}
\thanks{Mohit Lamba, Kranthi~Kumar~Rachavarapu and Kaushik Mitra are with the Department
of Electrical Engineering, Indian Institute of Technology Madras, India (e-mail: ee18d009@smail.iitm.ac.in, ee18d004@smail .iitm.ac.in, kmitra@ee.iitm.ac.in).}
\thanks{This paper has a supplementary downloadable PDF report available at http://ieeexplore.ieee.org., provided by the author. Contact ee18d009@smail.iitm.ac.in for further questions about this work.}
\thanks{*M. Lamba and K. Kumar have contributed equally to the work}
}

%
%

\markboth{Journal of \LaTeX\ Class Files,~Vol.~14, No.~8, August~2015}%
{Shell \MakeLowercase{\textit{et al.}}: Bare Demo of IEEEtran.cls for IEEE Journals}
%



\maketitle

\begin{abstract}
Light Field (LF) offers unique advantages such as post-capture refocusing and depth estimation, but low-light conditions severely limit these capabilities.
To restore low-light LFs we should harness the geometric cues present in different LF views, which is not possible using single-frame low-light enhancement techniques. 
We, therefore, propose a deep neural network architecture for Low-Light Light Field (L3F) restoration, which we refer to as L3Fnet. The proposed L3Fnet not only performs the necessary visual enhancement of each LF view but also preserves the epipolar geometry across views. We achieve this by adopting a two-stage architecture for L3Fnet. Stage-I looks at all the LF views to encode the LF geometry. This encoded information is then used in Stage-II to reconstruct each LF view. 

To facilitate learning-based techniques for low-light LF imaging, we collected a comprehensive LF dataset of various scenes. For
each scene, we captured four LFs, one with near-optimal exposure
and ISO settings and the others at different levels of low-light
conditions varying from low to extreme low-light settings. The effectiveness of the proposed L3Fnet is supported by both visual and numerical comparisons on this dataset.
 To further analyze the performance of low-light reconstruction methods, we also propose an L3F-wild dataset that contains LF captured late at night with almost zero lux values. No ground truth is available in this dataset. To perform well on the L3F-wild dataset, any method must adapt to the light level of the captured scene. \textcolor{black}{To do this we use a pre-processing block that makes L3Fnet robust to various degrees of low-light conditions.} Lastly, we show that L3Fnet can also be used for low-light enhancement of single-frame images, despite it being engineered for LF data. We do so by converting the single-frame DSLR image into a form suitable to L3Fnet, which we call as \textit{pseudo-LF}.
\end{abstract}


%
\IEEEpeerreviewmaketitle

\section{Introduction}

\begin{figure}[!t]
\captionsetup[subfigure]{labelformat=empty}
\centering
\subfloat[Captured LF]{\includegraphics[width=0.48\linewidth]{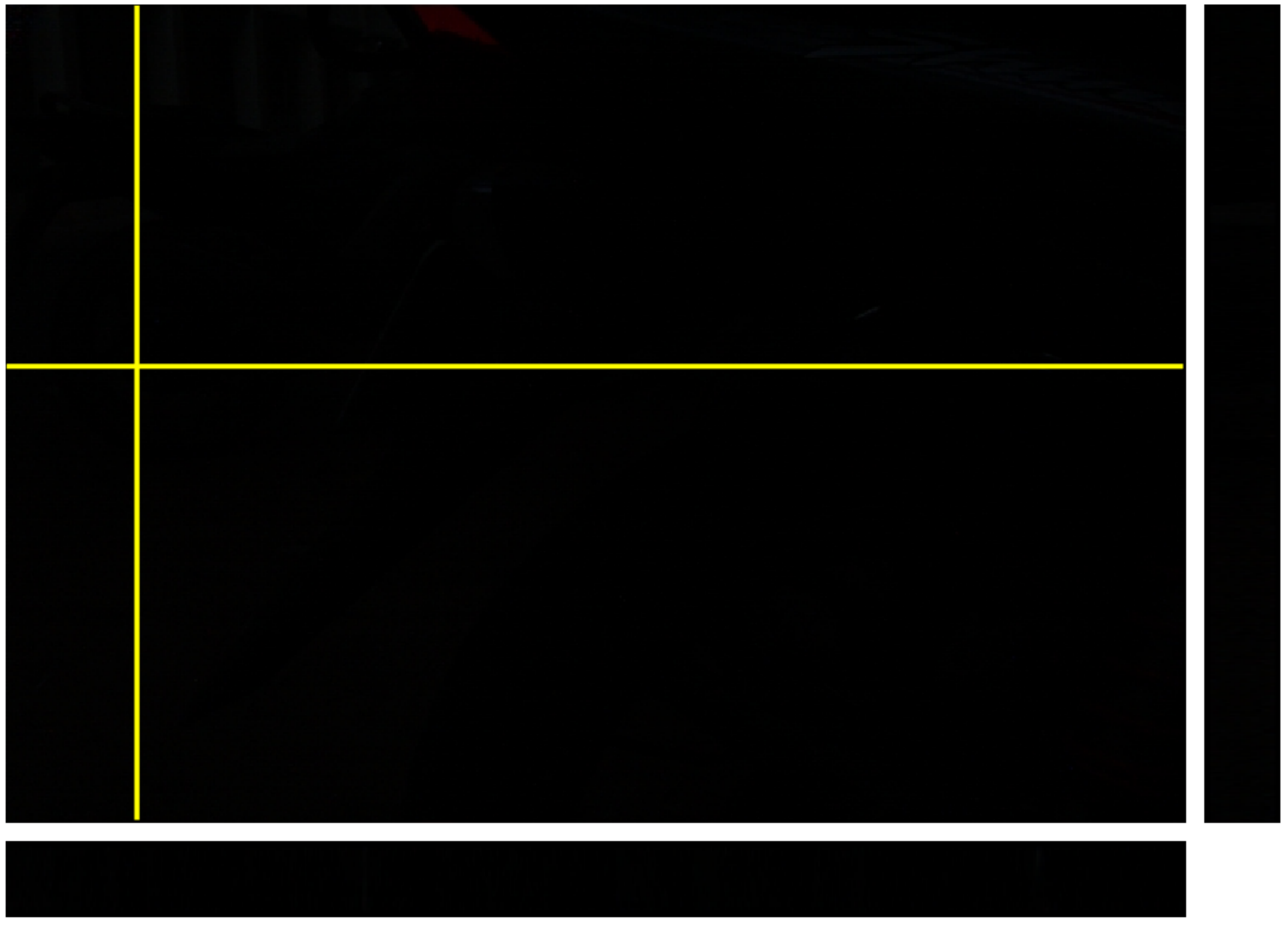}}
\hfil 
\subfloat[Depth from captured LF]{\includegraphics[width=0.48\linewidth, height=0.345\linewidth]{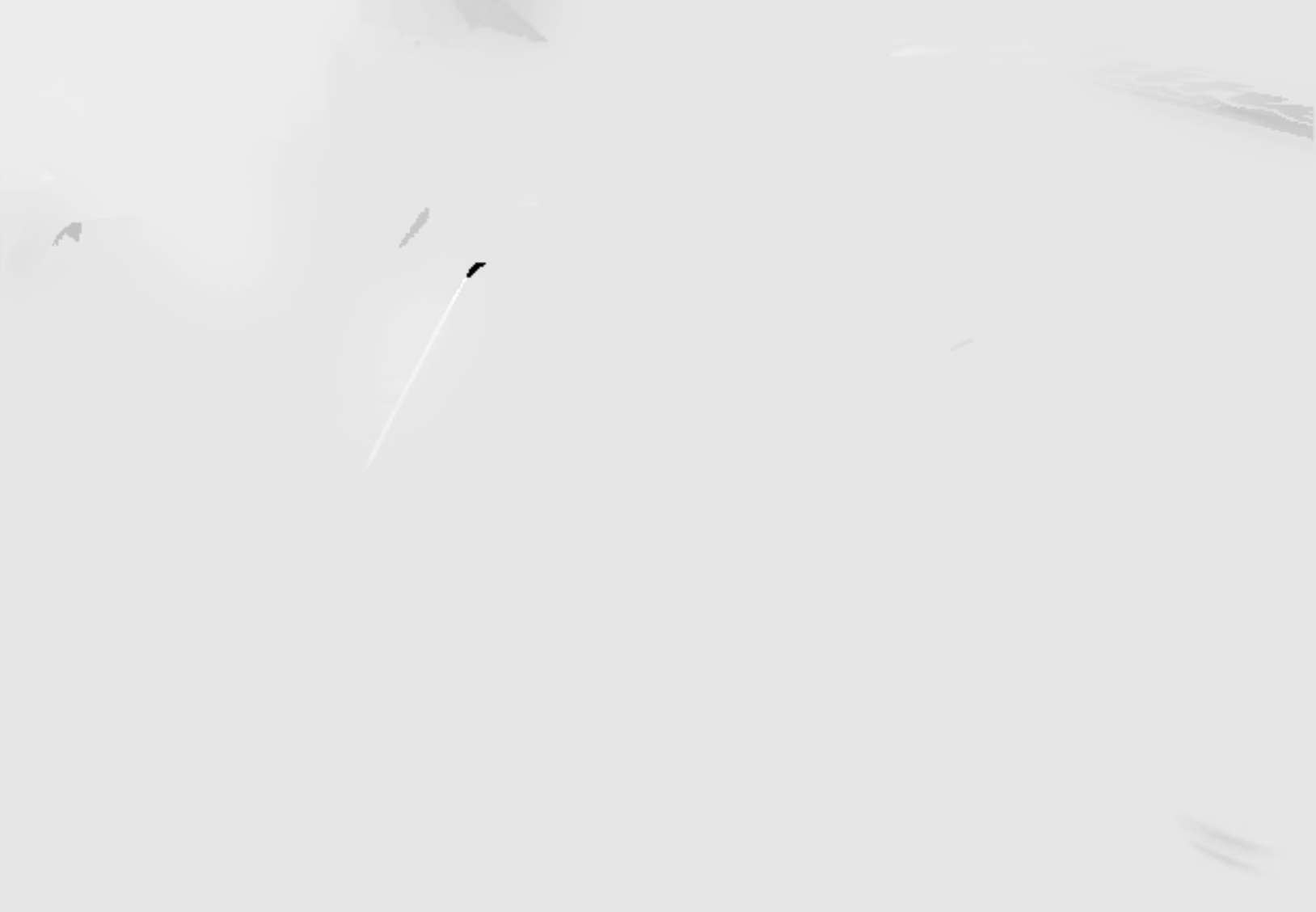}} \vspace{-2mm}\\
\subfloat[Histogram equalized LF]{\includegraphics[width=0.48\linewidth]{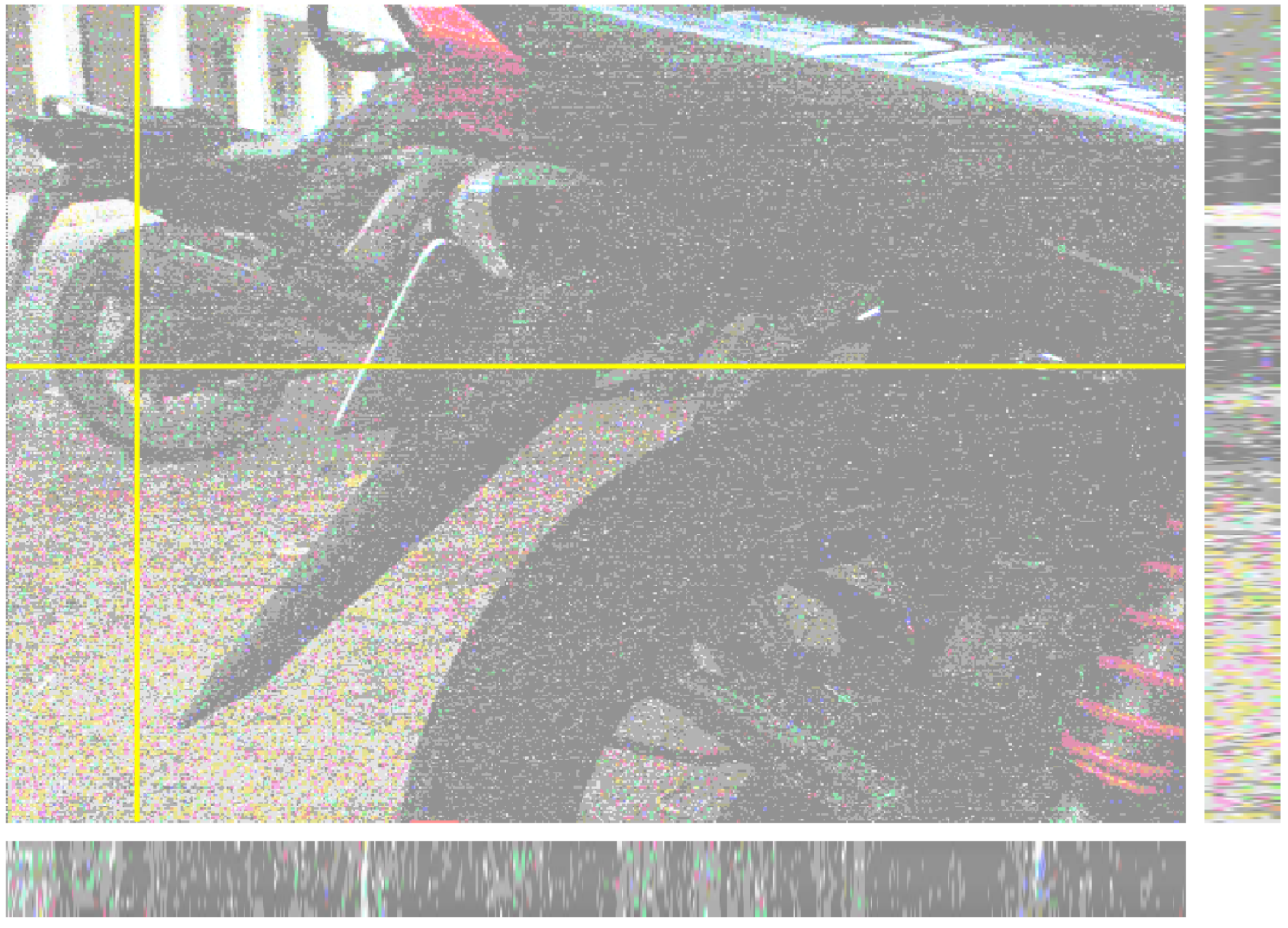}}
\hfil
\subfloat[Depth from histogram equalized LF]{\includegraphics[width=0.48\linewidth, height=0.345\linewidth]{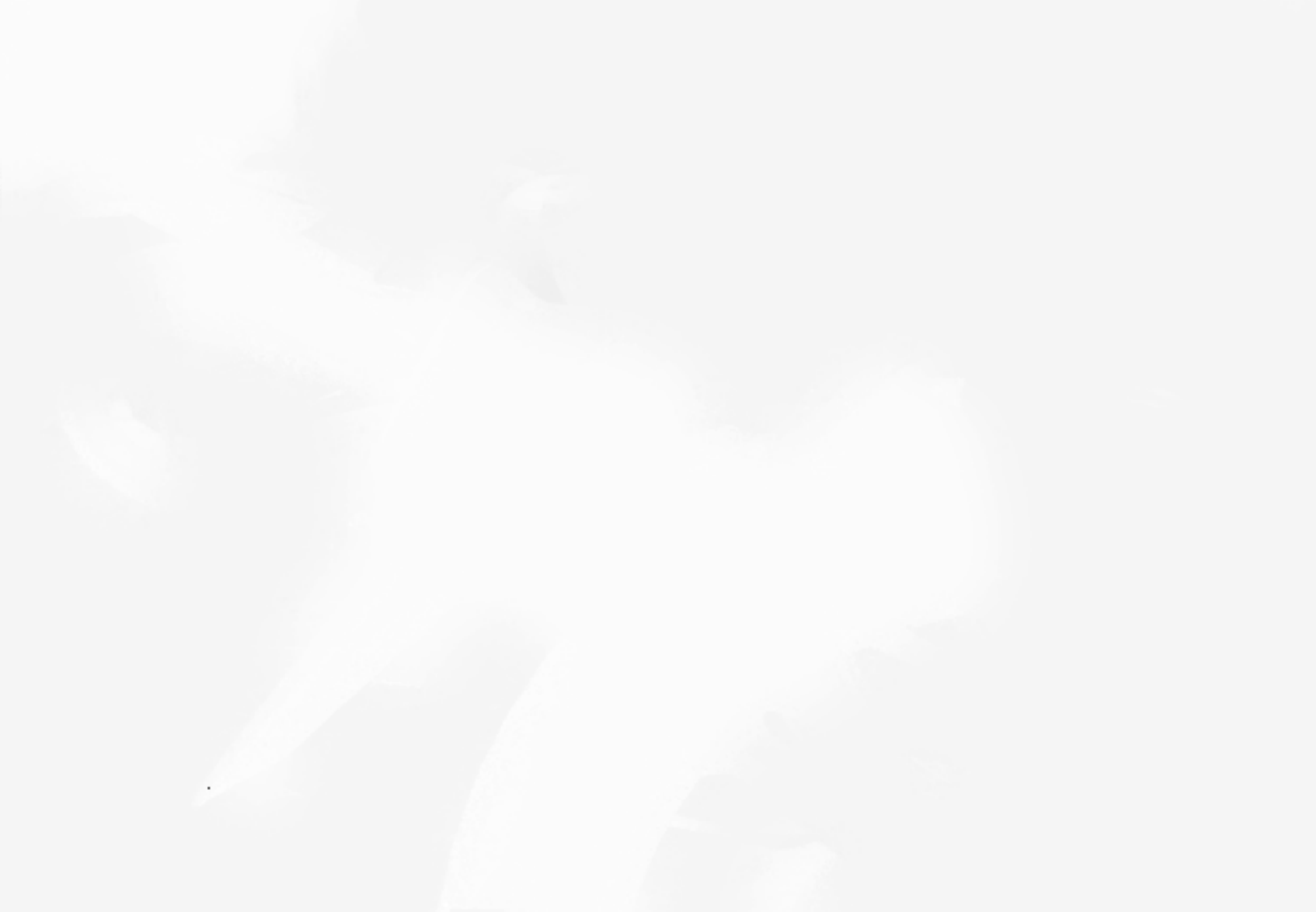}} \vspace{-2mm} \\
\subfloat[Our restored LF]{\includegraphics[width=0.48\linewidth]{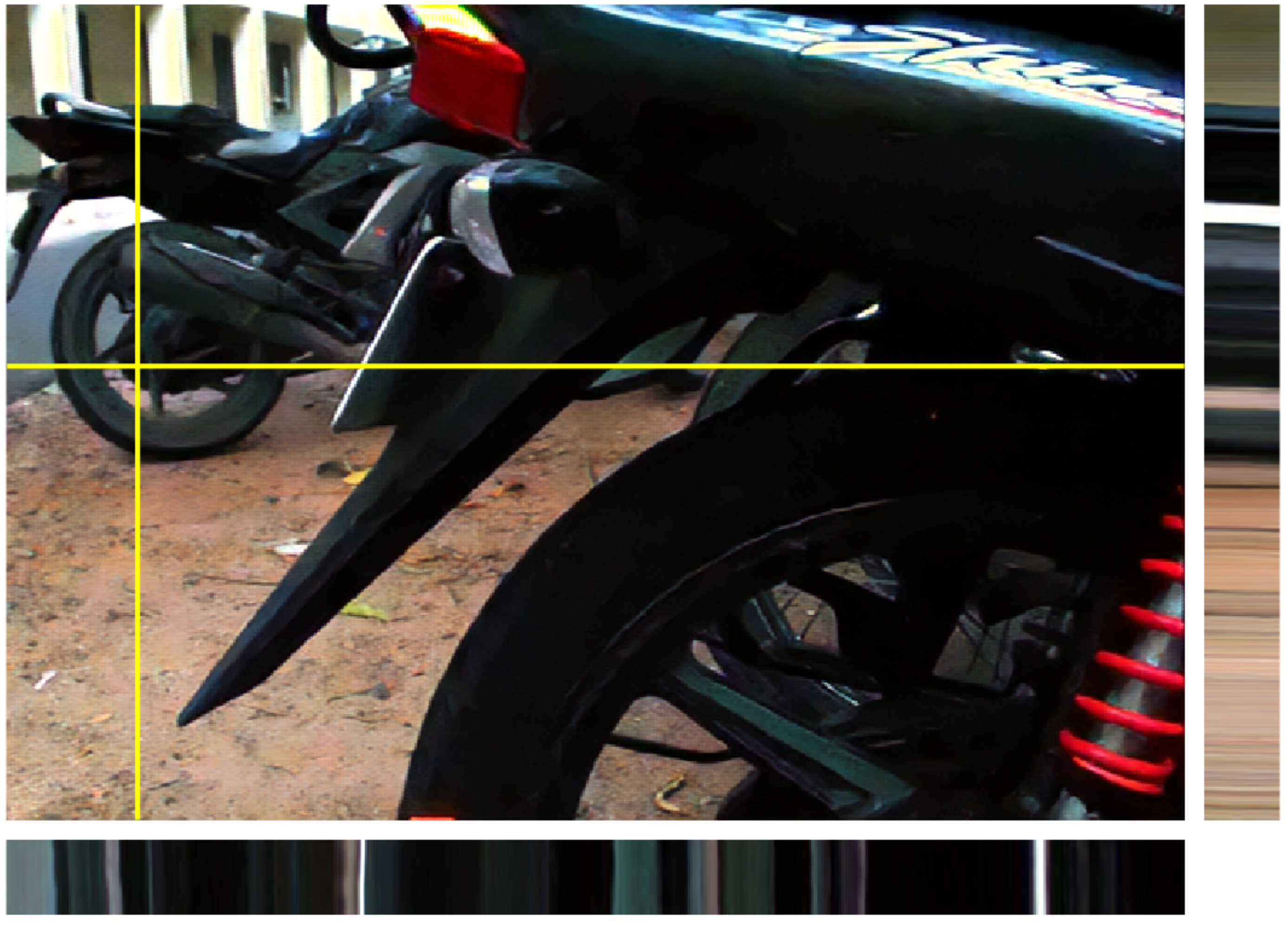}}
\hfil
\subfloat[Depth from our restored LF]{\includegraphics[width=0.48\linewidth, height=0.345\linewidth]{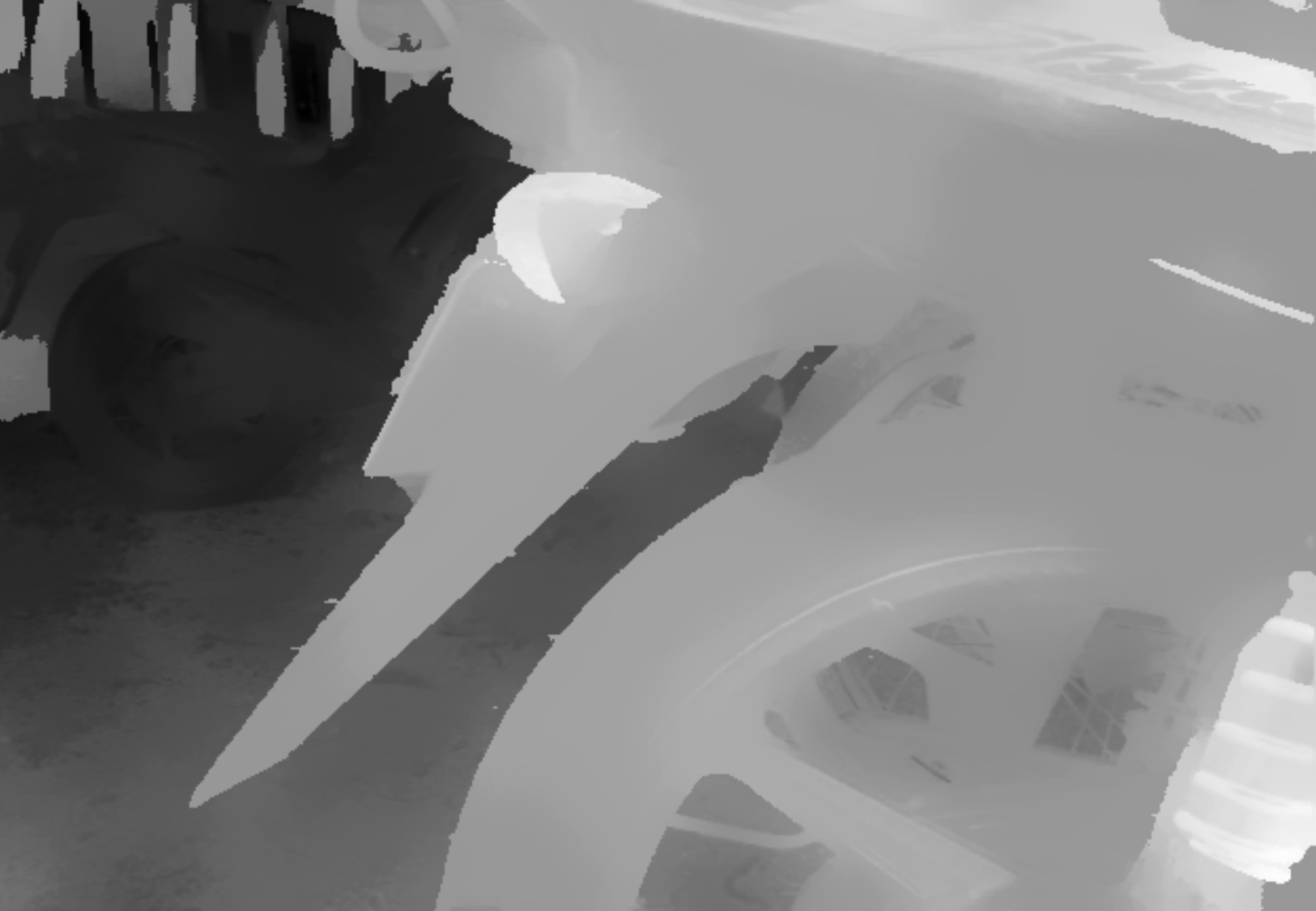}}
\caption{
Low-light is a severe bottleneck to Light Field (LF) applications. For example, the depth estimate of LF captured in low
light is very poor. Our proposed method not only visually restores each of the LF views but also preserves the LF geometry for faithful depth estimation.}
\label{fig:title_image}
\vspace{-0.5cm}
\end{figure}

 \IEEEPARstart{I}{mages} captured in low-light such as in the dark or night-time, not only lack the pleasing visual aesthetics but are also corrupted by high amount of noise and color distortions. This forthrightly hinders the performance of many computer vision algorithms \cite{low_light_degenrate}. 
This plight of low-light conditions is not unique to conventional 2D photographs, but is also shared by Light Fields (LFs) \cite{adelson1992single, ng2005TechReport, levoy1996LF, gortler1996lumigraph}.  
In contrast with a conventional camera, which captures only 2-D spatial information, a light field camera captures both 2-D spatial and 2-D angular information about the scene.
Capturing the full light field allows us to perform post-capture controls such as digital refocusing and aperture control \cite{ng2005TechReport,ng2005fourierslice}. It also enables easy scene depth estimation \cite{depthFind, depth2, depth3}. This has resulted in many LF applications such as view synthesis \cite{levoy1996LF, gortler1996lumigraph, LF_view_synthesis}, structure from motion \cite{SfM_LF}, pedestrian identification \cite{person_identification_LF}, reflection removal \cite{reflection_removal_LF}, and various other real-world application like autonomous driving and plant monitoring \cite{raytrix}. But, as stated previously, these LF applications are not immune to the challenges of low-light imaging. This is depicted in Fig. \ref{fig:title_image} where the depth estimation capability of LF is foiled due to low-light conditions. Commonly used techniques such as histogram equalization or having higher ISO does not help much in this regard as they boost the noise levels and introduce unwanted artifacts. Our goal, therefore, is to design a low-light LF restoration technique to mitigate these problems. 

Low-light restoration is an ill-posed problem because of the large amount of noise present in the low-light signal. Also, color information is not present adequately. An LF, however, by capturing multiple views of the low-light scene, contains rich geometric cues about the scene. Harnessing the complimentary information spread across the LF views should help in reducing the ill-posedness of the problem and result in better visual reconstruction. In addition to this, the restoring process should also preserve the LF epipolar constraints to allow for subsequent tasks like depth estimation. The existing works on the low-light enhancement \cite{guo2017lime,park2017lowGamma,2019underexposed,ren2019lowRetouch,chen2018learning2seeindark, lore2017llnet} are, however, not designed to keep these points in consideration and so are not suitable candidates for enhancing low-light LFs. 

We propose a two-stage deep neural network for Low-Light Light Field (L3Fnet) reconstruction. Stage-I of the L3Fnet operates on the full LF to first encode the LF geometry. This encoded information is then used as an auxiliary information in Stage-II for actual view restoration. Adopting such a two-staged architecture helps us to restore low-light LFs both aesthetically and with geometric correctness.

Training a learning-based model such as L3Fnet requires a low-light LF dataset but, unfortunately, there is no such dataset. An easy way out would have been to create a synthetic dataset using gamma correction, noise addition, or even retouching the images in software such as Adobe Photoshop and GIMP as was done for the single frame case \cite{2019underexposed, gammaarxiv,park2017lowGamma,lore2017llnet,guo2019pipelineGamma}. But these are only proxy solutions, and as pointed out by Pl{\"o}tz and Stefan \cite{plotz2017benchmarking}, benchmarking algorithms on synthetic dataset may not correlate with their performance on real-world data. We, therefore, collected our own Low-Light Light Field (L3F) dataset using commercially available Lytro Illum LF camera. L3F dataset was captured in the evenings when the visibility was below normal level. For each scene, $4$ shots were taken. For the first shot, optimal camera exposure and ISO settings were used to get the ground truth LF. Lytro Illum's exposure was then reduced in definite proportions to capture three more low-light LFs. While taking these shots, much care has been taken to avoid moving objects such as passing vehicles and wavering leaves to prevent any registration issues. Camera shake is another cause of incurring registration problems and is much more pronounced for Lytro Illum. Lytro Illum, unlike modern single-frame DSLR cameras, does not allow remote connectivity for capturing data. So the only way to capture LF was by physically pressing the shutter button causing unintentional camera shake. To limit this we captured multiple sets of four LFs for each scene and employed additional measures to keep the camera rigidly fixed. We then did a manual check to identify the set exhibiting the least amount of camera shake. Using these techniques the maximum pixel shift in the captured LFs is about $3-4$ pixels.

While capturing the L3F dataset we were constrained to capture the ground truth for each scene. This was essential to train our network. Because of this constraint, we were unable to capture extremely low-light LFs. We, therefore, decided to forsake this restriction and collected a separate low-light LF dataset captured late in the night with near $zero$ lux conditions at the camera lens. We call this dataset L3F-wild, which we use only for evaluation. L3F-wild LFs were captured with typical camera exposure of $1/5$ second and nominal ISO levels. Note that we do not use the standard practice of capturing images with a long exposure, of say $10$ seconds, in low-light conditions. This is to avoid possible motion blur and is a step towards fast low-light imaging. The L3F-wild dataset has a good amount of variation in scene brightness level, portraying a real-world scenario. We, consequently, \textcolor{black}{ use a pre-processing block which when appended to L3Fnet makes it robust to such variations in light levels by automatically estimating an appropriate amplification factor.}

We have already discussed why single-frame techniques are not suitable for enhancing low-light LF. And we now address the reverse question, \textit{can LF methods be used for single-frame DSLR images?} The proposed L3Fnet has been specifically engineered for LF and so cannot be directly operated on single-frame images. However, we propose a novel pixel shuffling mechanism, which can convert any DSLR image into a \textit{pseudo-LF}. Pseudo-LF has a form suitable to L3Fnet, and so can be enhanced using it. Later on, the enhanced pseudo-LF is transformed back into a single-frame DSLR image in a lossless fashion. This gives L3Fnet architecture a universal appeal for both LF and single-frame image low-light enhancement with restoration being more optimized for LF but maintaining a decent recovery for single-frame images.


To summarize, we make the following contributions: 
\begin{itemize}

    \item We propose a two-stage deep neural network, L3Fnet, for restoring extremely low-light LFs. 
    \item {We collected L3F dataset consisting of real LFs with varying levels of low-light, which can be used for training and evaluation of data driven methods.} 
    \item {\textcolor{black}{We use a pre-processing block that automatically adapts L3Fnet to changing light levels.}}
    \item Our proposed Pseudo-L3Fnet framework enables L3Fnet to process even single-frame DSLR images for better enhancement in several cases.

\end{itemize}

\section{Related Work}

\textbf{LF processing algorithm: } Many techniques and models have been proposed for several LF related tasks. This includes tasks such as spatial super-resolution \cite{wanner2012LF,wang2018lfnetTIP,zhang2019residual,LFviewCoherence}, deblurring \cite{ravi_arxiv_deblurring,2017lfdeblurring,2018lfdeblurring}, denoising \cite{2012LFdenoising,bm5d,2013LFdenoisng,lfdenosing2016,dansereau2013decoding}, and depth estimation \cite{depthFind, depth2, depth3}. But, to the best of our knowledge, no prior work has considered solving the challenges involved in low-light LF imaging. We, therefore, outline some of the important works on LF denoising as denoising is a crucial part of low-light restoration. The easiest approach to LF denoising is to individually denoise each LF view using standard techniques like BM3D \cite{bm3d}. This, however, does not capture the 4D structure of LF and was addressed by  Mitra and Veeraraghavan \cite{2012LFdenoising}. By operating on LF patches, they modeled each 4D patch using Gaussian Mixture Models (GMM) subject to the patch disparity information. 
This way they provided a common framework for several LF tasks such as super-resolution and denoising. 
Dansereau \etal \cite{2013LFdenoisng} observed that the LF of a Lambertian surface has a hyperfan-like shape in the frequency domain. By appropriately choosing the passband they tried to remove noise. Sepas-Moghaddam \etal \cite{lfdenosing2016} treating LF akin to video frames, first converted LF into an epipolar sequence and then applied video-based techniques to denoise the LF. This 3D stacking of epipoles, however, has limitations in capturing the 4D nature of LF and was consequently addressed by LFBM5D \cite{bm5d}, a popular LF denoising technique. LFBM5D does a realistic 4D LF modeling and is a natural extension to the then state-of-the-art denoising method BM3D. However, LFBM5D does not address the problem with low-light LFs because of its inability to enhance color. 

\textbf{Single-frame low-light enhancement: } A significant amount of literature exists for low-light single-frame image enhancement. We briefly review some of them. LIME \cite{guo2017lime} used a non deep learning optimization framework for low-light enhancement. It utilized the retinex theory \cite{retinex_1997} to decouple the captured image into reflectance and illumination components for subsequent enhancement. A similar idea was used by Park \etal \cite{park2017lowGamma} and proposed another variational optimization-based retinex model. Ying \etal \cite{ying2017new} also used the retinex model but combined it with the camera response model to preserve the naturalness of the enhanced image. Deep learning based methods employing encoder-decoder architecture have also been recently used for low-light enhancement \cite{lore2017llnet,2019underexposed,ren2019lowRetouch,chen2018learning2seeindark}. Amongst all these recent works, the work by Chen \etal~ \cite{chen2018learning2seeindark} is a landmark paper on extreme low-light enhancement and is closest to our work. 
\textcolor{black}{Following this work by Chen \etal, \cite{gu2019self,2019DID} are some of the more recent methods which show results on extreme low-light images.}
Other recent works \textcolor{black}{\cite{2019underexposed,ren2019lowRetouch,lore2017llnet,r3_2018_bmvc,r1_2018_acm,r1_2016,r1_2019_kind,r3_2018lightennet,r3_2020_unsupervised}} also aim at low-light but their main objective is the enhancement of dim images. These images already had a good representation of the target scene and only lacked in aesthetics and contrast but had a decent amount of scene visibility. This is an interesting problem to solve but in our work we target image with much lower visibility, see Fig. \ref{fig:title_image}, \ref{fig:vis_compare} and \ref{fig:wild}. 
    \begin{figure}[t!]
	\centering
    \includegraphics[width=\linewidth]{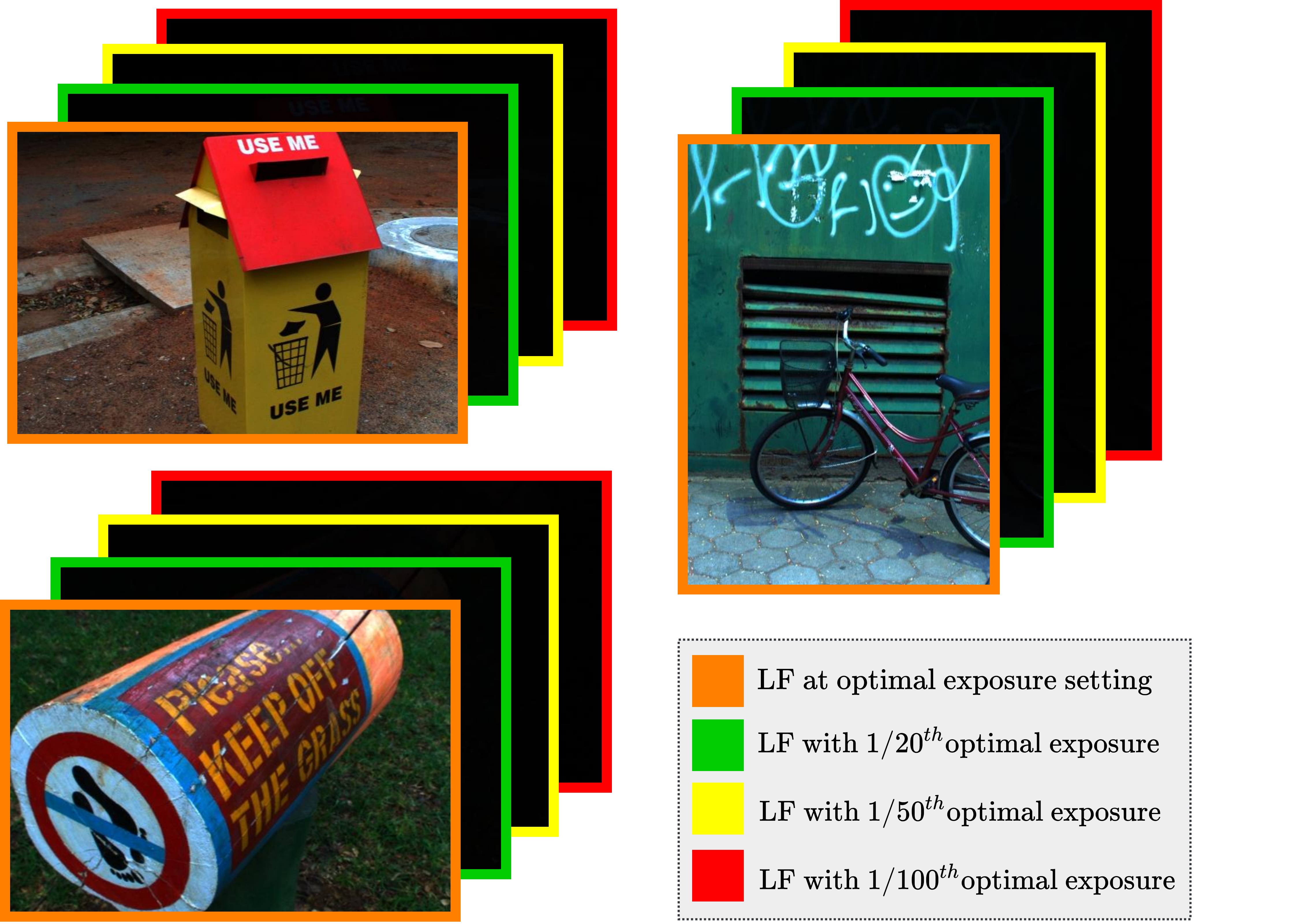}
    \caption{Sample center-view images from the Low-Light Light Field (L3F) dataset consisting of $27$ scenes. For each scene, we capture a LF at near optimal exposure setting and at three other exposure settings with the exposures being $1/20^{th}, 1/50^{th}$ and $1/100^{th}$ of the optimal setting. Refer supplementary material to view the full dataset.}
    \label{fig:sample_L3F_images}
\vspace{-0.5cm}
\end{figure}
 

\section{L3F Dataset}
\label{sec:l3fdataset}
Creating a low-light LF enhancement dataset is a major challenge in designing learning-based solutions. Synthetic low-light LF data can be created using gamma correction and noise addition \cite{gammaarxiv,park2017lowGamma,lore2017llnet,guo2019pipelineGamma}, however, such techniques act globally and so do not mimic a real low-light situation which severely affects some regions more than others. Also, in synthetic data noise is generally modeled as Gaussian or Poisson distribution which may not hold true in real low-light data. The other popular technique is to retouch the low-light images in photo-editing software by trained photographers to obtain the ground-truth \cite{2019underexposed,ren2019lowRetouch}. 
Such proxy solutions are not suitable for evaluating low-light LFs because any local edit in a LF view needs to be propagated to all the other SAIs which is difficult to enforce manually.
Thus, different from recent methods that adopt such techniques, we introduce a Low-Light Light Field (L3F) dataset containing both low-light LF and the corresponding ground-truth LF. To the best of our knowledge, this is the first dataset available for training and benchmarking low-light LF enhancement techniques. 

\begin{figure*}[t]
    \centering
    \includegraphics[width=\linewidth]{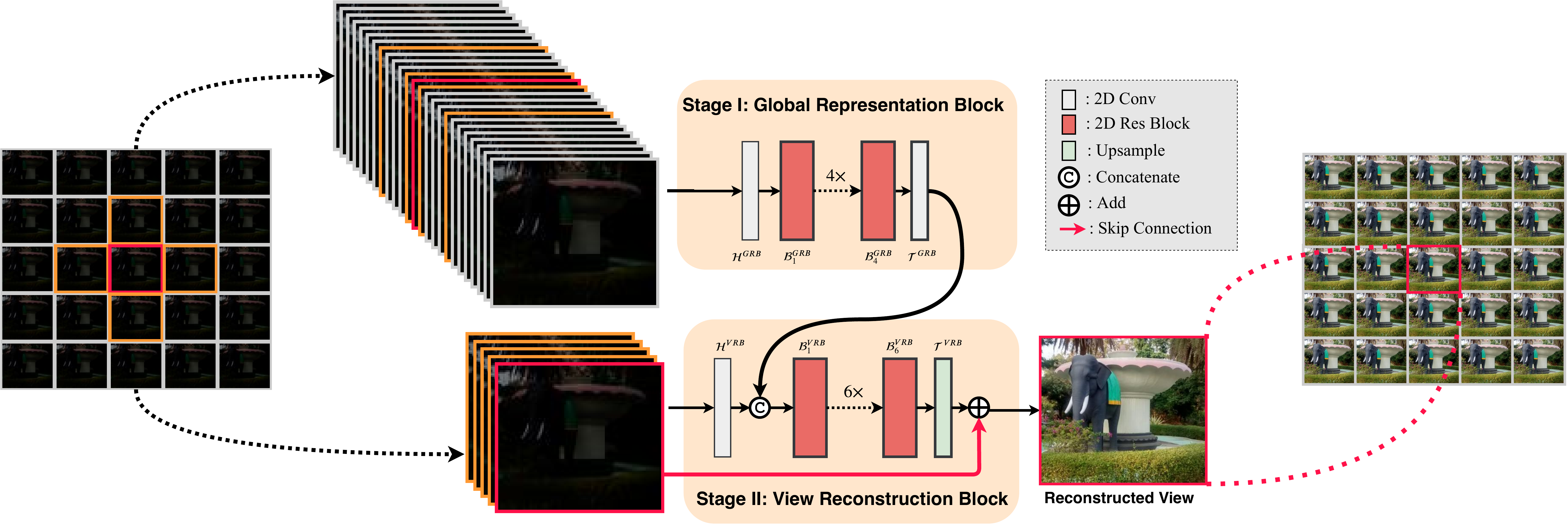}
    \caption{L3Fnet architecture: The proposed architecture consists of a Global Representation Block (Stage-I) and a View Reconstruction block (Stage-II). Stage-I operates on the full low-light LF to obtain a latent representation that encodes the LF geometry. This latent representation is then used by Stage-II to restore each LF view.
    }
    \label{fig:architecture1}
    \vspace{-0.5cm}
\end{figure*}

All the LFs in the proposed L3F dataset were captured in outdoor situations. The scene content predominantly includes an outdoor urban campus environment with lambertian and non-lambertian surfaces at varying depth and occlusion levels. The captured LFs do not include any dynamic objects in the scene such as moving people and vehicles. All the LFs were captured in the evening with limited natural light where the typical illuminance was between $0.1$ Lux to $20$ Lux.  

We use the Lytro Illum Camera to capture all the LFs. For each scene we captured four LF data. One of them was captured using the least ISO and appropriate exposure settings to make it look like a well-lit image. This we use as the ground truth reference LF for that scene. The other three LFs were captured by reducing the exposure setting to $1/20^{th}$, $1/50^{th}$ and $1/100^{th}$ of the reference LF exposure time. For brevity, we refer them as L3F $-$ 20, 50 and 100 datasets. Chen \etal \cite{chen2018learning2seeindark}~limited their work to approximately $1/20^{th}$ setting but we go even more low-light with L3F-100 dataset for better understanding and contrasting the effect of low-light on LF reconstruction. 

Motion blur caused due to camera shake is a common artifact when capturing long-exposure sequences. To limit this we mount the Lytro camera on tripod and capture the LFs using timer mode. Unlike DSLR cameras the  Lytro  Illum  camera does not allow remote capture which makes the data capturing process harder and laborious. Consequently, the four LFs captured for the same scene exhibits small spatial misalignment. To contain this we capture multiple sets of these four LFs for a target scene and choose the set exhibiting the least alignment problem.  
\textcolor{black}{We could have further used matching algorithms, but they had a lot of difficulty in finding the correct correspondences between well-lit and extremely dark images.} Thus, no subsequent alignment operation was performed.
\textcolor{black}{In the supplementary we quantitatively show that for most scenes the misalignment is $\le1$ pixel.}
This small misalignment in L3F dataset is taken care by our proposed method.

The images were captured in the Light Field Raw (LFR) format of the Lytro Illum camera. The resolution of each LFR file is $5368 \times 7728$ pixels. The raw images are captured in the Bayer sensor pattern where the pixels are hexagonally packed. We use Light Field Matlab Toolbox \cite{dansereau2013decoding} to demosaic, decode, devignetize and color-correct the LFR files. The decoded $4$D LF data have a resolution of $15\times15\times434\times625\times3$, where $15\times15$ represents the angular resolution, $434 \times 625$ represents the spatial resolution of each view and $3$ corresponds to the RGB channels.

The proposed dataset contains a total of $108$ LFs organized into $27$ sets. Each set corresponds to a unique scene with $4$ LFs captured at various exposure settings. Fig.~\ref{fig:sample_L3F_images} shows a few sample LF center-views from the proposed dataset. $33\%$ of the dataset i.e., $9$ sets ($36$ LFs) forms the test set. The dataset is available for download from the \href{https://mohitlamba94.github.io/L3Fnet/}{\textit{project page}}.

While capturing L3F$-$20, 50, and 100 datasets, we were constrained to capture the well-lit LF image also, which is required for training the L3Fnet. Forsaking this restriction, we captured even darker low-light LFs taken late in the night. While capturing this dataset, the Lux measure at Lytro Illum's lens was almost nil, and so no ground truth was possible for these LF images. We call this dataset \textit{L3F-wild}. Although L3F-wild cannot be used for training, the performance of methods trained on L3F$-$20, 50, and 100 dataset can be checked by evaluating on L3F-wild dataset. L3F-wild dataset is a step forward towards fast low-light imaging because it does not adopt the standard practices of low-light imaging such as prolonged exposure or high ISO. Long exposures like $5-10$ seconds cause a lot of motion blur, and high ISO boosts the noise. L3F-wild was, however, captured with a typical exposure time of $1/4 - 1/15$ second and low ISO values around $100$. In the experimental section and supplementary, we demonstrate the effectiveness of L3Fnet on both L3F-100 and L3F-wild datasets. 

\section{Low-Light Light Field Network (L3Fnet)}
\subsection{Important Features of L3Fnet}

Our proposed method is designed to enhance LFs captured in low-light. But before describing the details of the proposed solution, we highlight some of the desired characteristics that the proposed solution should have. These characteristics are not task-specific but essential to any LF architecture. These features were cataloged after studying different architectures for various LF related tasks. Firstly, L3Fnet should \textit{not make strong assumptions} about the scene, so that, given any low-light LF we should be able to restore it. This is contrary to works like  \cite{wanner2012LF,2015LFgeoprior}, which required explicit geometric information of the scene, such as depth-map, to perform the given task. Secondly, L3Fnet should not restore the LF views independent of each other. This may lead to visually pleasing reconstruction, but the LF epipolar geometry would not preserved. This was noticed in works such as \cite{LFviewCoherence}, where only pairs of LF views were fed to the CNN model, failing to capture the high \textit{view coherence} of the LF \cite{wang2018lfnetTIP}. The third desired characteristics is  \textit{view parallelization}, which is the ability to restore each view parallelly. This is contrary to the approach taken by LFNet \cite{wang2018lfnetTIP}, which used bidirectional recurrent units to model the inter-view dependencies. But as pointed out in \cite{zhang2019residual}, the dependencies were not modeled well, and the algorithm was slow because of sequential processing over time. So if possible, recurrent units should not be used in L3Fnet for potential \textit{view parallelization}. View parallelization also has the advantage that we can selectively reconstruct only some of the desired LF, without reconstructing the full Light Field, saving time and memory. Lastly, we do not want different model architectures with multiple sets of weights for different views. This was adopted by \cite{zhang2019residual} to retrain their network for each Sub-Aperture Image (SAI) and consequently have a separate set of weights for each LF view. We, on the other hand, would like to have a \textit{single architecture with shared weights} for all LF views. We have summarized this discussion in Table. \ref{table:guidlines}.
\begin{table}[t!]
\caption{Comparison of architectural features present in L3Fnet.}
\resizebox{1\linewidth}{!}{
\setlength{\tabcolsep}{1.5pt}
\def\arraystretch{1.4}
\begin{tabular}{c|c|c|c|c}
 \hline & \textbf{No strong} & \textbf{View} & \textbf{View} & \textbf{Shared weights} \\
 &\textbf{priors}&\textbf{coherence} & \textbf{Parallelization} & \textbf{\& architecture} \\ \hline \hline
\textbf{Wanner \etal} \cite{wanner2012LF} & \xmark & \xmark  & \cmark & \cmark \\ \hline
\textbf{Yoon \etal} \cite{LFviewCoherence} & \cmark & \xmark & \cmark & \cmark \\ \hline
\textbf{LFNet} \cite{wang2018lfnetTIP} & \cmark & \cmark & \xmark & \cmark \\ \hline
\textbf{Zhang \etal} \cite{zhang2019residual} & \cmark & \cmark & \cmark & \xmark \\ \hline
\textbf{Ours L3Fnet} & \cmark & \cmark & \cmark & \cmark \\ \hline

\end{tabular}
}
\label{table:guidlines}
\end{table}

We have incorporated these features in the design of L3Fnet, see Fig. \ref{fig:architecture1}, to the extent possible without harming the primary purpose of enhancing low-light LFs. L3Fnet admits no strong assumption about the target scene because it requires no extra information other than the low-light LF for enhancement. To achieve view-coherence, we have a Global Representation Block (GRB), which operates on the entire LF to obtain a latent representation that encodes the LF epipolar geometry. 
The View Reconstruction Block (VRB) of L3Fnet then uses the global information encoded by GRB to restore each LF view independent of other LF view. Thus we incorporate the view parallelization property. 

\textcolor{black}{Splitting the L3Fnet architecture into two stages is very beneficial. If all the SAIs are pushed into a CNN model, the model may be hard to train and test because of huge computational requirements. This is because LFs are much larger than even HD DSLR images. We, therefore, decomposed the L3Fnet architecture into two stages. Stage-I looks at all the SAIs and provides a latent representation to Stage-II, to aid individual restoration of each SAI. This architecture thereby prevents the parameter and computational complexity explosion.}

\textcolor{black}{Restoring low-light light fields subsumes the task of denoising, color restoration and preserving the LF geometry. The proposed GRB leverages information from multiple views to aid denoising and in preserving  high-frequency details and the LF geometry. We shall verify these claims in the experimental section where we show that by using GRB depth estimation is benefitted, PSNR/SSIM improves, and the high-frequency details are preserved well.}

\subsection{Network Architecture}
The architecture of L3Fnet is shown in Fig. \ref{fig:architecture1}. We now describe each of its component in detail below.

\textbf{Stage-I Global Representation Block (GRB):} Given an input LF $\mathcal{L}^{low}\in \mathbf{R}^{U\times V\times W\times H\times 3}$, we first stack all the views across channels to obtain $\mathcal{\hat{I}}\in \mathbf{R}^{W\times H\times 3UV}$. Such a represenation can be processed using a 2D Convolutional Neural Network (CNN) to obtain a low-dimensional LF representation useful for any down-stream task.

In the Global Representation Block (GRB) a convolutional layer $\mathcal{H}^{GRB}$ is used to reduce the input channel dimensions of $\mathcal{\hat{I}}$, see Table \ref{Table:architecturevalues}.
\begin{equation}
    {J}_{0} = \mathcal{H}^{GRB}\left(\mathcal{\hat{I}} \right)
\end{equation}
To now extract useful information, we further process this representation using $M$ (fixed at $4$) residual blocks~\cite{ResidualBlock} to obtain the global representation
\begin{equation}
    {J}_{m} = \mathcal{B}^{GRB}_{m}\left({J}_{m-1} \right), \quad m\in\{1,\hdots,M\},
\end{equation}
where $\mathcal{B}^{GRB}_{m}$ denotes the m$^{th}$ residual block in the GRB.
This feature map is then fed to the final convolutional layer $\mathcal{T}^{GRB}$
as a post-processing step
\begin{equation}
    {J}_{M+1} = \mathcal{T}^{GRB}\left({J}_{M} \right).
\end{equation}
By processing all the views together using a CNN architecture the network captures the implicit LF structure such as disparity, sub-pixel information etc., relevant for the final task at hand.
This global representation is then used by the view reconstruction block as an augmented information to reconstruct any of the input view.

\textbf{Stage-II View Reconstruction Block (VRB):}
While the GRB representation is too coarse and common to all SAIs, for better reconstruction of each SAI we explicitly use its immediate neighbours. Formally, to reconstruct a particular view $\mathcal{I}(u,v) \in \mathbf{R}^{W\times H\times 3}$, its neighbours $\{\mathcal{I}(u,v-1),\mathcal{I}(u-1,v),\mathcal{I}(u,v+1),\mathcal{I}(u+1,v)\}$ are stacked across channels to obtain $\mathcal{\Tilde{I}}\in \mathbf{R}^{W\times H\times 15}$. The view restoration process begins by processing $\mathcal{\Tilde{I}}$ using a convolutional layer $\mathcal{H}^{VRB}$ to obtain the feature map $C_0$:
\begin{equation}
    {C}_{0} = \mathcal{H}^{VRB}\left(\mathcal{\Tilde{I}} \right).
\end{equation}
The global feature map is concatenated with this before being fed to $N$ (fixed at $6$) residual blocks:
\begin{align}
    {C}_{1} &= \mathcal{B}^{VRB}_{1}\left({C}_{0} \oplus	 {J}_{M+1} \right) \\
    {C}_{n} &= \mathcal{B}^{VRB}_{n}\left({C}_{n-1} \right), \quad\quad n\in\{2,\hdots,N\},
\end{align}
where $\mathcal{B}^{VRB}_{n}$ denotes the n$^{th}$ resblock in the view reconstruction block and $\oplus$ denotes the concatenation operation. The resblocks in GRB and VRB branches share the same structure. 
Finally, the output feature map is passed through a transposed convolution block $\mathcal{T}^{VRB}$ followed by a long skip connection from the input SAI to restore the SAI
\begin{equation}
    {I}_{out}(u,v) = \mathcal{T}^{VRB}\left({C}_{N} \right) + I(u,v).
\end{equation}

Many works \cite{rcan,rir} have reported difficulty in training very deep networks by simply stacking residual blocks. Therefore, to stabilize the training and optimization of deep networks, adding a long/global skip connection has become a standard practice\cite{rcan,attention_2019_CVPR}. Further, unlike the short skip connections which help in propagating finer details, long skip connection helps to transmit coarse level details which are crucial for image restoration tasks.

\begin{figure}
     \centering
     \includegraphics[width=\linewidth]{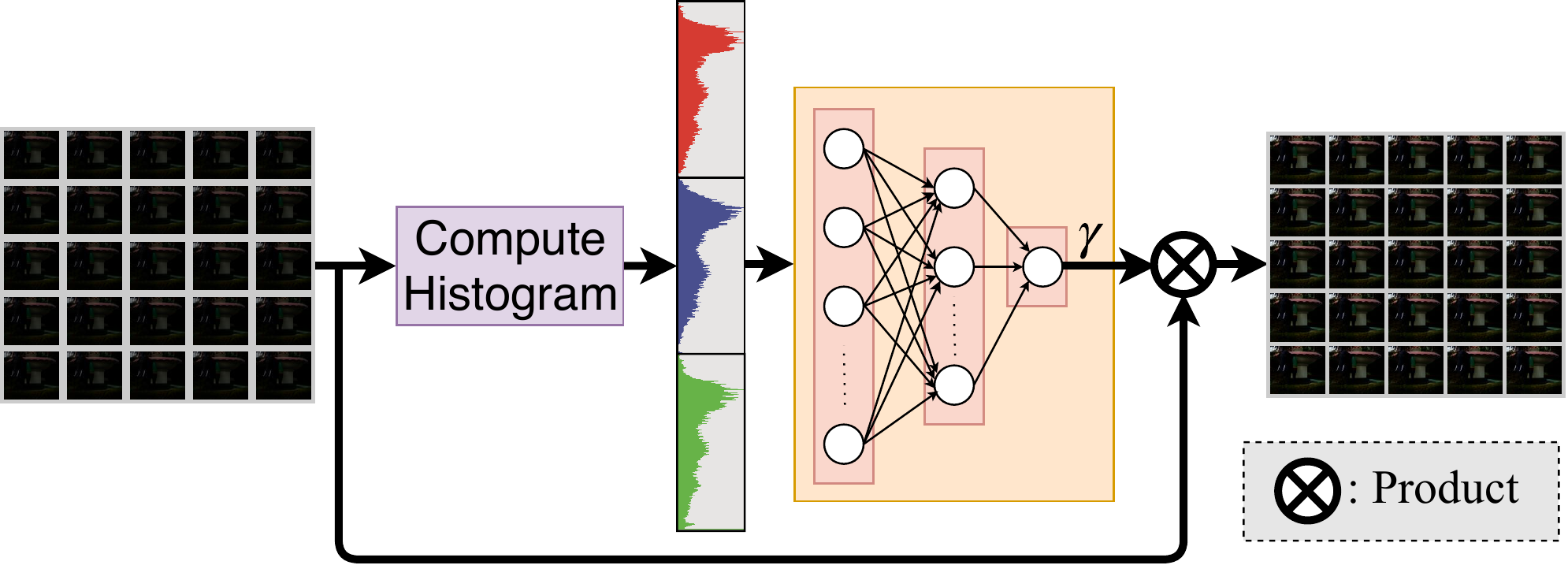}
     \caption{The Histogram Module computes the RGB histogram of a low-light LF and outputs an amplification factor $\gamma$. Normalizing the dark LF with this module allows L3Fnet to process images of varying low-light levels mitigating the over/under saturation problem.}
     \label{fig:histogram_module}
     \vspace{-0.5cm}
 \end{figure}

\begin{table}[t!]
    \centering
\caption{L3Fnet architecture summary. $K$ stands for Kernel Size, $S$ for Stride, Out for number of channels in Convolutional layers.}
\label{tab:my-table}
\bgroup
\def\arraystretch{1.4}
\begin{tabular}{|c|c|c|c|c|c|}
\hline
\textbf{Branch} & \textbf{Name} & \textbf{Type} & \textbf{K} & \textbf{S} & \textbf{Out} \\ \hline \hline
\multirow{4}{*}{\textbf{\bgroup
\def\arraystretch{1} \begin{tabular}[c]{@{}c@{}}\\ \\ \\ Stage-I\\ \\ Global Representation \\ Block\end{tabular}\egroup}} & \multirow{2}{*}{$\mathcal{H}^{GRB}$} & Conv-2D & 7 & 1 & 64 \\ 
 &  & Conv-2D & 3 & 2 & 128 \\ \cline{2-6} 
 & \multirow{4}{*}{$\mathcal{B}^{GRB}$} & Res-Block & 3 & 1 & 128 \\ 
 &  & Res-Block & 3 & 1 & 128 \\
 &  & Res-Block & 3 & 1 & 128 \\
 &  & Res-Block & 3 & 1 & 128 \\ \cline{2-6} 
 & $\mathcal{T}^{GRB}$ & Conv-2D & 1 & 1 & 64 \\ \hline \hline
\multirow{5}{*}{\textbf{\bgroup
\def\arraystretch{1} \begin{tabular}[c]{@{}c@{}}\\ \\ \\ \\ \\ Stage-II\\ \\ View Reconstruction \\ Block\end{tabular}\egroup}} & \multirow{2}{*}{$\mathcal{H}^{VRB}$} & Conv-2D & 7 & 1 & 15 \\ 
 &  & Conv-2D & 3 & 2 & 64 \\ \cline{2-6} 
 & \multirow{6}{*}{$\mathcal{B}^{VRB}$} & Res-Block & 3 & 1 & 128 \\
 &  & Res-Block & 3 & 1 & 128 \\
 &  & Res-Block & 3 & 1 & 128 \\
 &  & Res-Block & 3 & 1 & 128 \\
 &  & Res-Block & 3 & 1 & 128 \\
 &  & Res-Block & 3 & 1 & 128 \\ \cline{2-6} 
 & \multirow{2}{*}{$\mathcal{T}^{VRB}$} & Transpose-2D & 2 & 2 & 128 \\ 
 &  & Conv-2D & 3 & 1 & 3 \\ \hline \hline
\multirow{4}{*}{\bgroup
\def\arraystretch{1} \textbf{\begin{tabular}[c]{@{}c@{}} Histogram\\ Module\end{tabular}\egroup}} & \multirow{4}{*}{$\mathcal{H}$} & FC & - & - & 200 \\ 
 &  & FC & - & - & 100 \\ 
 &  & FC & - & - & 50 \\ 
 &  & FC & - & - & 1 \\ \hline
\end{tabular}
\egroup
\label{Table:architecturevalues}
\vspace{-0.5cm}
\end{table}

\textbf{Histogram Module:} The typical illuminance of the scenes in our dataset varies from 0.1 lux$-$20 lux. This causes a lot of variation in the input data distribution. A simple workaround is to scale the input image with an appropriate amplification factor $\gamma$ which controls the brightness of the image. Chen \etal~\cite{chen2018learning2seeindark} requires a manual input for $\gamma$. \textcolor{black}{Contrary to this, we automate the process by pre-processing the low-light LF with the Histogram Module shown in Fig. \ref{fig:histogram_module}}.
 
To estimate the appropriate amplification factor $\gamma$, we make use of the RGB histogram of the input low-light LF image $\mathcal{L}^{low}$. Formally, let  $h^c \in \mathbf{R}^L$ be the normalized $L$-bin histogram of the input image corresponding to the color channel $c \in \{R,G,B\}$.
This histogram is then fed to $3$ fully connected layers $\mathcal{H}$ to compute the amplification factor $\gamma$. $\gamma$ is then used to appropriately scale the input low-light LF:
\begin{align}
\gamma &= \mathcal{H}\left(h^R \oplus h^G \oplus h^B\right) \\
\mathcal{L}^{norm} & = \gamma \times \mathcal{L}^{low},
\end{align}
where $\oplus$ denotes concatenation.
\textcolor{black}{During the training phase, we only need to update the weights of the MLP to obtain the required  $\gamma$ and this requires back-propagation to happen only till the first layer of MLP and not beyond.} Scaling the input image with the amplification factor acts as an important pre-processing step. Without this pre-scaling step the restored LF would sometimes tend to be over or under-saturated for light levels of different gradations.

\textbf{Loss Function:}
For the proposed solution we try to minimise the following loss function:
\begin{align}
     Loss = &  ~\alpha_1 \sum_{k=1}^{K} || I^{out}(u_k,v_k) - I^{GT}(u_k,v_k)||_1 \nonumber \\ 
     &~+~\alpha_2  \sum_{k=1}^{K}  CX\left(I^{out}(u_k,v_k) , I^{GT}(u_k,v_k)\right)  \nonumber \\
     &~+~\lambda  ||w||_1,
     \label{eq:loss}
\end{align}
where $CX(\cdot,\cdot)$ is the contextual loss \cite{contextualloss}, $w$ denotes the model parameters and $I^{GT}(u_k,v_k)$ is the ground-truth $(u_k,v_k)^{th}$ view of the LF having $K$ views.
The first component in our loss function performs a dense matching (i.e., pixel-to-pixel) between the restored and the ground truth LF to recover the structure and color information. Since recovering the precise color information from a low-light signal is a  hard problem, a small shift from actual hue information is expected, no matter how long the network is trained. So, in a bid to prevent our loss function from having large errors due to this mismatch, we chose the sum-of-absolute-deviation ($L1$ loss ) over other alternatives such as $L2$ loss to perform the dense pixel matching. But as our dataset has small misalignment, directly using the $L1$ loss would lead to blurred outputs. To prevent this we additionally used a low weighted contextual loss to handle this problem. Our ablation studies nicely demonstrate the importance of both components in our loss function.



\section{Pseudo-LF: Extending L3Fnet to single-frame DSLR images}
\label{sec:pseudolf}

\begin{figure}[t!]
    \centering
    \includegraphics[width=\linewidth]{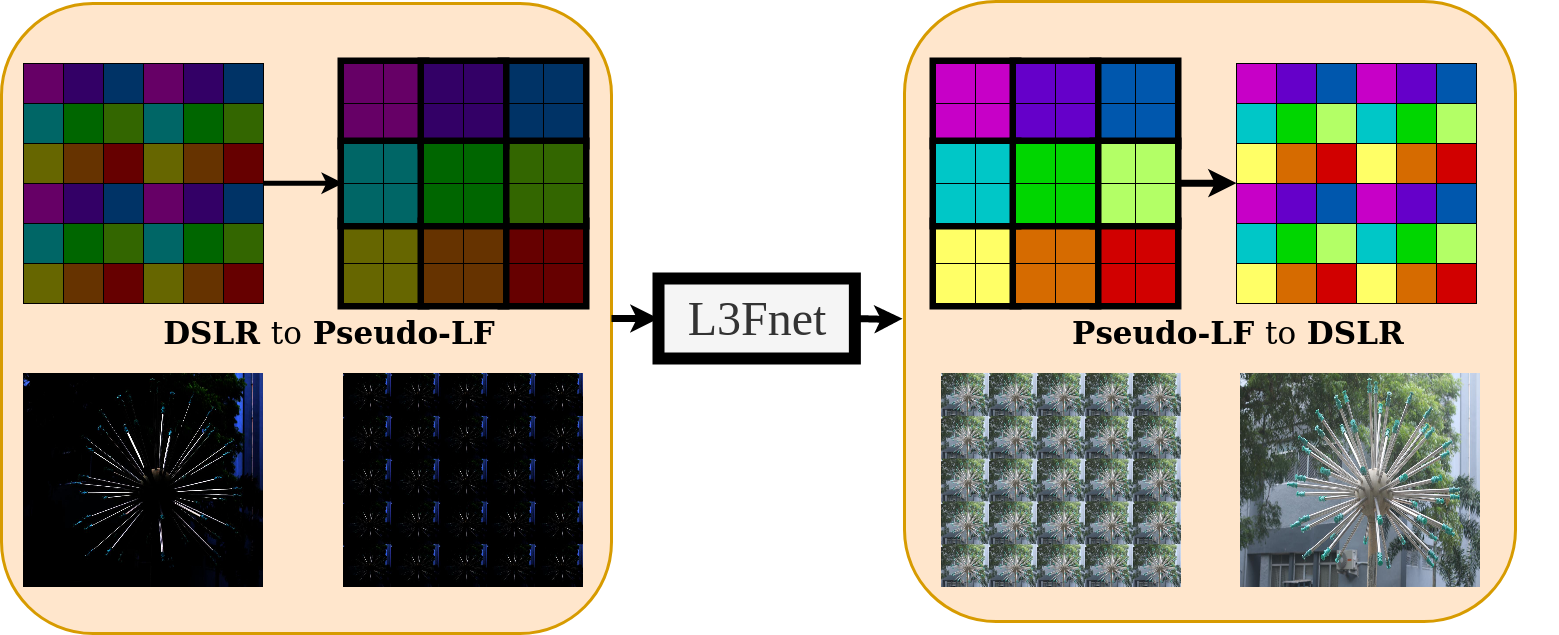}
    \caption{Pseudo-LF: Extending the proposed L3Fnet to single-frame DSLR images. The proposed Pseudo-LF transformation coverts a 2D DSLR image into a 4D pseudo-LF. Pseudo-LF transformation allows a CNN designed for LF to also process single-frame DSLR images.}
    \label{fig:pseudoLFarchi}
    \vspace{-0.5cm}
\end{figure}

\begin{figure*}[t!]
	\scriptsize
	\centering
	\renewcommand{\arraystretch}{1.5}
	\setlength{\tabcolsep}{1pt}
			\begin{tabular}{ccccccc} \includegraphics[width=0.14 \linewidth, ]{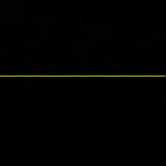} & \includegraphics[width=0.14 \linewidth, ]{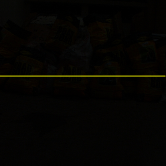} & \includegraphics[width=0.14 \linewidth, ]{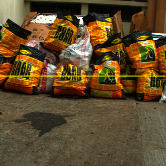} && \includegraphics[width=0.14 \linewidth, ]{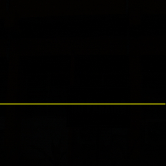} & \includegraphics[width=0.14 \linewidth, ]{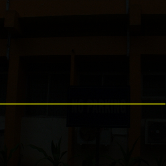} & \includegraphics[width=0.14 \linewidth, ]{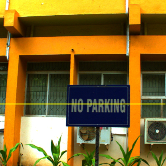} \\	\includegraphics[width=0.14 \linewidth, ]{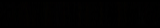} & \includegraphics[width=0.14 \linewidth, ]{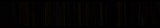} & \includegraphics[width=0.14 \linewidth, ]{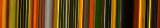} && \includegraphics[width=0.14 \linewidth, ]{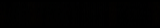} & \includegraphics[width=0.14 \linewidth, ]{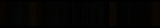} & \includegraphics[width=0.14 \linewidth, ]{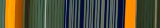} 
			\vspace{-0.2cm}
			 \\
			 L3F-100 input & L3F-20 input & Ground Truth && L3F-100 input & L3F-20 input & Ground Truth
			 \vspace{0.1cm}
			 \\	\multicolumn{7}{c}{\textbf{\normalsize{--- --- --- --- Results on L3F-100 Dataset --- --- --- ---}}} \\
			 
			 LFBM5D \cite{bm5d} & PBS \cite{r1_2018_acm} & RetinexNet \cite{r3_2018_bmvc} & SGN \cite{gu2019self} & DID \cite{2019DID} & SID \cite{chen2018learning2seeindark} & Ours \\
			 \includegraphics[width=0.14 \linewidth, ]{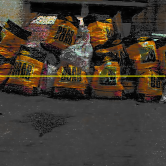} &
			 \includegraphics[width=0.14 \linewidth, ]{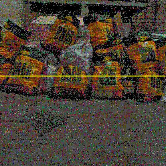} &
			 \includegraphics[width=0.14 \linewidth, ]{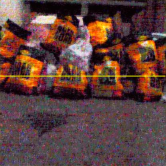} &
			 \includegraphics[width=0.14 \linewidth, ]{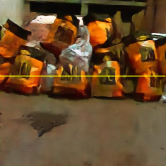} &
			 \includegraphics[width=0.14 \linewidth, ]{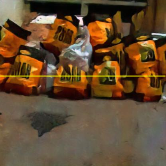} &
			 \includegraphics[width=0.14 \linewidth, ]{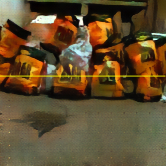} &
			 \includegraphics[width=0.14 \linewidth, ]{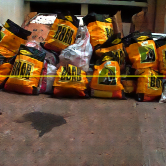}\\
			 \includegraphics[width=0.14 \linewidth, ]{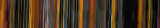} &
			 \includegraphics[width=0.14 \linewidth, ]{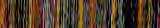} &
			 \includegraphics[width=0.14 \linewidth, ]{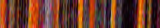} &
			 \includegraphics[width=0.14 \linewidth, ]{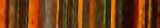} &
			 \includegraphics[width=0.14 \linewidth, ]{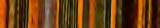} &
			 \includegraphics[width=0.14 \linewidth, ]{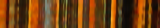} &
			 \includegraphics[width=0.14 \linewidth, ]{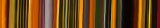} \vspace{-0.2cm}\\
			 18.52/0.47 &14.50/0.32&18.64/0.41&21.06/0.53&21.12/0.54& 21.22/0.50 & \textbf{21.44/0.62}
			 \\
			 
			 \includegraphics[width=0.14 \linewidth, ]{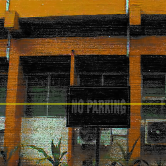} &
			 \includegraphics[width=0.14 \linewidth, ]{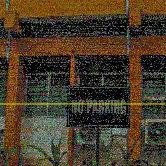} &
			 \includegraphics[width=0.14 \linewidth, ]{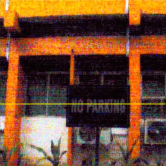} &
			 \includegraphics[width=0.14 \linewidth, ]{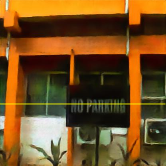} &
			 \includegraphics[width=0.14 \linewidth, ]{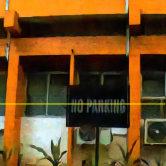} &
			 \includegraphics[width=0.14 \linewidth, ]{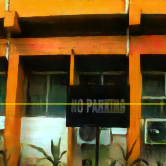} &
			 \includegraphics[width=0.14 \linewidth, ]{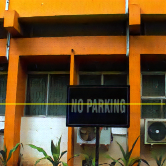}\\
			 \includegraphics[width=0.14 \linewidth, ]{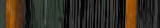} &
			 \includegraphics[width=0.14 \linewidth, ]{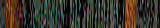} &
			 \includegraphics[width=0.14 \linewidth, ]{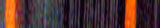} &
			 \includegraphics[width=0.14 \linewidth, ]{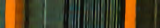} &
			 \includegraphics[width=0.14 \linewidth, ]{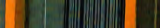} &
			 \includegraphics[width=0.14 \linewidth, ]{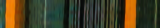} &
			 \includegraphics[width=0.14 \linewidth, ]{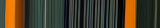}
			 \vspace{-0.2cm}\\
			 13.61/0.32 &12.13/0.30&17.09/0.39&18.39/0.62&17.70/0.60& 18.44/0.59 & \textbf{19.41/0.68}
			 
			 \vspace{0.1cm}
			 \\	\multicolumn{7}{c}{\textbf{\normalsize{--- --- --- --- Results on L3F-20 Dataset --- --- --- ---}}} \\
			 
			 LFBM5D \cite{bm5d} & PBS \cite{r1_2018_acm} & RetinexNet \cite{r3_2018_bmvc} & SGN \cite{gu2019self} & DID \cite{2019DID} & SID \cite{chen2018learning2seeindark} & Ours \\
			 \includegraphics[width=0.14 \linewidth, ]{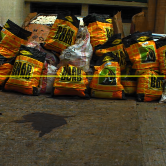} &
			 \includegraphics[width=0.14 \linewidth, ]{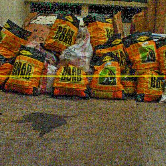} &
			 \includegraphics[width=0.14 \linewidth, ]{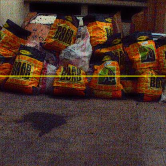} &
			 \includegraphics[width=0.14 \linewidth, ]{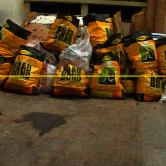} &
			 \includegraphics[width=0.14 \linewidth, ]{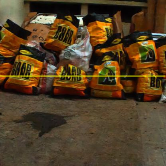} &
			 \includegraphics[width=0.14 \linewidth, ]{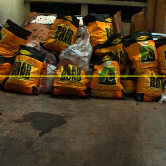} &
			 \includegraphics[width=0.14 \linewidth, ]{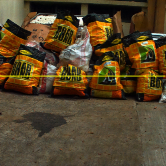}\\
			 \includegraphics[width=0.14 \linewidth, ]{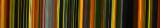} &
			 \includegraphics[width=0.14 \linewidth, ]{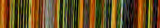} &
			 \includegraphics[width=0.14 \linewidth, ]{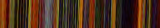} &
			 \includegraphics[width=0.14 \linewidth, ]{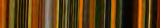} &
			 \includegraphics[width=0.14 \linewidth, ]{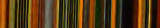} &
			 \includegraphics[width=0.14 \linewidth, ]{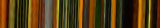} &
			 \includegraphics[width=0.14 \linewidth, ]{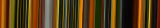}
			 \vspace{-0.2cm}\\
			 21.64/0.67 &16.92/0.42&18.99/0.50&22.21/0.65&21.93/0.66&22.22/0.64 & \textbf{24.80/0.71}
			 \\
			 
			 \includegraphics[width=0.14 \linewidth, ]{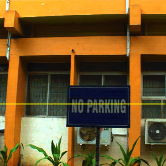} &
			 \includegraphics[width=0.14 \linewidth, ]{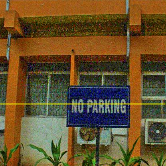} &
			 \includegraphics[width=0.14 \linewidth, ]{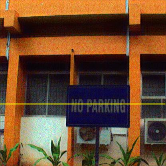} &
			 \includegraphics[width=0.14 \linewidth, ]{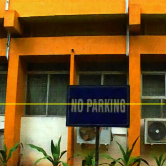} &
			 \includegraphics[width=0.14 \linewidth, ]{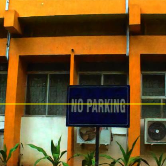} &
			 \includegraphics[width=0.14 \linewidth, ]{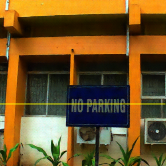} &
			 \includegraphics[width=0.14 \linewidth, ]{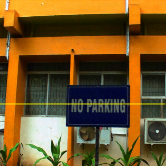}\\
			 \includegraphics[width=0.14 \linewidth, ]{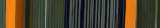} &
			 \includegraphics[width=0.14 \linewidth, ]{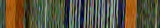} &
			 \includegraphics[width=0.14 \linewidth, ]{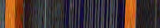} &
			 \includegraphics[width=0.14 \linewidth, ]{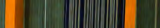} &
			 \includegraphics[width=0.14 \linewidth, ]{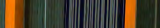} &
			 \includegraphics[width=0.14 \linewidth, ]{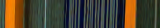} &
			 \includegraphics[width=0.14 \linewidth, ]{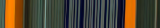}
			 \vspace{-0.2cm}\\
			 21.28/0.75 &14.30/0.45&17.49/0.53&20.45/0.76&19.79/0.75& 20.86/0.76 & \textbf{23.11/0.83}

\end{tabular}
\caption{\textcolor{black}{(\textit{Best viewed when zoomed in}) Visual and epipolar comparisons of the restored Light Fields. For the very low light case of the L3F-100 dataset, all the existing methods obtain very blurry restorations with much loss in details. Additionally, they sometimes have a large amount of noise. The proposed L3Fnet, however, overcomes these limitations to a good extent.}}
\label{fig:vis_compare}
\end{figure*}

The proposed L3Fnet has been specifically engineered for low-light Light Fields and thus cannot enhance low-light single-frame DSLR images. On the contrary, we will show that (see Fig. \ref{fig:vis_compare}) methods designed for single-frame DSLR images are also inappropriate for low-light LF. Thus, we introduce a very simple and light-weight transformation called \textit{pseudo-LF transform}, which allows L3Fnet to enhance even DSLR images. This is illustrated in Fig. \ref{fig:pseudoLFarchi}.

The first step is to convert the the DSLR image into \textit{pseudo-LF}. To do this, the high-resolution DSLR image is divided into blocks of $B \times B$ pixels. The $i^{th} $ pseudo SAI, where $i \in [1,\hdots, B^2]$, is then obtained by collecting together the $i^{th}$ pixel from each $B \times B$ block.
We use the term \textit{pseudo} because the resulting pseudo-LF SAIs have no real disparity and only happen to be shifted subsampled versions of the input high-resolution DSLR image.
The converted pseudo-LF, and not the original DSLR image, is then processed by the L3Fnet to obtain well-lit pseudo-LF. The resulting well-lit LF is finally converted back to well-lit DSLR image by reversing the sampling process described just now. Recently Gu \etal~\cite{gu2019self} and Shi \etal~\cite{shi2016real} have also used a similar pixel shuffling technique for better performance on DSLR images but do not allow CNNs designed for LF to process DSLR images.
The proposed pseudo-L3Fnet pipeline may appear a workaround to fit L3Fnet in the DSLR image framework, but it has some inherent advantages. \textcolor{black}{It allows a large receptive field which helps the restoration by gathering more contextual information. For example, modern CNN architectures mostly use a 3$\times$3 convolution kernel that operates directly over a DSLR image and hence it has a receptive field of just 3$\times$3 for one convolution layer. However, we first convert the DSLR image into $B^2$ pseudo-SAIs. Stage-I of L3Fnet then collects all the $B^2$ pseudo-SAIs and subsequently performs 3$\times$3 convolution with a stride of 1. This is equivalent to a receptive field of $3B \times 3B$ and with a stride of $B$ in the DSLR image. In our experiments, we use $B=10$ and so the effective receptive field is $30 \times 30$. To have the same effect by convolving directly on the DSLR image, requires a $30 \times 30$ convolution kernel with a stride of $10$. However, such large kernel sizes and strides have become obsolete in modern CNN architectures and deep learning libraries are mostly optimized for $3\times 3$ kernel size \cite{shufflenetv2}.
Thus, even though L3Fnet has been designed for LF, by using the pseudo-LF transformation it can be effectively used for single image enhancement with the advantage of having a  large receptive field.}

\begin{table}[t!]
\centering
\caption{PSNR(dB)/SSIM comparison of the proposed L3Fnet with recent works on the L3F-20, L3F-50 and L3F-100 datasets. The effectiveness of the proposed L3Fnet is more evident for the extremely low light images in the L3F-100 dataset. \textbf{Bold} indicates best value and \underline{underline} indicates second-best.
}
\renewcommand{\arraystretch}{1.5}
\begingroup
\setlength{\tabcolsep}{8pt}
    
\begin{tabular}{c|ccc}
\hline
 &
\textbf{L3F-20} & \textbf{L3F-50} & \textbf{L3F-100} \\ \hline \hline
\textbf{LFBM5D \cite{bm5d}} & $24.48/\underline{0.79}$ & $20.94/0.64$ & $18.14/0.46$\\
\textcolor{black}{\textbf{PBS \cite{r1_2018_acm}}}
& $20.80/0.68$ & $16.48/0.53$ & $13.94/0.38$\\
\textcolor{black}{\textbf{RetinexNet \cite{r3_2018_bmvc}}} & $21.82/0.72$ & $18.98/0.59$ & $17.8/0.41$\\
\textcolor{black}{\textbf{DID \cite{2019DID}}} & 24.09/0.78 & 22.63/\underline{0.68} & 20.68/\underline{0.61}\\
\textcolor{black}{\textbf{SGN \cite{gu2019self}}} & $24.10/0.76$ & $22.18/0.67$ & $20.70/0.59$\\
\textbf{SID \cite{chen2018learning2seeindark}} & $\underline{24.53}/0.76$ & $\underline{22.87}/0.66$ & $\underline{20.75}/0.58$\\ \hline
\textbf{Our L3Fnet} & $\mathbf{25.25/0.82}$ & $\mathbf{23.67/0.74}$ & $\mathbf{22.61/0.70}$ \\ \hline
\end{tabular}

\endgroup
\label{table:quant_comp}
\end{table}

\section{Experimental Results}

\subsection{Implementation details} 
\textbf{Data Pre-processing: }Recent works on single-frame low-light image restoration \cite{burstphoto, chen2018learning2seeindark}  chose to work with raw format. This is because the raw data directly stores the signal measured by the sensor and so is immune to any loss of information due to the Image Signal Processing pipeline of the camera. 
However, there are few difficulties with directly working on raw LF obtained from Lytro Illum. The raw LF requires a significant amount of pre-processing \cite{dansereau2013decoding} which involves tasks such as alignment rectification, hexagonal to orthogonal lattice conversion, de-vignetisation, and demosaicing \cite{dansereau2013decoding}. We wanted L3Fnet to focus on low-light restoration rather than learning these difficult geometric transformations for which already efficient techniques exist. So we chose to work with decoded JPEG images with the highest quality factor. From the training time and resource utilization perspective also this is beneficial because the JPEG compression reduced the decoded image size form massive $400-500 MB$ to $40-50 MB$, which is still much larger than raw single-frame DSLR data.

Lytro Illum captures 225 views of a scene arranged in a $15 \times 15$ grid. As the peripheral views are affected by ghosting and vignetting effect \cite{zhang2019residual, wang2018lfnetTIP}, we restore only the central $8\times 8$ views. \textcolor{black}{ For the SAIs lying at the boundary of this $8 \times 8$ grid, the adjoining SAIs are taken from the $9^{th}$ row and column.}

\begin{figure}[t!]
	\scriptsize
	\centering
	\setlength{\tabcolsep}{1pt}
	\resizebox{.5\textwidth}{!}{
		\begin{tabular}{cccc}
			Scene I &  & SID \cite{chen2018learning2seeindark} & Proposed   \\
			\includegraphics[width=0.25\linewidth]{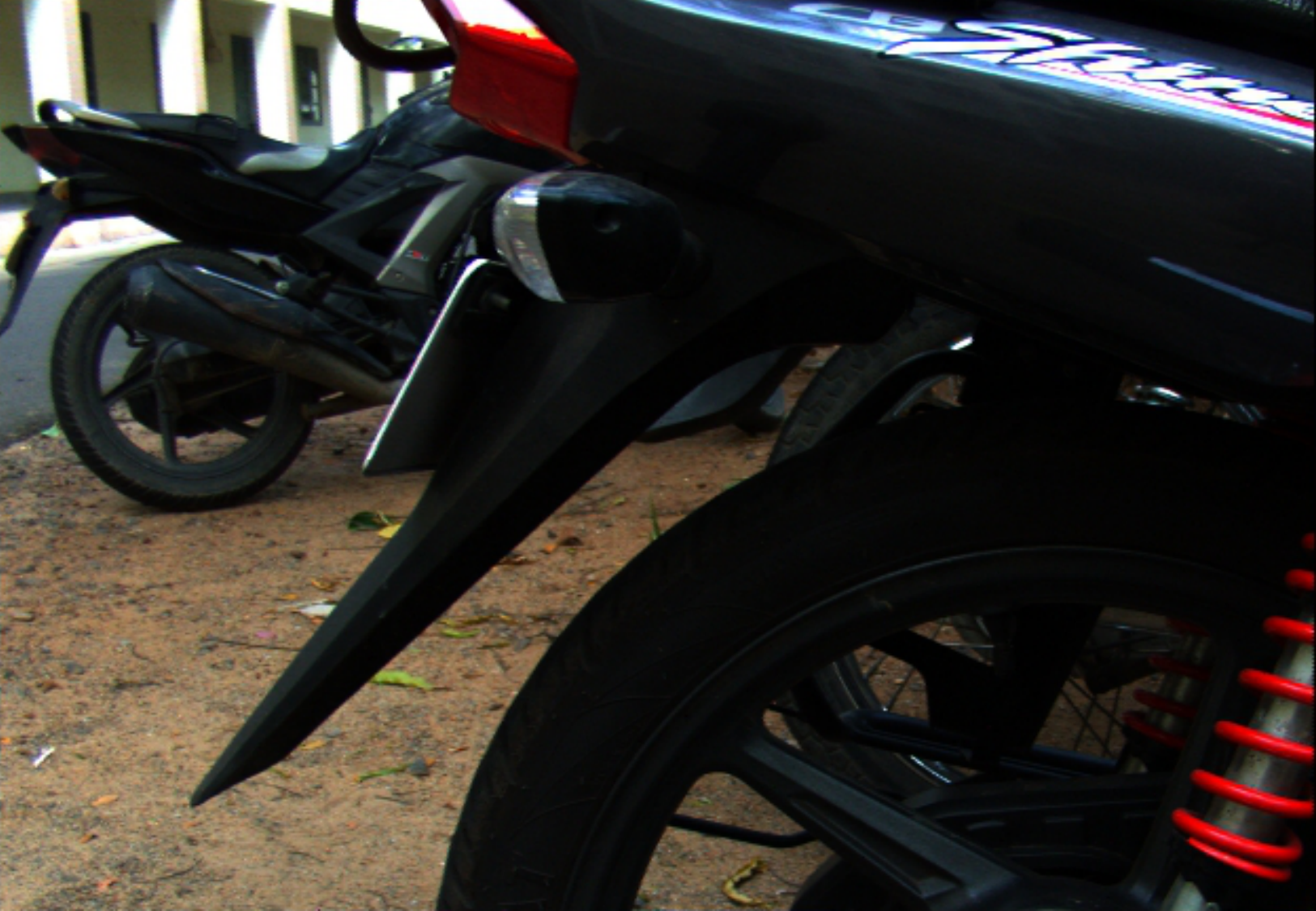} &
			\rotatebox[origin=lt]{90}{L3F-20}  & 
			\includegraphics[width=0.25\linewidth]{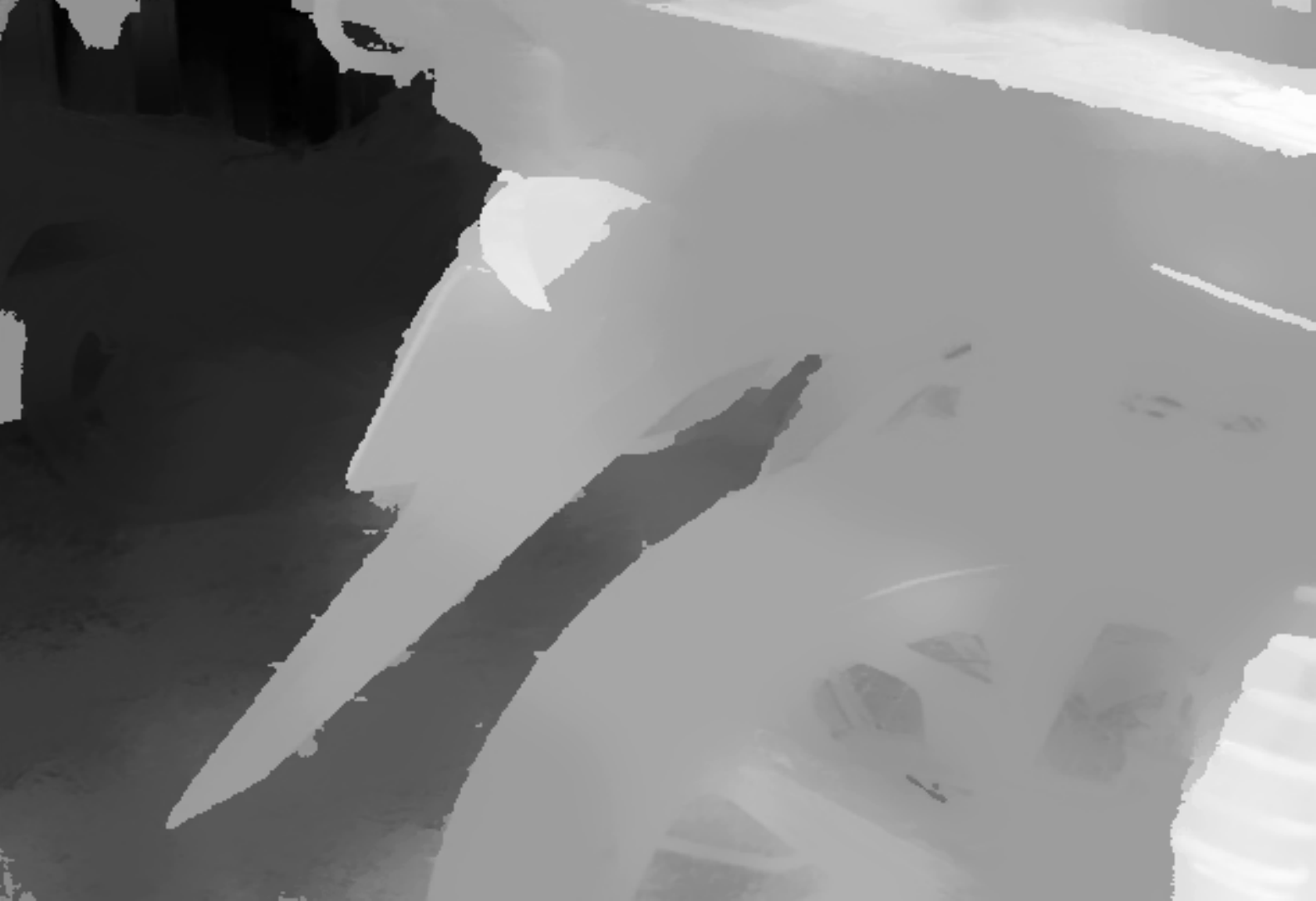} \hspace{-0.15cm} &
			\includegraphics[width=0.25\linewidth]{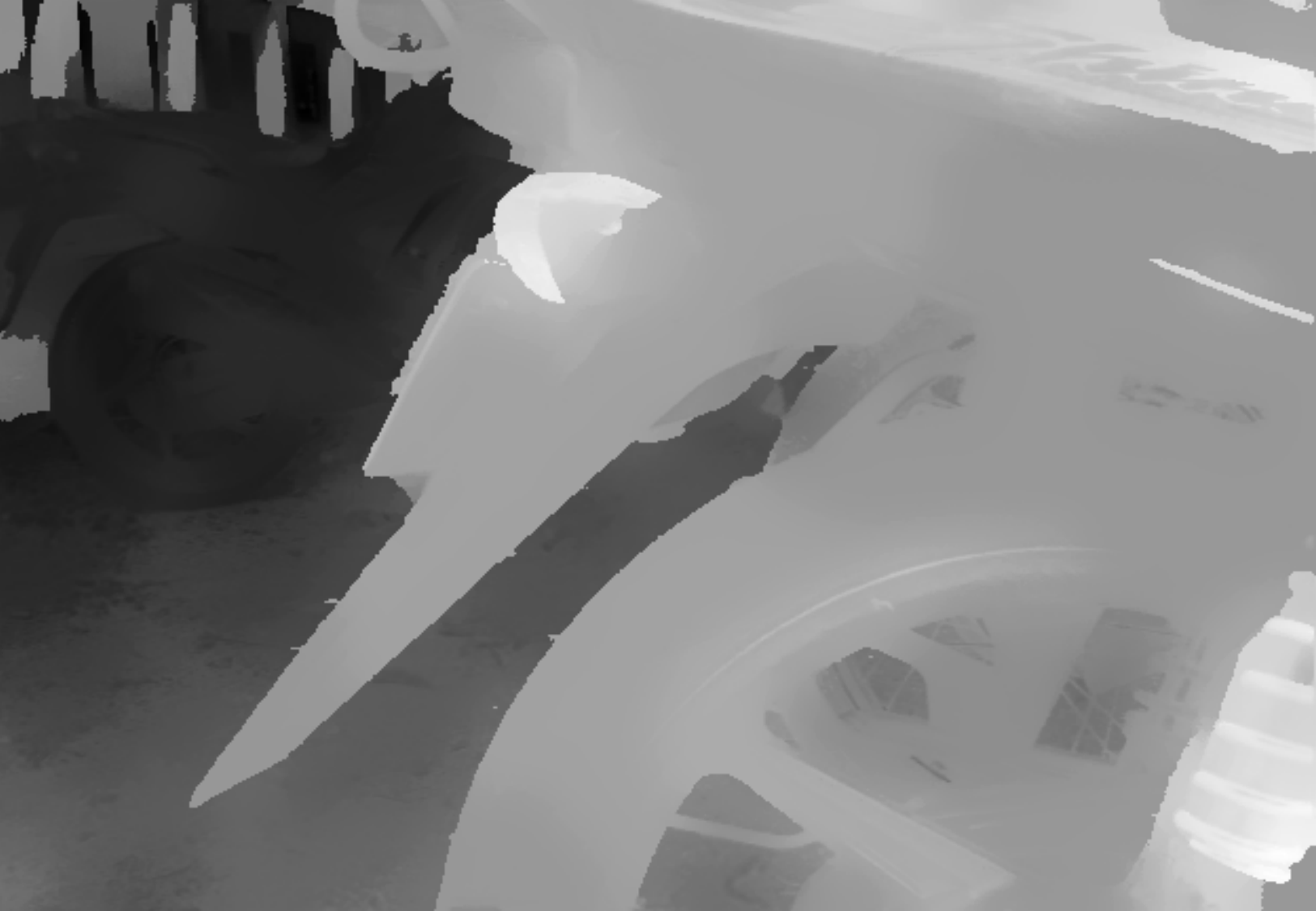}  \vspace{-0.05cm} \\
			
			\includegraphics[width=0.25\linewidth]{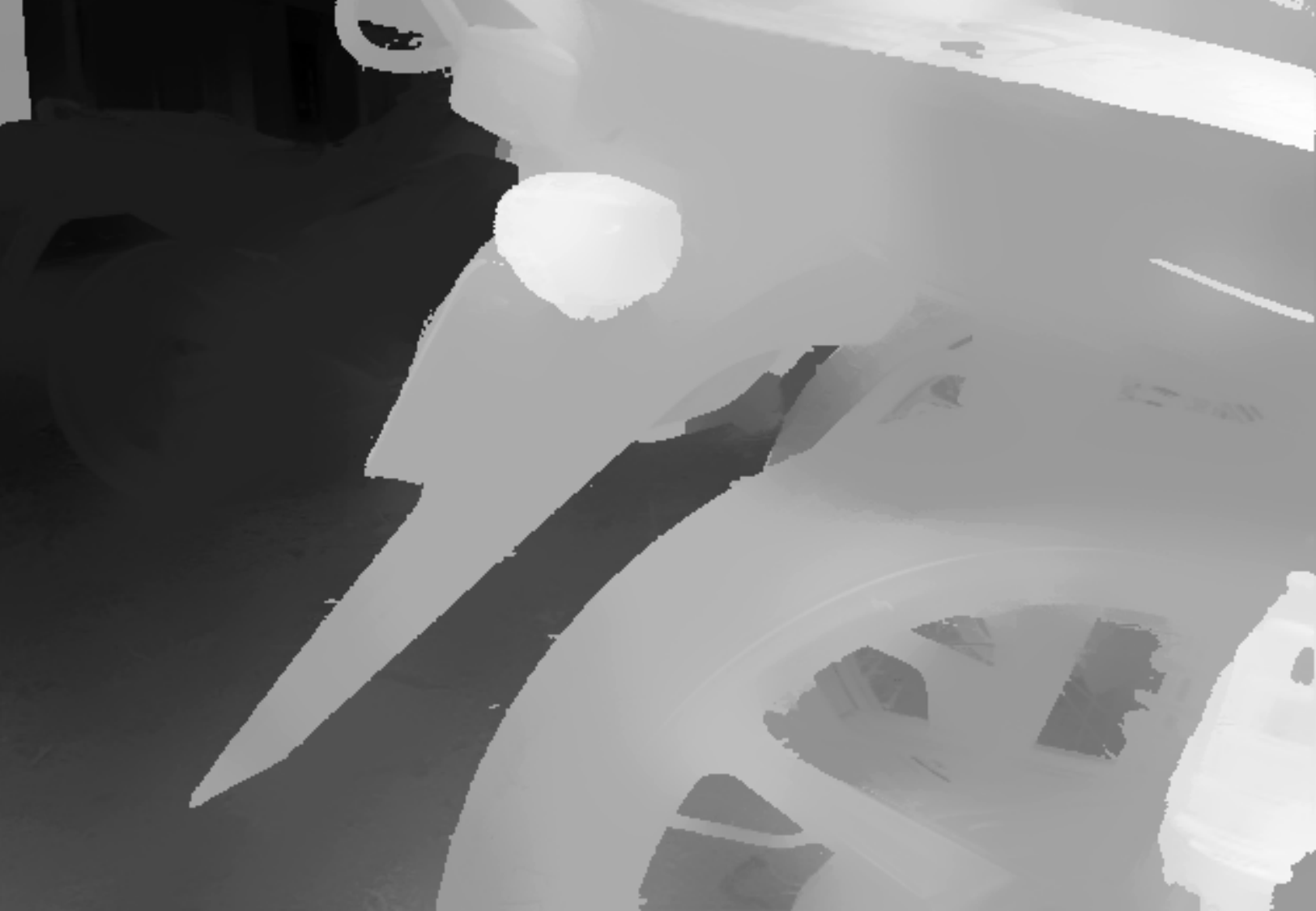}  &
            \rotatebox[origin=lt]{90}{L3F-100} &
			\includegraphics[width=0.25\linewidth]{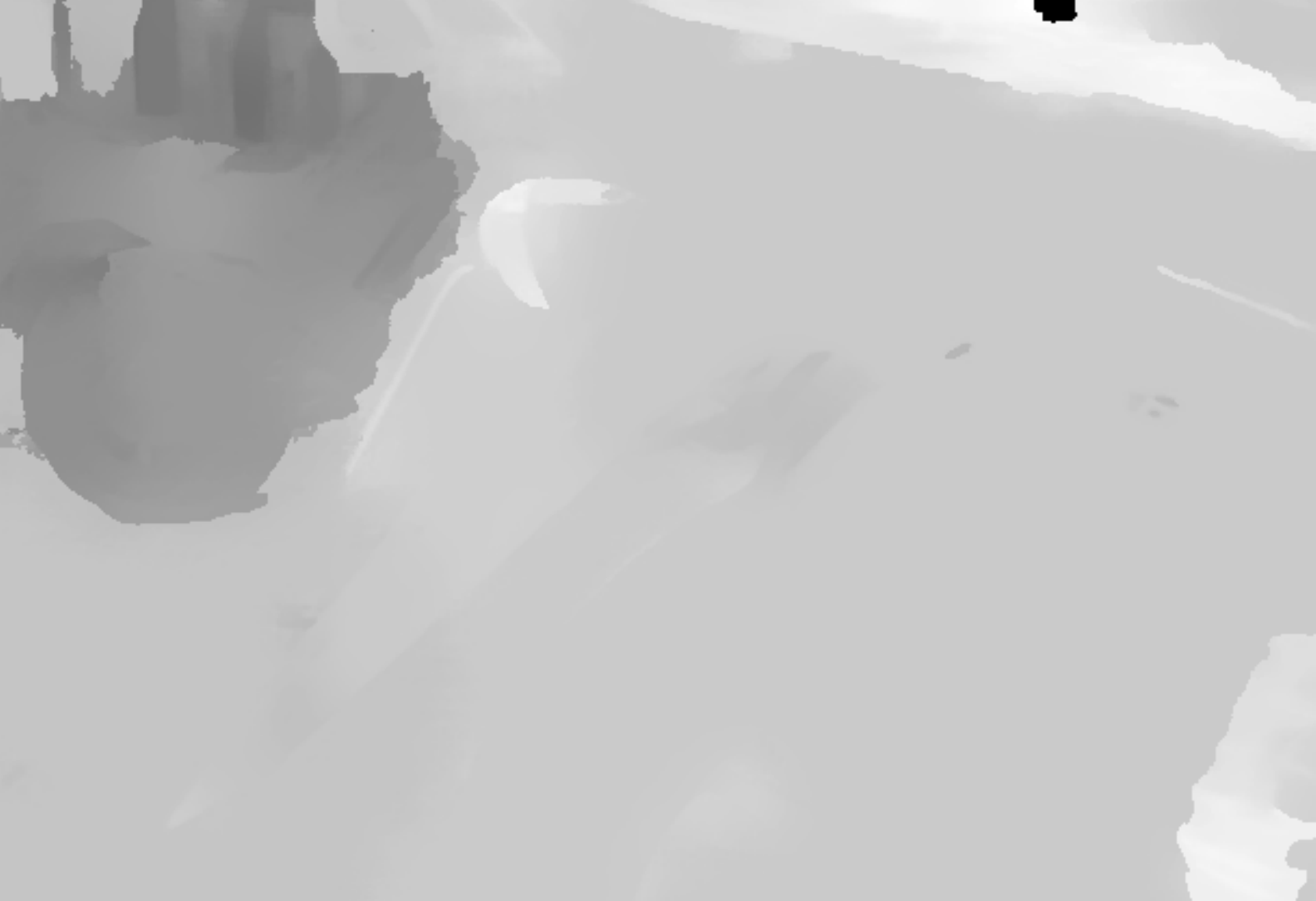} \hspace{-0.15cm} &
			\includegraphics[width=0.245\linewidth]{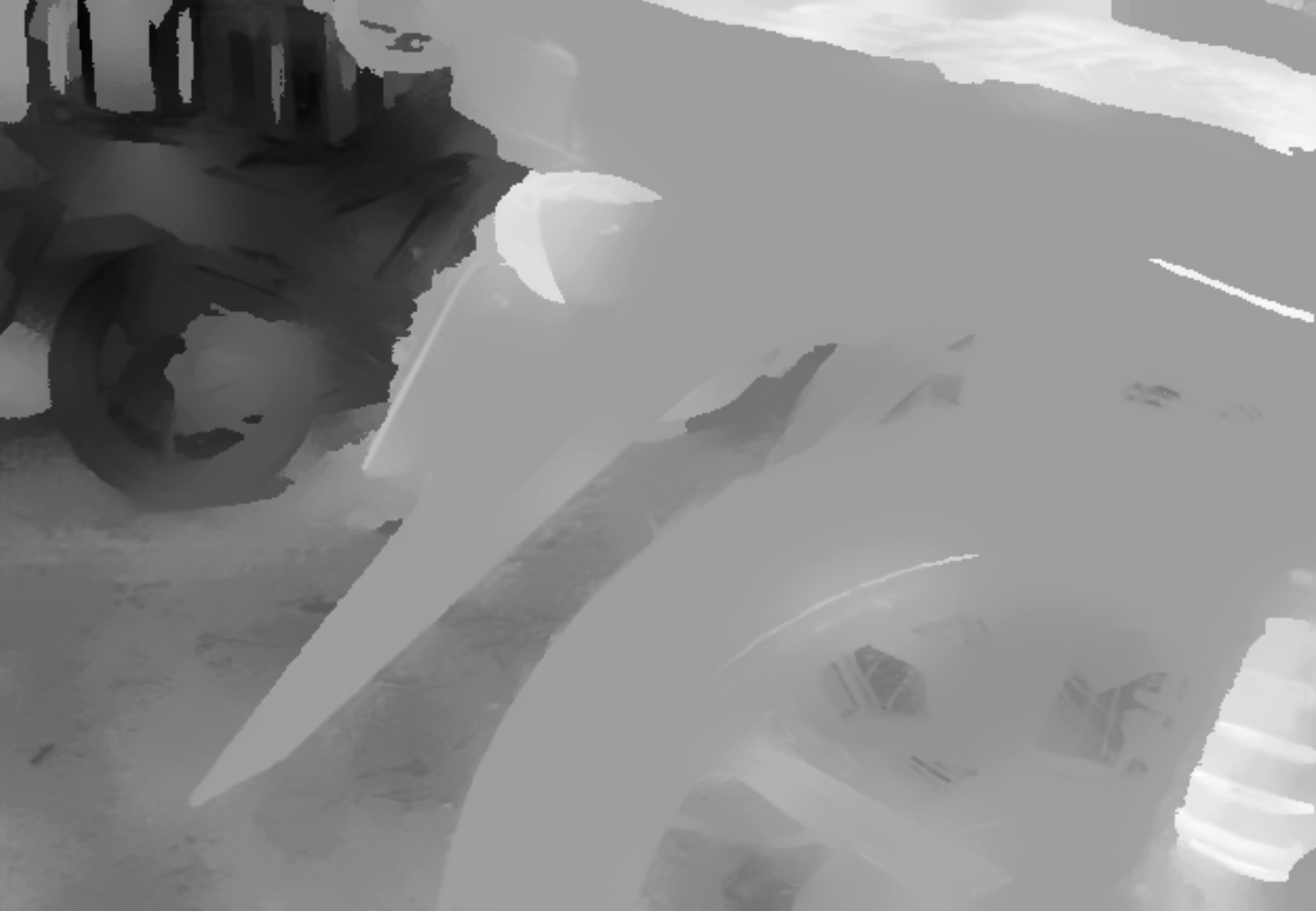}  \vspace{0.05cm}\\
			
			Scene II &  & SID \cite{chen2018learning2seeindark} & Proposed   \\
			\includegraphics[width=0.25\linewidth]{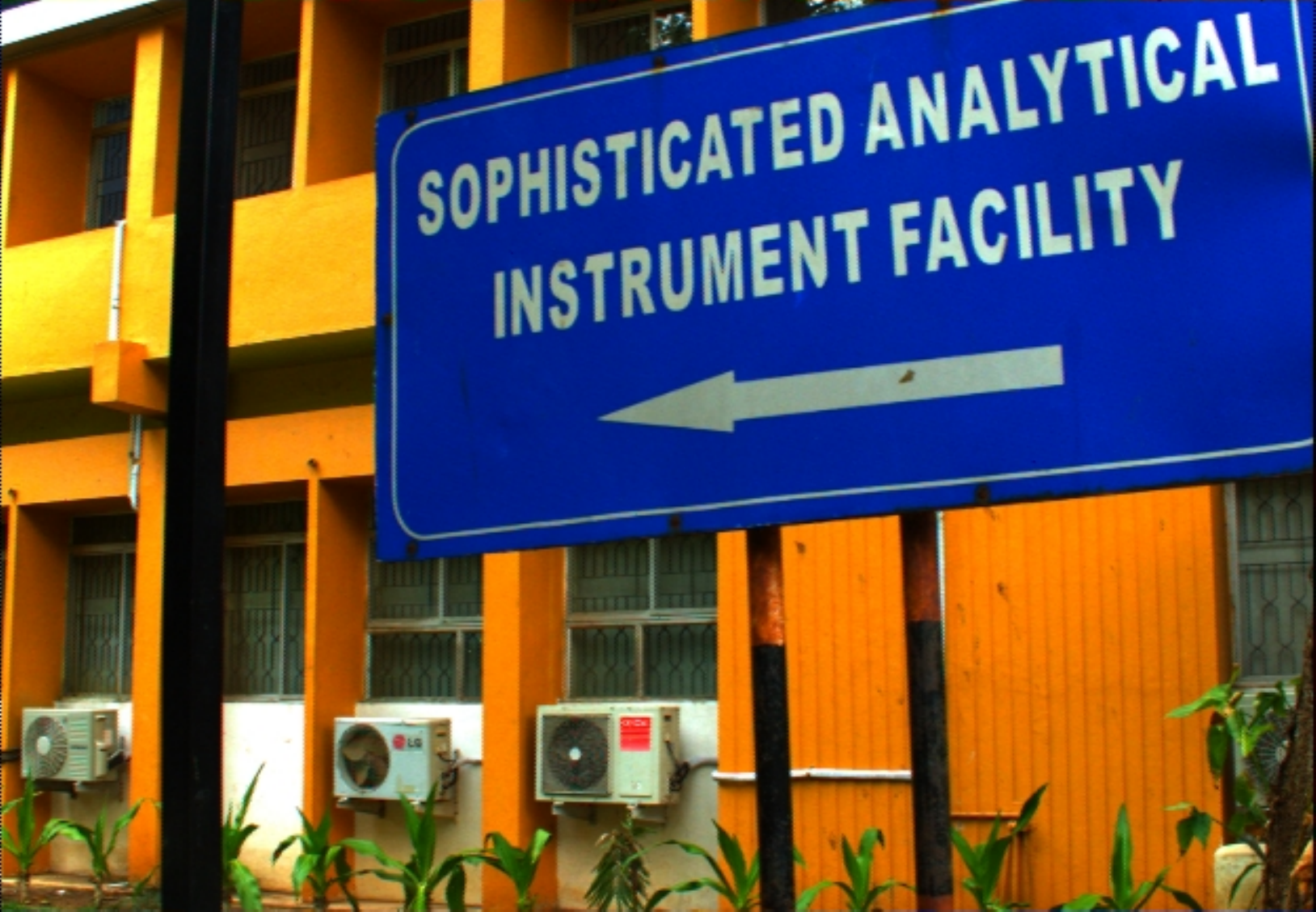} &
			\rotatebox[origin=lt]{90}{L3F-20}  & 
			\includegraphics[width=0.25\linewidth]{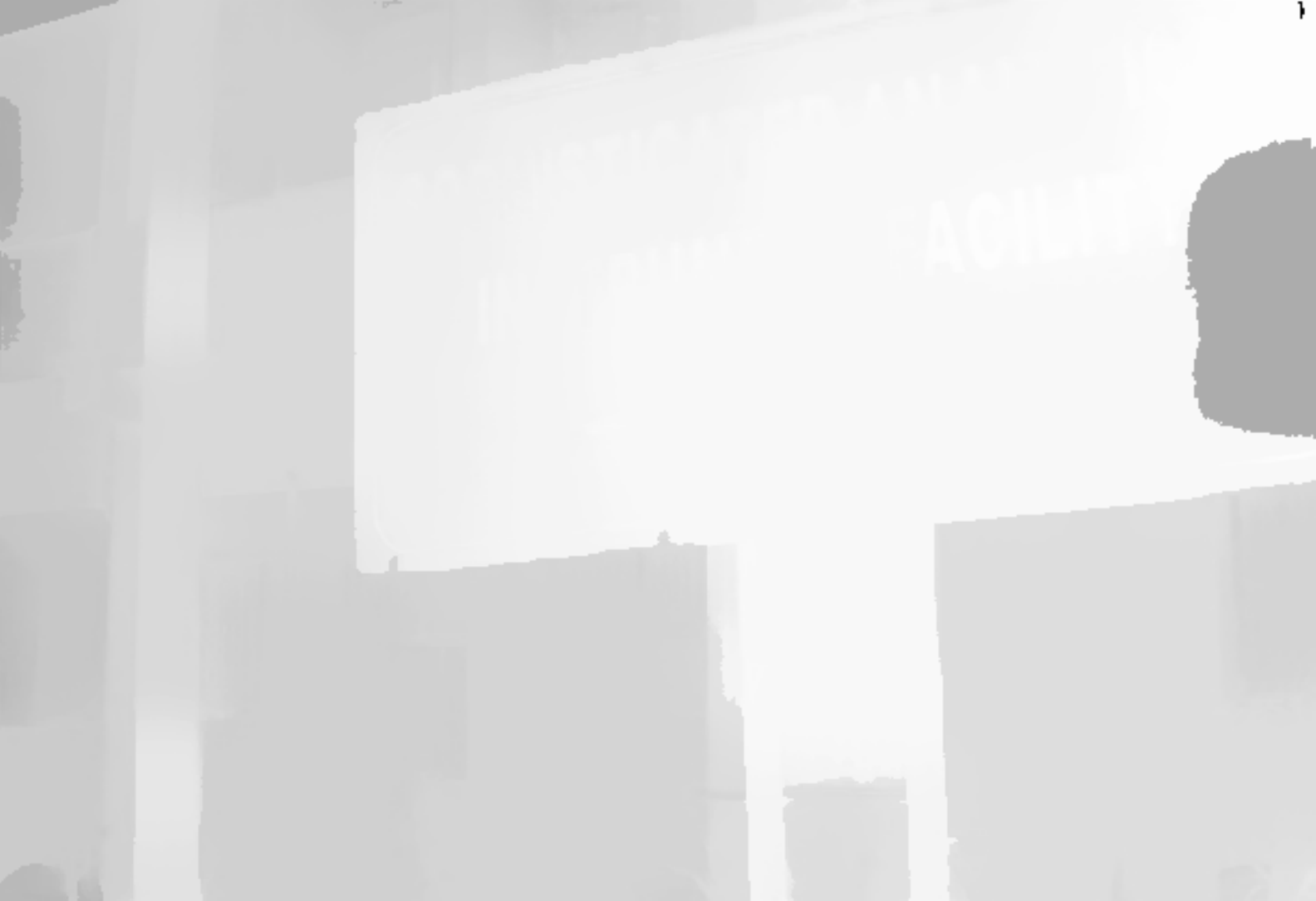} \hspace{-0.15cm} &
			\includegraphics[width=0.25\linewidth]{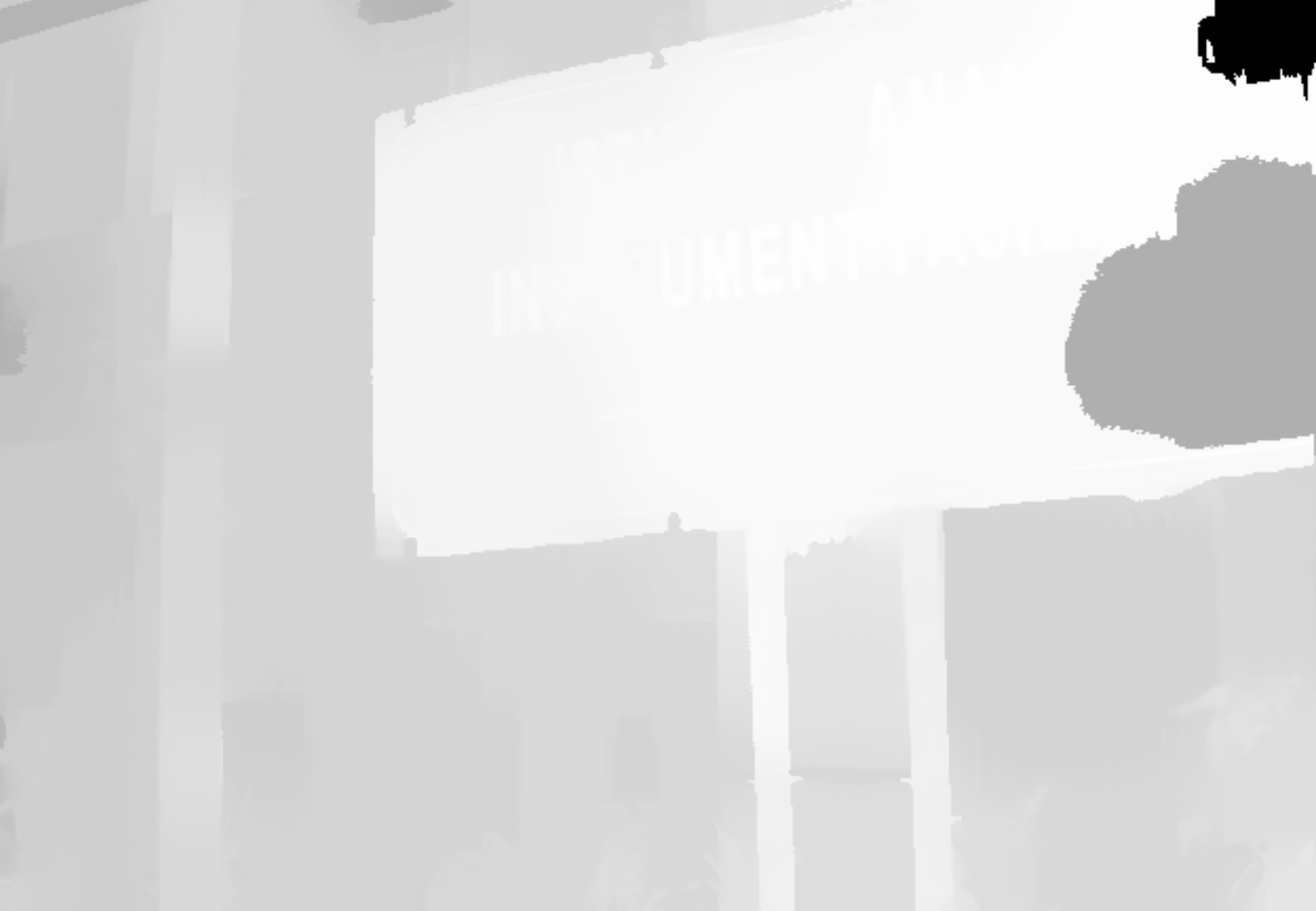}  \vspace{-0.05cm} \\
			
			\includegraphics[width=0.25\linewidth]{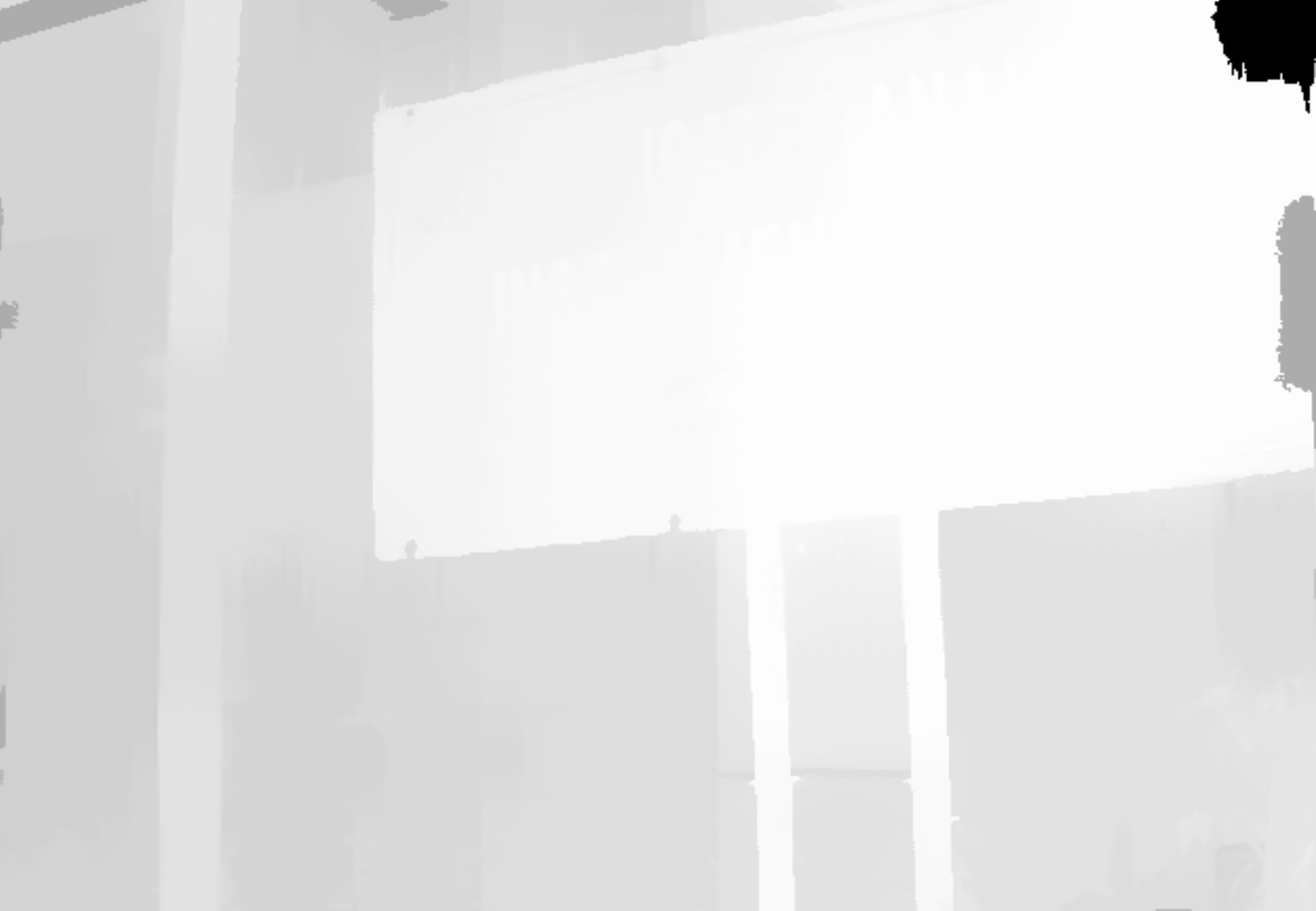}  &
            \rotatebox[origin=lt]{90}{L3F-100} &
			\includegraphics[width=0.25\linewidth]{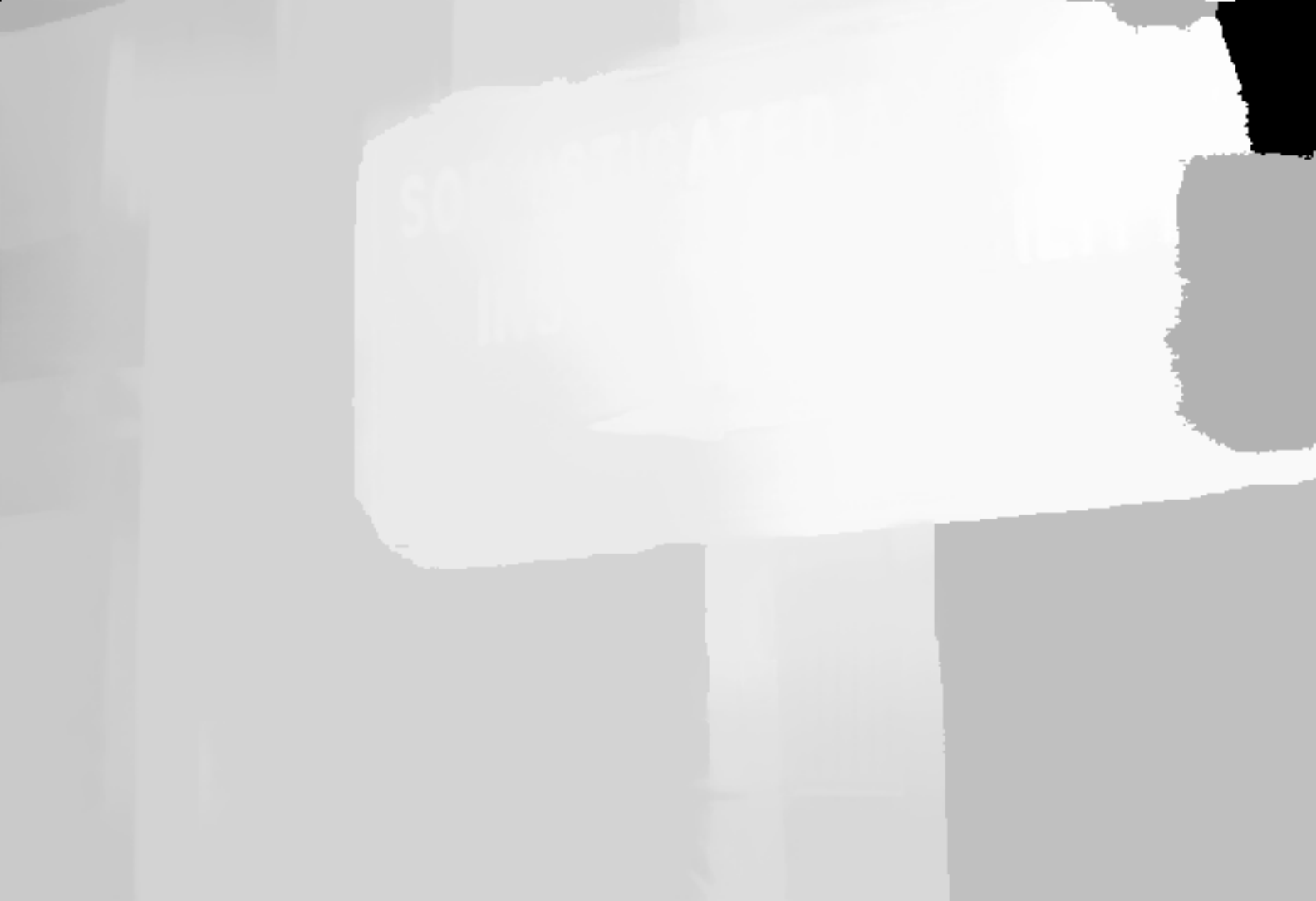} \hspace{-0.15cm} &
			\includegraphics[width=0.245\linewidth]{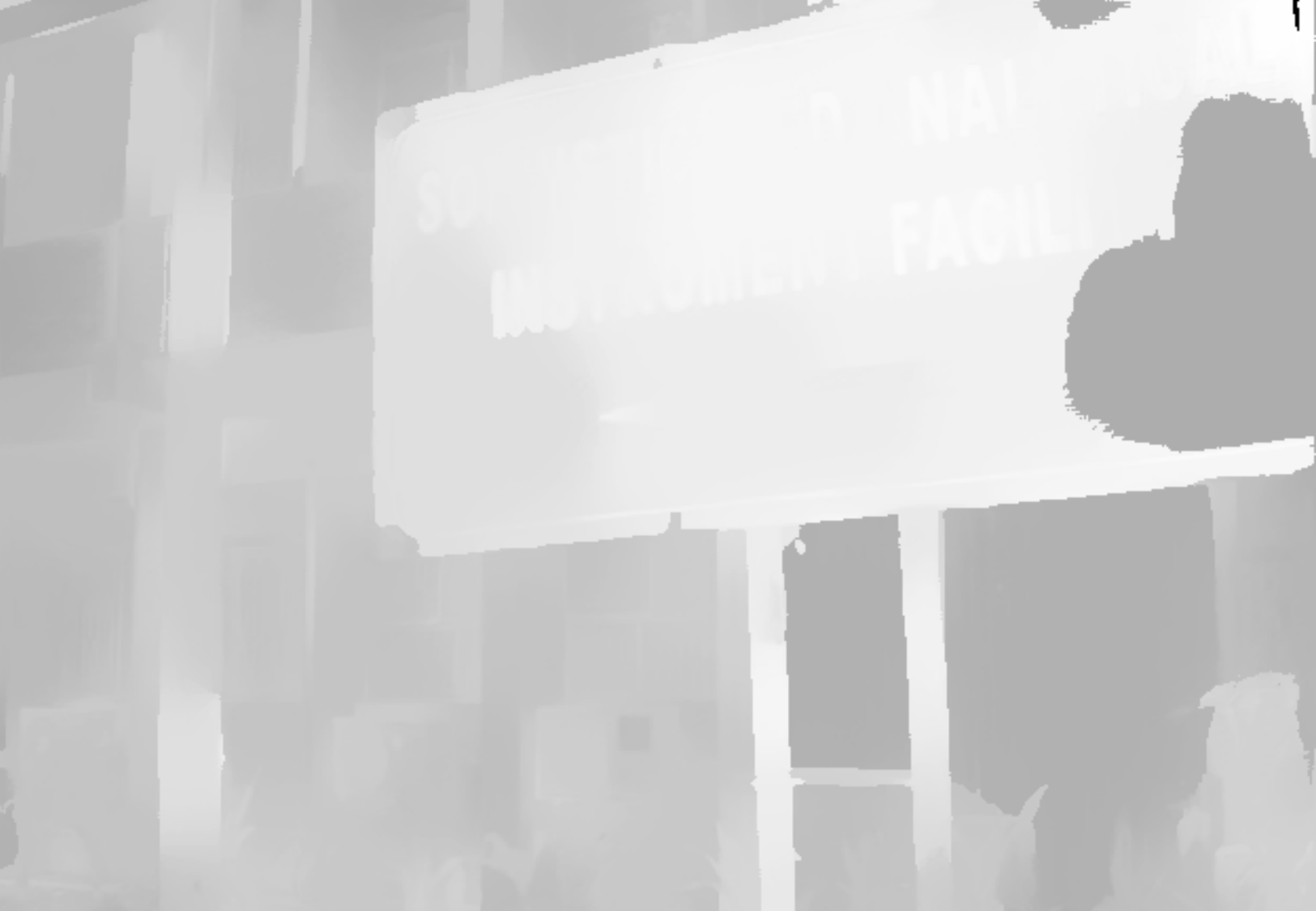}  \\
		\end{tabular}\hspace{-0.15cm}
	}
\caption{Depth estimation from reconstructed low-light LFs for low-light (L3F-20) and extreme low light (L3F-100) cases: 
Depth estimates from LFs restored using SID~are good for L3F-20 but very poor for the case of L3F-100 dataset because it is hardly able to preserve the epipolar geometry. On the other hand, our method is able to preserve the epipolar geometry even in the extreme low-light case.
}
\label{fig:depth}
\vspace{-0.5cm}
\end{figure}

\textbf{Reducing Memory Requirement: }\textcolor{black}{Stage-I computes the global representation by looking at all the 64 SAIs. This representation is then used by Stage-II to restore each SAI. But to speed-up the training and reduce the model size on GPU, we pass only $12$ randomly chosen SAIs to Stage-II and the loss is computed only over these $12$ SAIs. 
Note that during inference, however, we can together restore all the SAIs because during inference, the chain rule mechanism of PyTorch is disabled, which is responsible for most of the memory and computation.}

\textbf{Loss Function:} $L1$ loss turns out to be more crucial than the CX loss \cite{contextualloss} in our ablation studies (see Table \ref{table:new_ablation}). We thus set $\alpha_1=5$ and $\alpha_2=0.1$ for first $20k$ iterations. After this $\alpha_1$ is reduced to $1$. 
For CX loss, we used the feature maps at the $9$, $13$ and $18^{th}$ layer of VGG19 \cite{vgg}. 
$\lambda$ was fixed to $10^{-6}$ for all iterations. L3Fnet was trained for $100k$ iterations.

\textbf{Data augmentation:} To augment the data in the training phase, we use horizontal flipping, vertical flipping, and color augmentation. Color augmentation is achieved by swapping the color channels in random order. The training is done on patches of size $180 \times 180$ and full spatial resolution is used at the time of testing. In each iteration, the patch location is chosen randomly within each sub-aperture view. Since each view has a resolution of $625 \times 433$ and we use randomly selected patches of $180 \times 180$, along with data augmentation techniques we had sufficient training data.

\textbf{Baseline:} 
We compare our L3Fnet with a LF denoising technique LFBM5D \cite{bm5d} and single-frame low-light enhancement methods \textcolor{black}{SID \cite{chen2018learning2seeindark}, SGN \cite{gu2019self}, DID \cite{2019DID}, RetinexNet \cite{r3_2018_bmvc} and PBS \cite{r1_2018_acm}}. 
As LFBM5D is a denoising technique, we had to suitably scale it for better color restoration before proceeding for denoising. LFBM5D additionally requires an estimate of the noise variance. This was obtained from small texture-less patches of the low-light LF.

SID ~for single-frame images chose to work with raw format and performed an end to end training. It is however not suited for raw Lytro LF images. For reconstruction using SID, it has to be independently operated over each SAI which is like a single-frame image. The SAIs are however not readily available in raw LFR format and needs to decoded using the Light Field Matlab Toolbox \cite{dansereau2013decoding}. 

The existing methods such as SID were trained and evaluated using their own loss function comprising of just the $L1$ loss. We, however, also tried our loss function consisting of $L1$ + Contextual loss to train them. We found that our loss function gave slightly better results and so we use these results for comparison.

\subsection{Visual and Geometric Reconstruction Comparisons}

\begin{figure*}[t!]
\captionsetup[subfigure]{labelformat=empty}
\centering

\subfloat[Exposure Time:1/15 s, ISO:80]{\includegraphics[width=0.19\linewidth]{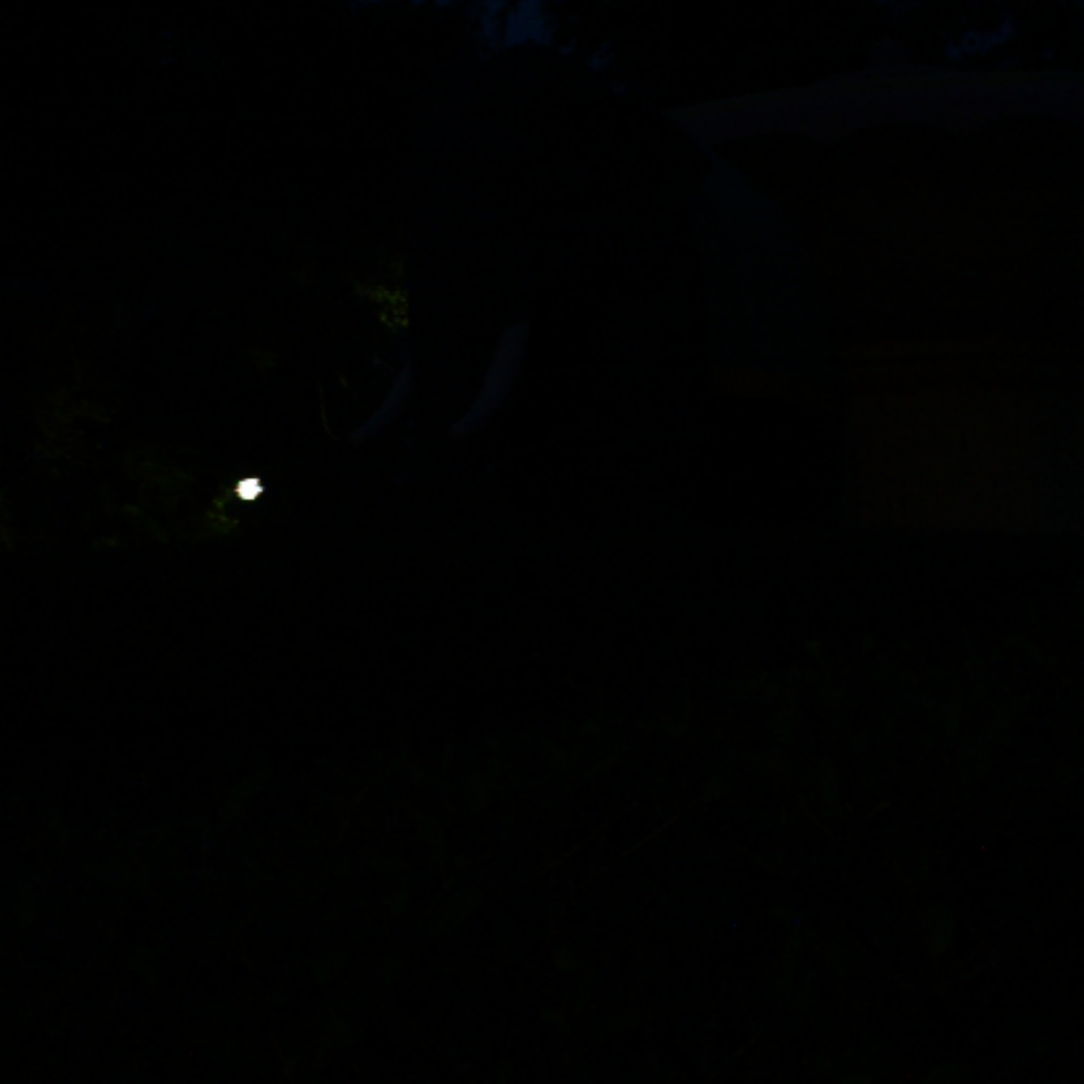}} 
\hfil
\subfloat[L3Fnet-20 restoration]{\includegraphics[width=0.19\linewidth]{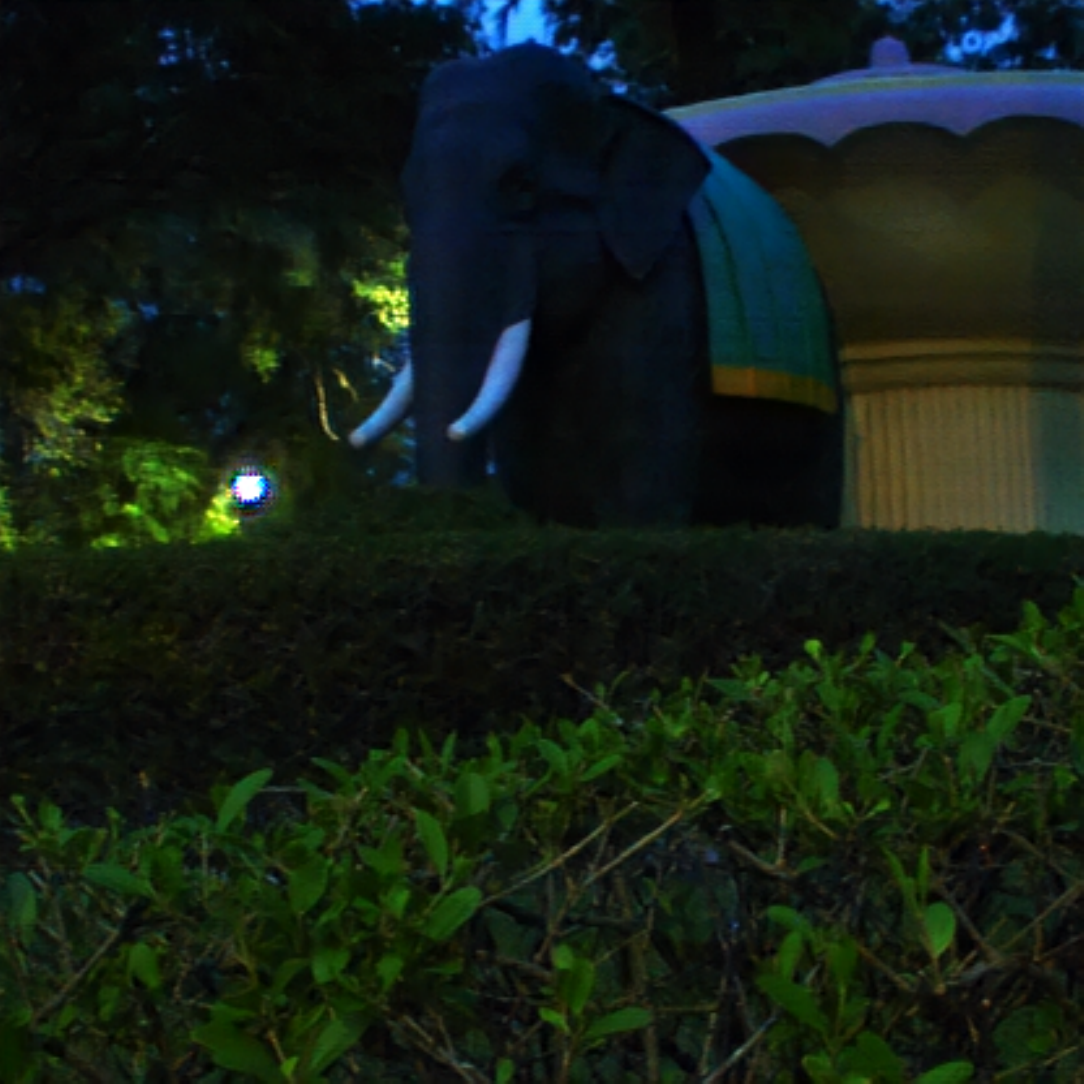}} 
\hfil
\subfloat[L3Fnet-50 restoration  ]{\includegraphics[width=0.19\linewidth]{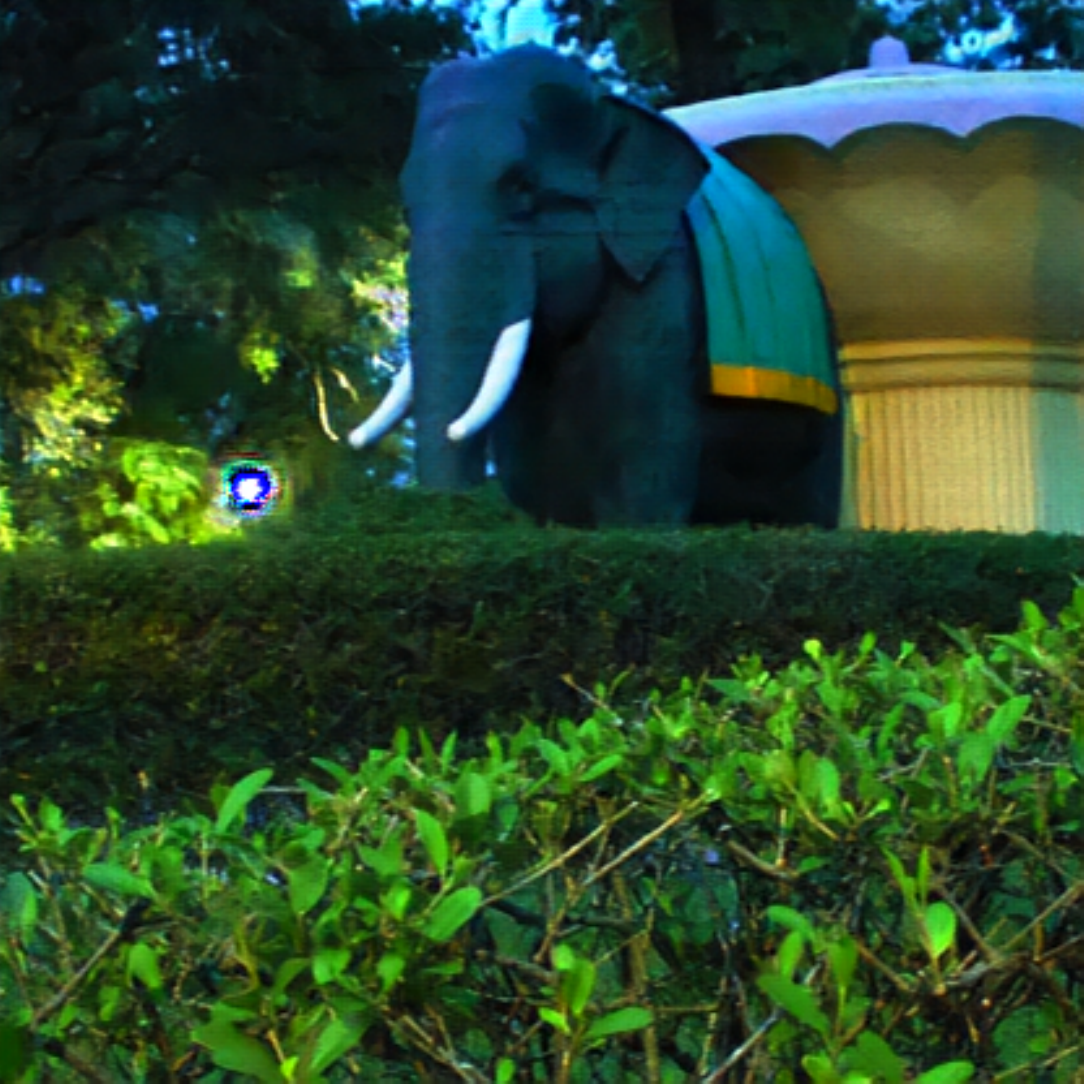}}
\hfil
\subfloat[L3Fnet-100 restoration  ]{\includegraphics[width=0.19\linewidth]{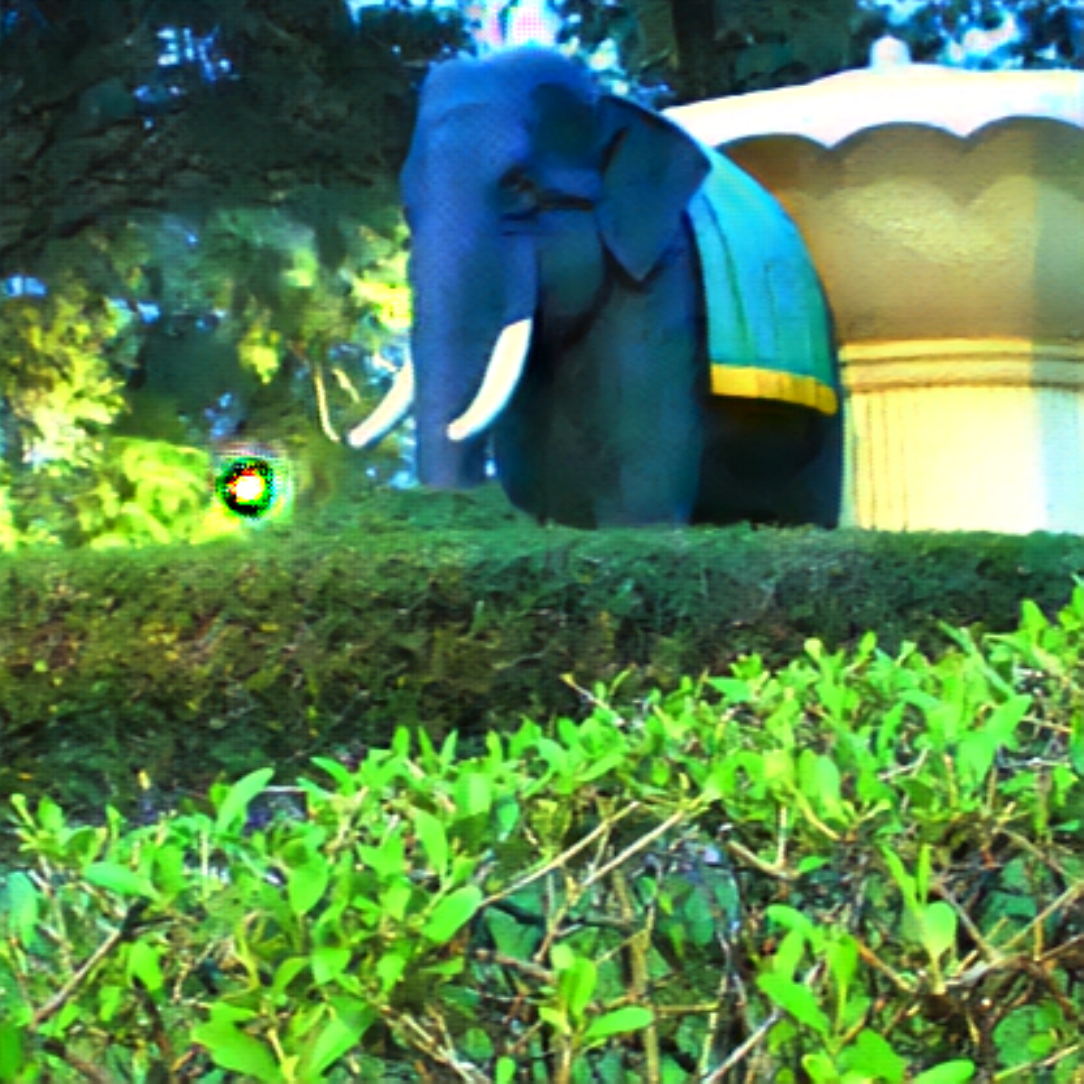}}
\hfil
\subfloat[L3Fnet-$\gamma$ restoration  ]{\includegraphics[width=0.19\linewidth]{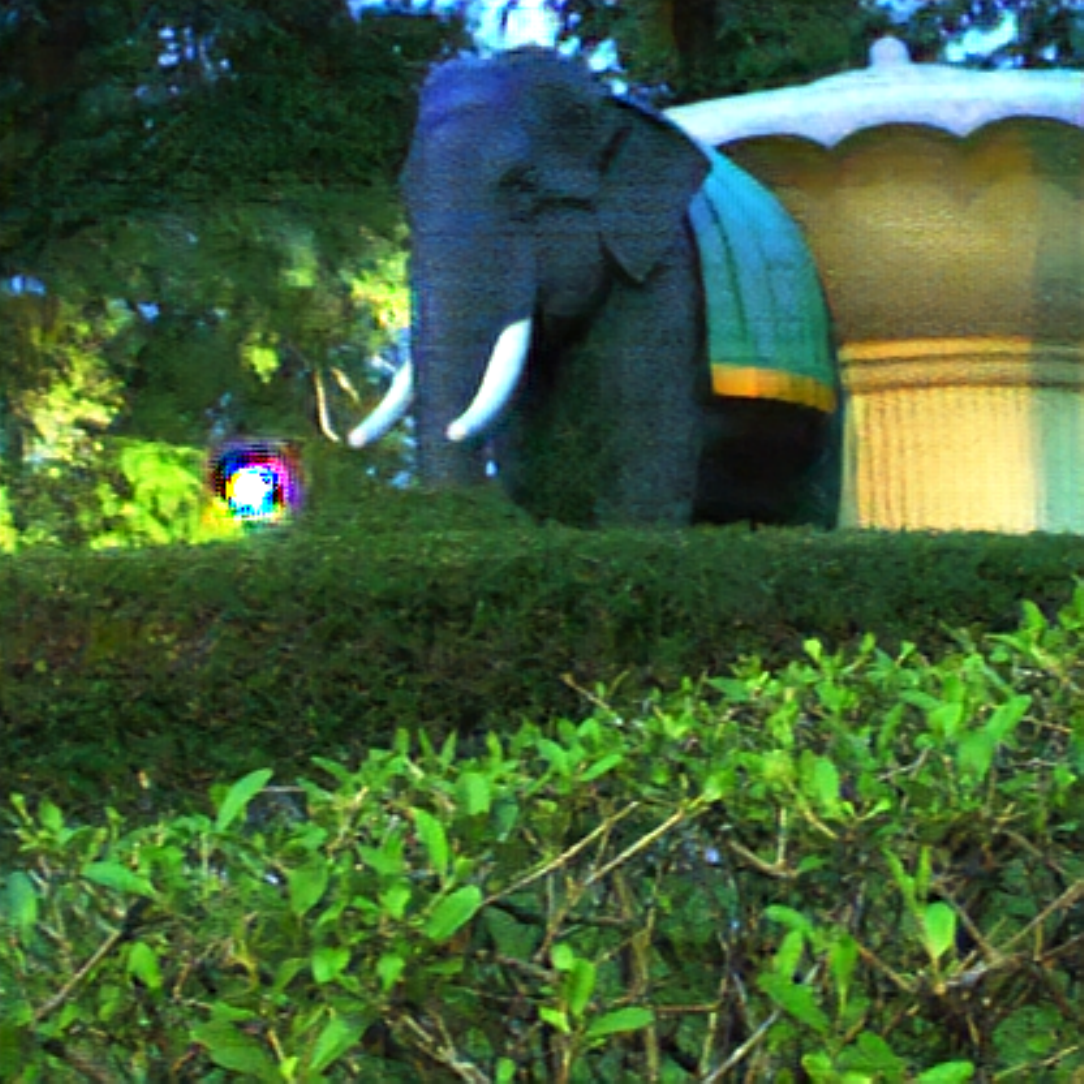}}\\
\subfloat[Exposure Time:1/5 s, ISO:100]{\includegraphics[width=0.19\linewidth]{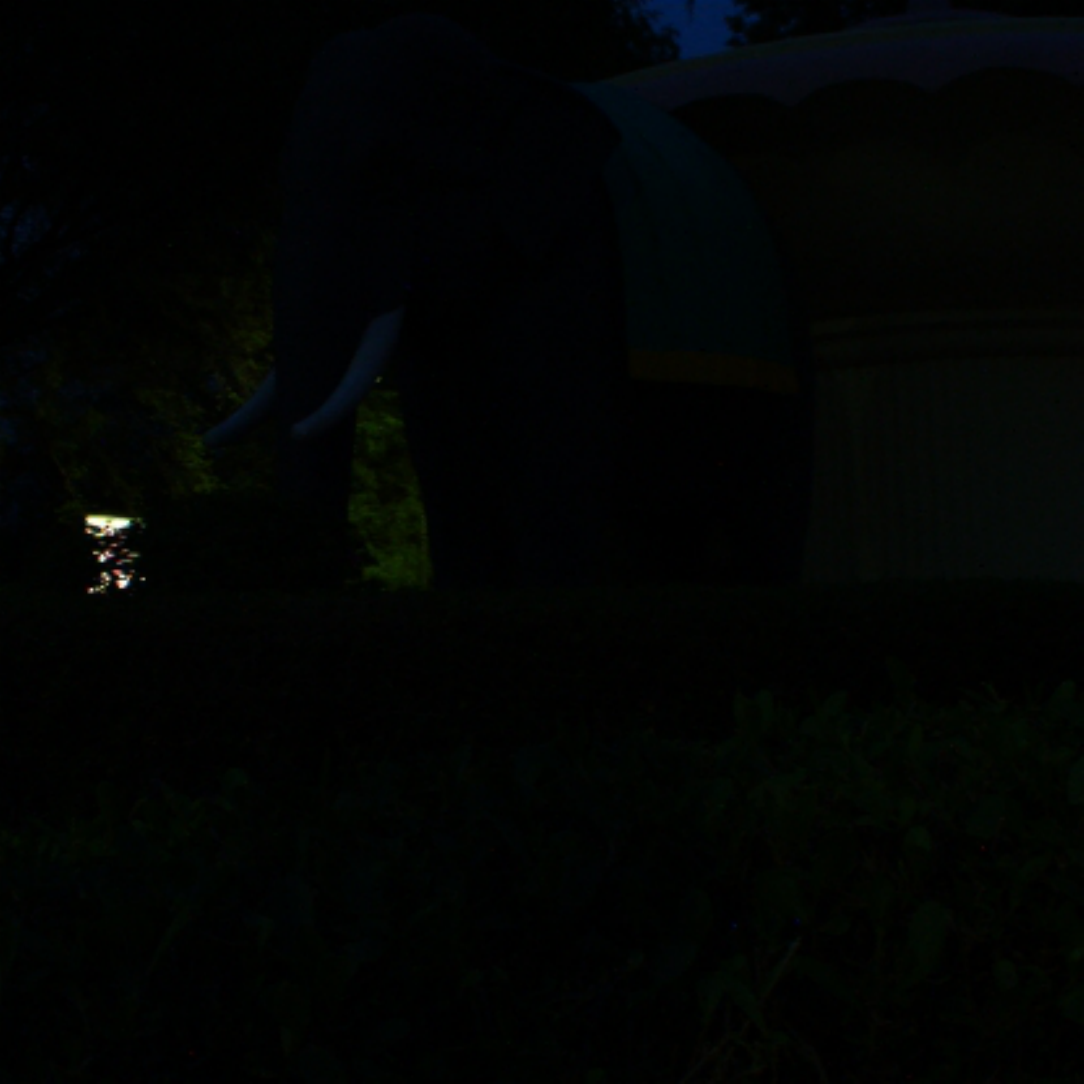}}
\hfil
\subfloat[L3Fnet-20 restoration  ]{\includegraphics[width=0.19\linewidth]{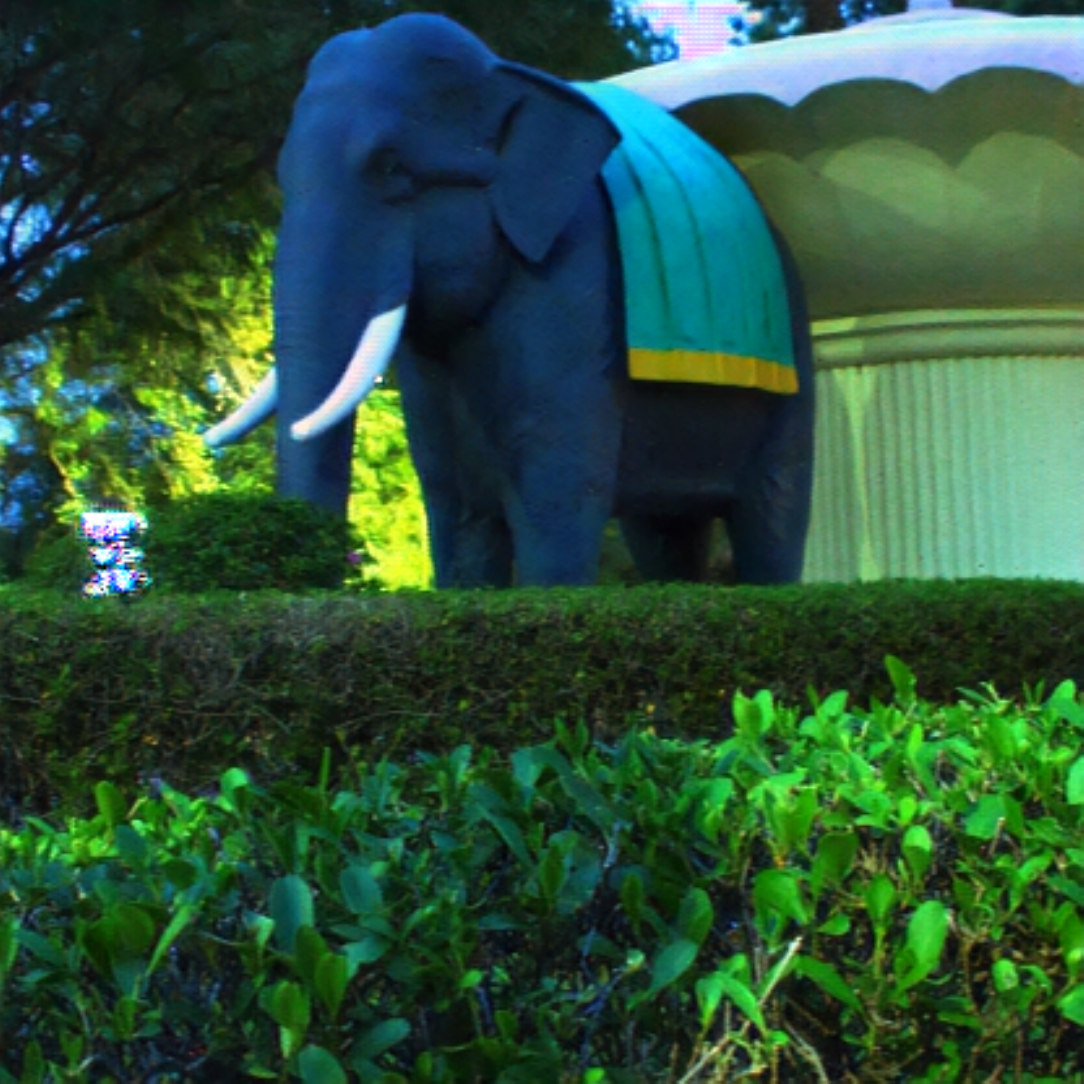}} 
\hfil
\subfloat[L3Fnet-50 restoration  ]{\includegraphics[width=0.19\linewidth]{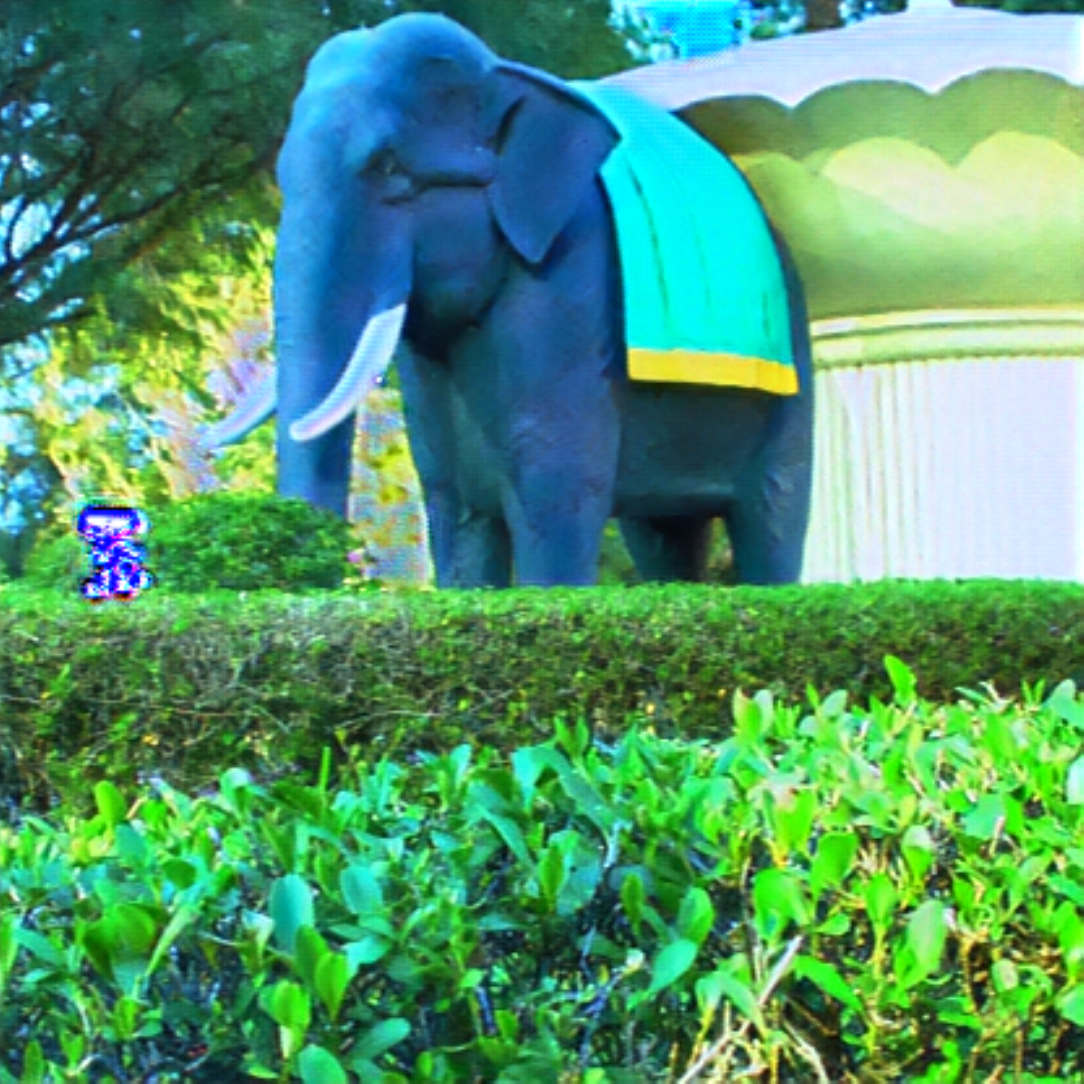}}
\hfil
\subfloat[L3Fnet-100 restoration  ]{\includegraphics[width=0.19\linewidth]{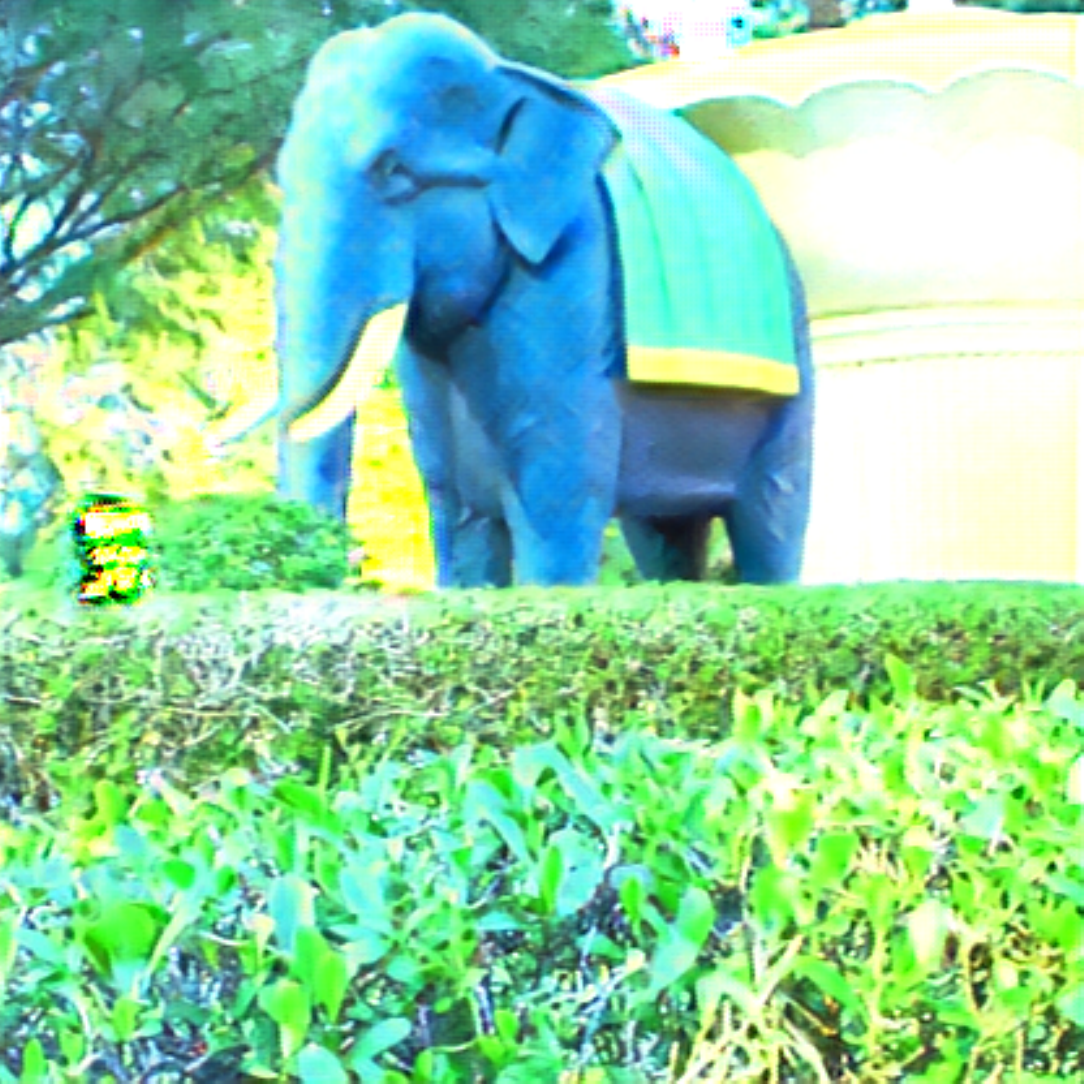}}
\hfil
\subfloat[L3Fnet-$\gamma$ restoration  ]{\includegraphics[width=0.19\linewidth]{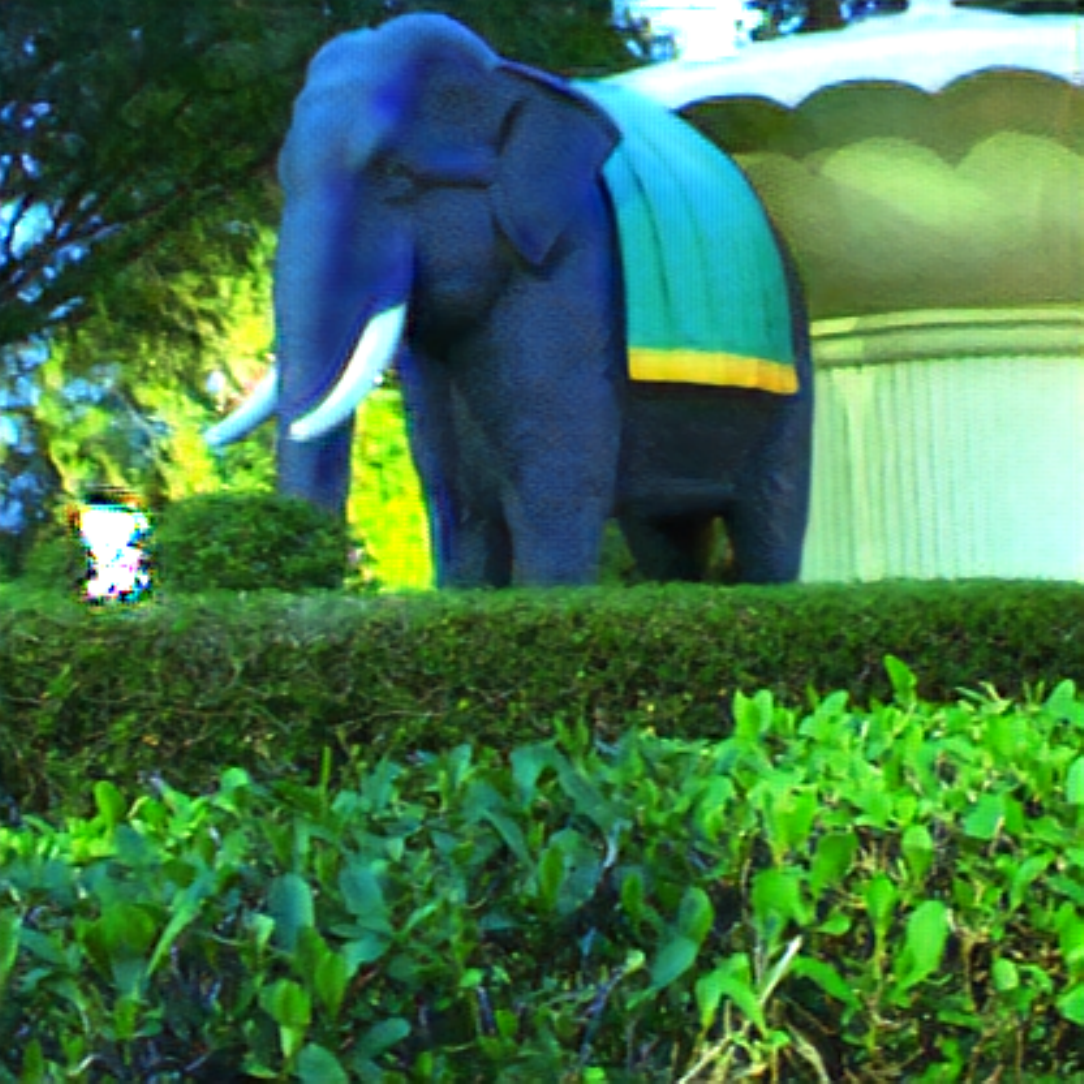}}
\caption{{Results on the L3F-wild dataset. 
\textcolor{black}{The dark LF images have been pre-processed with the Histogram Module allowing L3Fnet to adapt to different levels of low-light.} For example, in the top row, L3Fnet trained on L3F-20 dataset (denoted as L3Fnet-20) fails to achieve ambient brightness.  Likewise, when the light levels are increased for test image in the second row, L3Fnet-100 oversaturates. L3Fnet-$\gamma$, however, adjusts to both lighting conditions using the Histogram Module. Check supplementary for more visual results.}}
\label{fig:wild}
\vspace{-0.5cm}
\end{figure*}

\textcolor{black}{The proposed L3F-20, L3F-50 and L3F-100 datasets have very different light illumination levels, see Sec. \ref{sec:l3fdataset}. In our first  experiment, we train and test our method and the existing methods for each of the three datasets independently. That is, we train all the methods on the training-set of L3F-20 dataset and then test it on the testing-set of L3F-20. The same is repeated for L3F-50 ad L3F-100 datasets. 
Our proposed L3Fnet is equipped with an amplification module (which we call as the Histogram module) which can handle varying illumination levels. However, existing methods lack this feature. Thus, for a fair comparison with the exiting methods, we switched OFF the Histogram Module of L3Fnet for the above experiments. Anyway, since we are training the methods for a particular light level (for example L3F-20) and testing it for the same light level (L3F-20), there is no need for estimating the amplification factor.}

The results are shown in Fig. \ref{fig:vis_compare} and Table. \ref{table:quant_comp}. 
The proposed L3Fnet does better restoration for all three datasets. However, the difference between L3Fnet and the other methods is most evident for the extreme low-light case of L3F-100 dataset. For this dataset, all the existing methods struggle to restore the finer details and the results are very blurry. \textcolor{black}{L3Fnet harnesses information from all the views and hence is able to better restore each of the views. The single-frame methods such as SID \cite{chen2018learning2seeindark}, SGN \cite{gu2019self} and DID \cite{2019DID}, however, restore each SAI independent of other SAIs which causes performance degradation, including detail loss and blurry results. We further observe that for PBS \cite{r1_2018_acm} and RetinexNet \cite{r3_2018_bmvc}, which are also single-frame methods, the restoration additionally has a lot of noise. This is because these methods have been designed for enhancing under-exposed images, which have sufficient color information with moderate level of noise. However, for the challenging case of extreme low light imaging, color information is almost lost and noise is very high.}


The restored image may look aesthetically pleasing but the LF geometry might be destroyed. 
We thus additionally show the depth estimates by L3Fnet and SID for L3F-20 and L3F-100 dataset in Fig. \ref{fig:depth}. The method proposed by Jeon \etal \cite{depthFind} is used for depth estimation. Similar to visual restorations, depth estimation becomes more and more challenging for lower light levels and the depth estimates of L3Fnet are closer to the ground truth.

For the L3F-20 dataset, depth estimates from LFs restored using SID are good but lack the fine details. For example the spokes of the bike's tyre in scene I of Fig. \ref{fig:depth} are clearly demarcated in L3Fnet's depth estimates but not for the case of SID. Likewise, for scene II the sharpness of the board corners is much better preserved in our results. Similar observations hold true for the L3F-100 dataset.

\begin{table}[t!]
    \centering
    \caption{Variation caused in $\gamma$ values by the histogram block on the test dataset. The predicted amplification factor increases monotonically with decreasing light level.}
    \vspace{-0.2cm}
    \begingroup
\setlength{\tabcolsep}{6pt}

    \begin{tabular}{c|c}
      \hline \textbf{Dataset} & \textbf{$\gamma$ range} \\ \hline \hline
        L3F--20 & $0.3-0.4$\\ 
        L3F--50 & $0.8-0.9$\\
        L3F--100 & $1.4-1.7$\\ \hline
    \end{tabular}
    \endgroup
    
    \label{tab:gamma_values}
    \vspace{-0.1cm}
\end{table}

\subsection{L3Fnet for LF captured in Wild}
\label{sec:L3Fnet for LF captured in Wild}
\begin{table}[t!]
    \centering
\caption{Ablation study to examine the contribution of Stage-I in L3Fnet model and the two loss functions: L1 and CX.}
\vspace{-0.2cm}
\begingroup
\setlength{\tabcolsep}{2pt}
\resizebox{.45\textwidth}{!}{    \begin{tabular}{c|cccc|c}
    \hline
      & \textbf{Stage-I}& \textbf{Stage-II}& \textbf{L1 loss}& \textbf{CX loss}&\textbf{PSNR/SSIM}\\
      \hline \hline
      \textbf{Net-I}& \cmark&\cmark&\cmark&\xmark&\text{$21.71/0.63$}\\
      \textbf{Net-II}& \cmark&\cmark&\xmark&\cmark&\text{$13.68/0.18$}\\
      \textbf{Net-III}& \xmark&\cmark&\cmark&\cmark&\text{$22.21/0.65$}\\ \hline
      \textbf{Proposed}& \cmark&\cmark&\cmark&\cmark&{$\mathbf{22.61/0.70}$}\\
      \hline
    \end{tabular}
    }
    \label{table:new_ablation}
    \endgroup
    \vspace{-0.25cm}
\end{table}

\textcolor{black}{In this section, we evaluate the performance of L3Fnet on the L3F-wild dataset. This dataset was collected under near 0 lux conditions and hence neither the ground-truth is available nor the optimal exposure is known. Therefore, it is crucial to estimate an appropriate amplification factor to restore these LF images. To facilitate this, the Histogram Module of L3Fnet was switched ON. }

 Since ground truth is not available for these images it cannot be used for training L3Fnet. We instead re-train L3Fnet with L3F $-$ 20, 50, and 100 datasets merged together but pre-processed with the Histogram Module to estimate the desired amplification factor $\gamma$. The weights of the Histogram Module and L3Fnet were learnt together in a end-to-end fashion. We call this trained network as L3Fnet-$\gamma$. \textcolor{black}{In Table \ref{tab:gamma_values} we report the range of $\gamma$ values predicted by L3Fnet-$\gamma$ for different light levels.} For ease in notation and brevity, we refer L3Fnet trained on L3F $-$ 20, 50, and 100 datasets as L3Fnet-20, L3Fnet-50, and L3Fnet-100, respectively. 

In Fig. \ref{fig:wild}, we show a night time scene captured using different ISO and exposure settings and try to restore it using L3Fnet-20, L3Fnet-50, L3Fnet-100 and L3Fnet-$\gamma$. We can easily notice the over/under saturation artifacts in LFs restored by L3Fnet-20, L3Fnet-50 and L3Fnet-100. The reason is that these networks are agnostic to image statistics and hence can not adapt to different illumination condition. But L3Fnet-$\gamma$ avoids this problem to a large extent by estimating an appropriate amplification for input LF.
\textcolor{black}{We also tried using Gamma correction instead of linear amplification, but the restoration exhibited a lot of artifacts. More results and discussion can be found in the supplementary.}

\subsection{Ablation Study}

\begin{figure}
	\scriptsize
    \centering
    \setlength{\tabcolsep}{1pt}
    \begin{tabular}{cc}	
    \setlength{\tabcolsep}{1pt}
			\begin{tabular}{c}
				\includegraphics[width=0.26\linewidth]{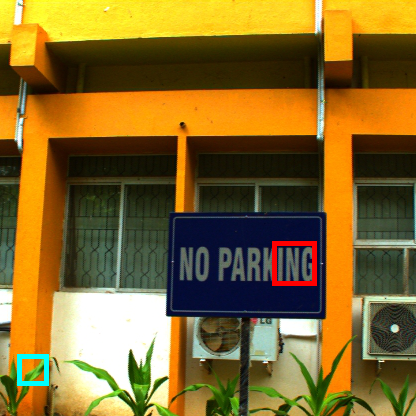}
				\\
				GT image\\
			\end{tabular}
		\hspace{-2mm}
		\setlength{\tabcolsep}{1pt}
			\begin{tabular}{cccccc}
				\includegraphics[trim={691 131 100 244},clip=True,width=0.135\linewidth]{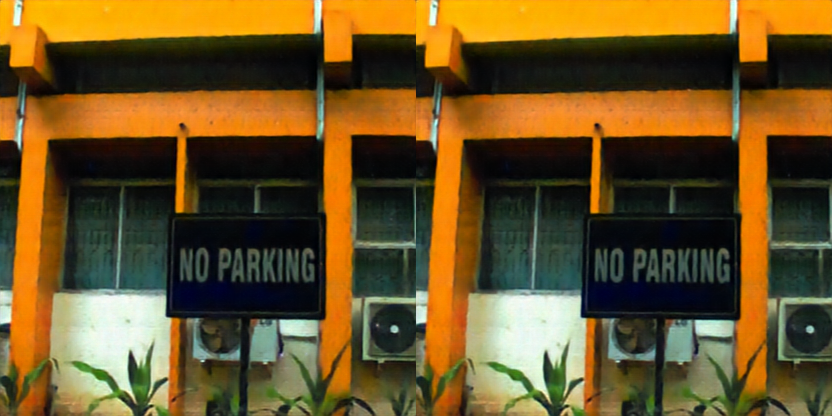}  &
				\includegraphics[trim={691 131 100 244},clip=True,width=0.135\linewidth]{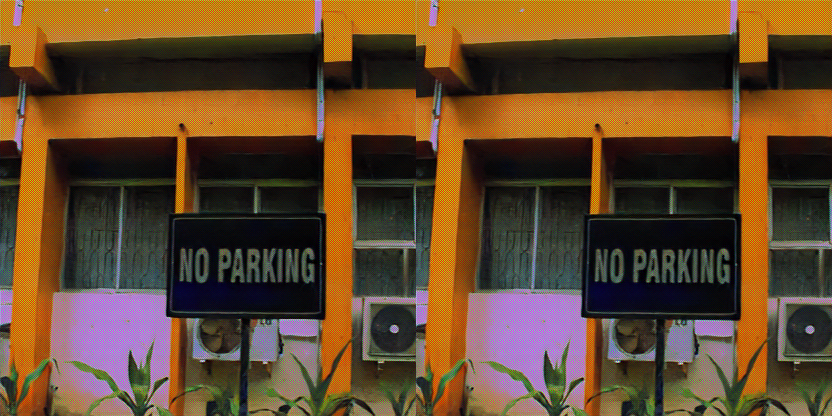}  &
				\includegraphics[trim={691 131 100 244},clip=True,width=0.135\linewidth]{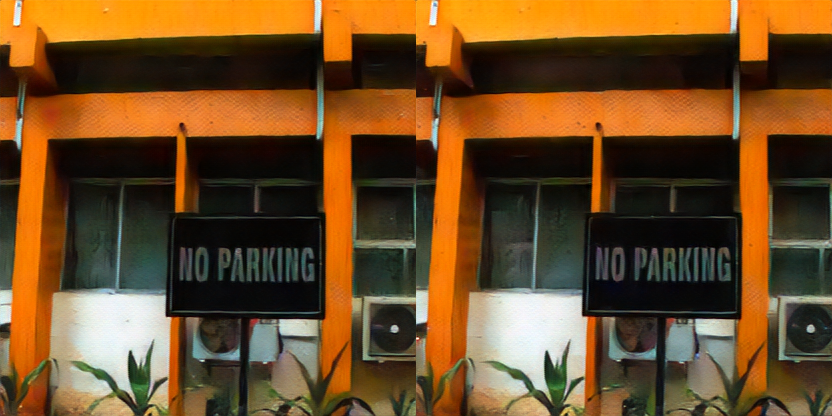}  &
				\includegraphics[trim={691 131 100 244},clip=True,width=0.135\linewidth]{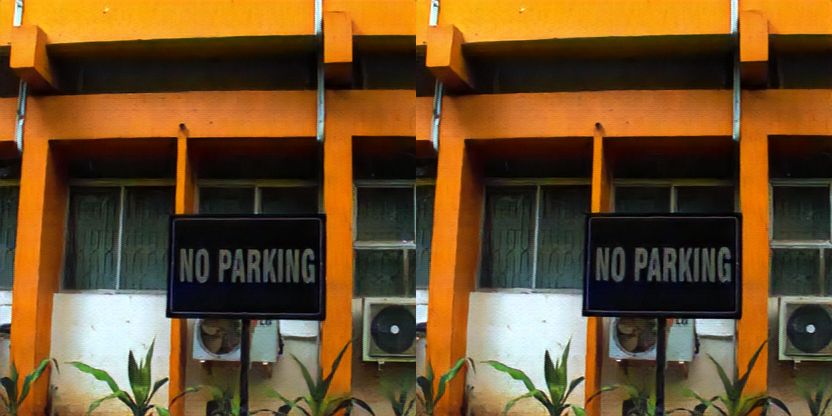}  &
				\includegraphics[trim={691 131 100 244},clip=True,width=0.135\linewidth]{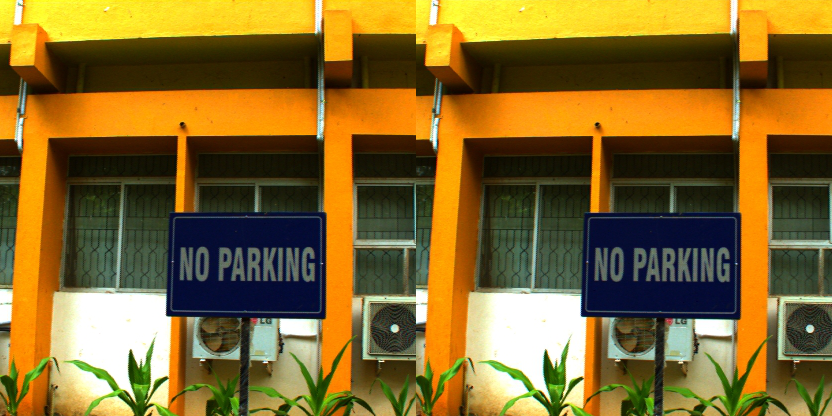}
				\\
				\includegraphics[trim={436 31 368 357},clip=True,width=0.135\linewidth]{ablation/4_onlyL1.png}  &
				\includegraphics[trim={436 31 368 357},clip=True,width=0.135\linewidth]{ablation/4_onlycx.png}  &
				\includegraphics[trim={436 31 368 357},clip=True,width=0.135\linewidth]{ablation/4_onlystage1.png}  &
				\includegraphics[trim={436 31 368 357},clip=True,width=0.135\linewidth]{ablation/4_ours.png}  &
				\includegraphics[trim={436 31 368 357},clip=True,width=0.135\linewidth]{ablation/4_GT.png}  
				\\
				Net-I & Net-II & Net-III & L3Fnet & GT patch
			\end{tabular}
	\end{tabular}
	
    \caption{\textcolor{black}{Visual reconstruction results for ablation study on L3Fnet. Net-I produces blurry results since it uses only $L1$ loss. Net-II restores the sharpness by instead using the contextual loss but introduces a large amount of color artifacts. Net-III solves both problems by incorporating both loss functions, but because of the absence of Stage-I it could not capture the LF geometry leading to inferior results than those obtained by the proposed L3Fnet. The difference is better corroborated by comparing depth estimations results in Fig. \ref{fig:depth_Ablation}.  Refer Table \ref{table:new_ablation} for PSNR/SSIM values.}}
    \label{fig:vis_ablation}
\end{figure}

\begin{figure}[!t]
	\scriptsize
	\centering
	\setlength{\tabcolsep}{1pt}
			\begin{tabular}{ccccc}
			 \includegraphics[width=0.24 \linewidth]{depth_images/2_vis} \hspace{-0.05cm} &
			 \includegraphics[width=0.24 \linewidth]{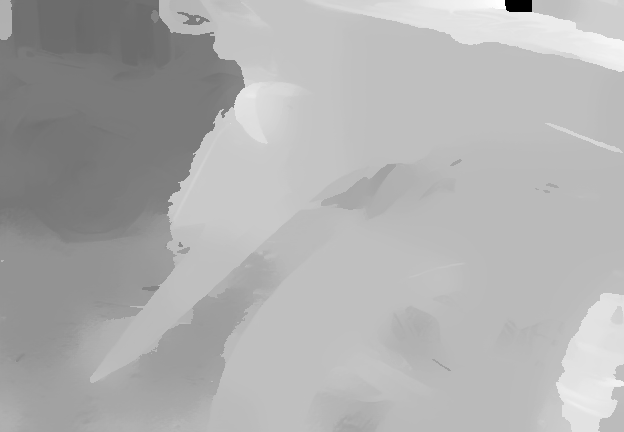} \hspace{-0.05cm} &
 			\includegraphics[width=0.24\linewidth]{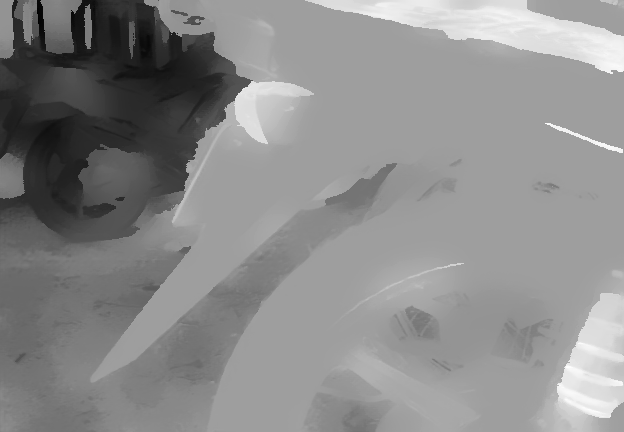} \hspace{-0.05cm}  &
 			\includegraphics[width=0.24\linewidth]{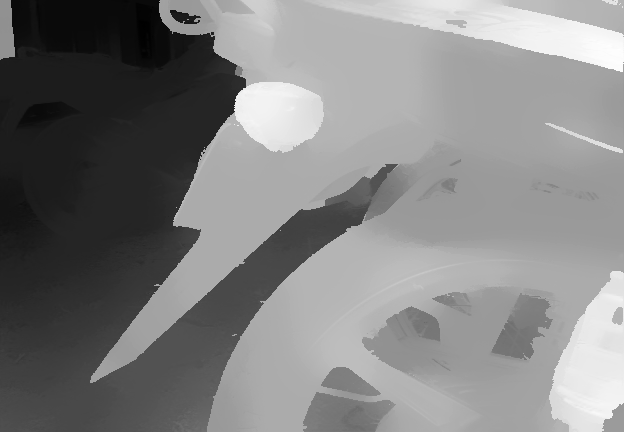} \hspace{-0.05cm} &
				\\
			  & Only Stage-II & Stage-I + Stage-II & GT \\ & (Net-III) & (L3Fnet) &
\end{tabular}
\caption{Depth estimates from reconstructed LF with and without Stage-I of the L3Fnet model. Stage-I captures the LF geometry and hence it helps in producing better depth maps.}
\label{fig:depth_Ablation}
\vspace{-0.5cm}
\end{figure}


\textbf{Effect of different loss functions: } We trained $2$ variants of the proposed L3Fnet on the L3F-100 dataset called Net-I and Net-II as shown in Table \ref{table:new_ablation}. Net-I, which was trained only with $L1$ loss, gave slightly blurry results. In contrast, Net-II was trained only with the Contextual Loss. While Net-I showed a small dip in performance due to blurriness, Net-II performed very poorly. The reason is without the $L1$ loss, the network could not learn the correct colors. We therefore have both $L1$ and Contextual loss for the proposed L3Fnet. Some visual results are also given in Fig. \ref{fig:vis_ablation}. Since the exclusion of $L1$ loss penalises the restoration far more than the Contextual loss, we first try to restore the color and basic geometry by giving higher importance to $L1$ loss in the first $20k$ iterations and subsequently reduce it to let contextual loss finetune the result. \textcolor{black}{Besides the $L1$ and Contextual loss we also tried to train with the SSIM loss along with $L1$ + Contextual losses.  Using  the  SSIM  loss, increases the SSIM value of the restored LF  marginally, but  the  decrease  in  PSNR value is  slightly  more, see supplementary. 
}

\textbf{Importance of Global Representation Block (Stage-I): } We trained a third variant of the proposed L3Fnet on the L3F-100 dataset by removing the Stage-I and instead added the 4 residual blocks of Stage-I to Stage-II so that the model parameters remains the same. Without Stage-I, L3Fnet was not able to preserve the epipolar constraints of LF. Fig. \ref{fig:depth_Ablation} shows that the depth map obtained from the LF restored by Net-III is inferior to that obtained from the LF restored when both stages are included in L3Fnet. Some more qualitative and quantitative results pertaining to Net-III can be found in Fig. \ref{fig:vis_ablation} and Table \ref{table:new_ablation} respectively.

\textbf{Effect of number of residual blocks and model parameters: }\textcolor{black}{We have fixed the total number of residual blocks in L3Fnet to $10$ for computational reasons. Out of the total 10 residual blocks, we allocate 4 residual blocks to Stage-I and 6 to Stage-II. We chose more residual blocks for Stage-II because our main goal is LF enhancement, which is being explicitly performed by Stage-II. Moreover, we want L3Fnet to give more weightage to the immediate neighbors of the SAI being restored and this is only possible in Stage-II.
In the supplementary, we report several experiments that validate our intuition. In these experiments, we tried allocating a different number of blocks to Stage I and II and found that the proposed allocation is better. In the supplementary, we have also analyzed the performance of L3FNet by varying the model parameter count and found the performance to be quite robust to change in model parameters.}



\subsection{Pseudo-LF for single-frame image reconstruction}
 
\textcolor{black}{In this section, we evaluate the proposed pseudo-L3F transformation for processing single-frame low-light DSLR images. The purpose here is not to surpass low-light DSLR reconstruction methods but to highlight the universality of L3Fnet to enhance both LF and single-frame images.} 

Similar to the L3F-100 dataset, we capture $50$ extreme low-light DSLR images of which $14$ were reserved for testing. We process JPEG images only as the raw format did not give much improvement but instead had much slower convergence because of additionally learning the Bayer demosaicing. 
We used all the data augmentation techniques mentioned for L3Fnet with patch-wise training.

\textcolor{black}{We compare our pseudo-L3Fnet pipeline  with SID \cite{chen2018learning2seeindark}. We also create a new baseline with the same number of residual blocks present in pseudo-L3Fnet. Pseudo-L3Fnet has 10 residual blocks (4 in Stage-I and 6 in Stage-II), and so the new baseline has 10 residual blocks stacked end-to-end with a long skip connection from input to output. The new baseline has approximately the same capacity and number of layers present in pseudo-L3Fnet. The difference is that input to the new baseline is the actual DSLR image, while the input to pseudo-L3Fnet is DSLR image converted to pseudo-LF.}
\begin{figure}
	\scriptsize
	\centering
	\setlength{\tabcolsep}{1pt}	
	\begin{tabular}{cc}	
		\begin{adjustbox}{valign=t}
			\begin{tabular}{cccccc}
				\textbf{Ground Truth} &  \textbf{Baseline} & \textbf{SID~\cite{chen2018learning2seeindark}} & \textbf{Pseudo-L3Fnet} \\
				\includegraphics[width=0.24 \linewidth]{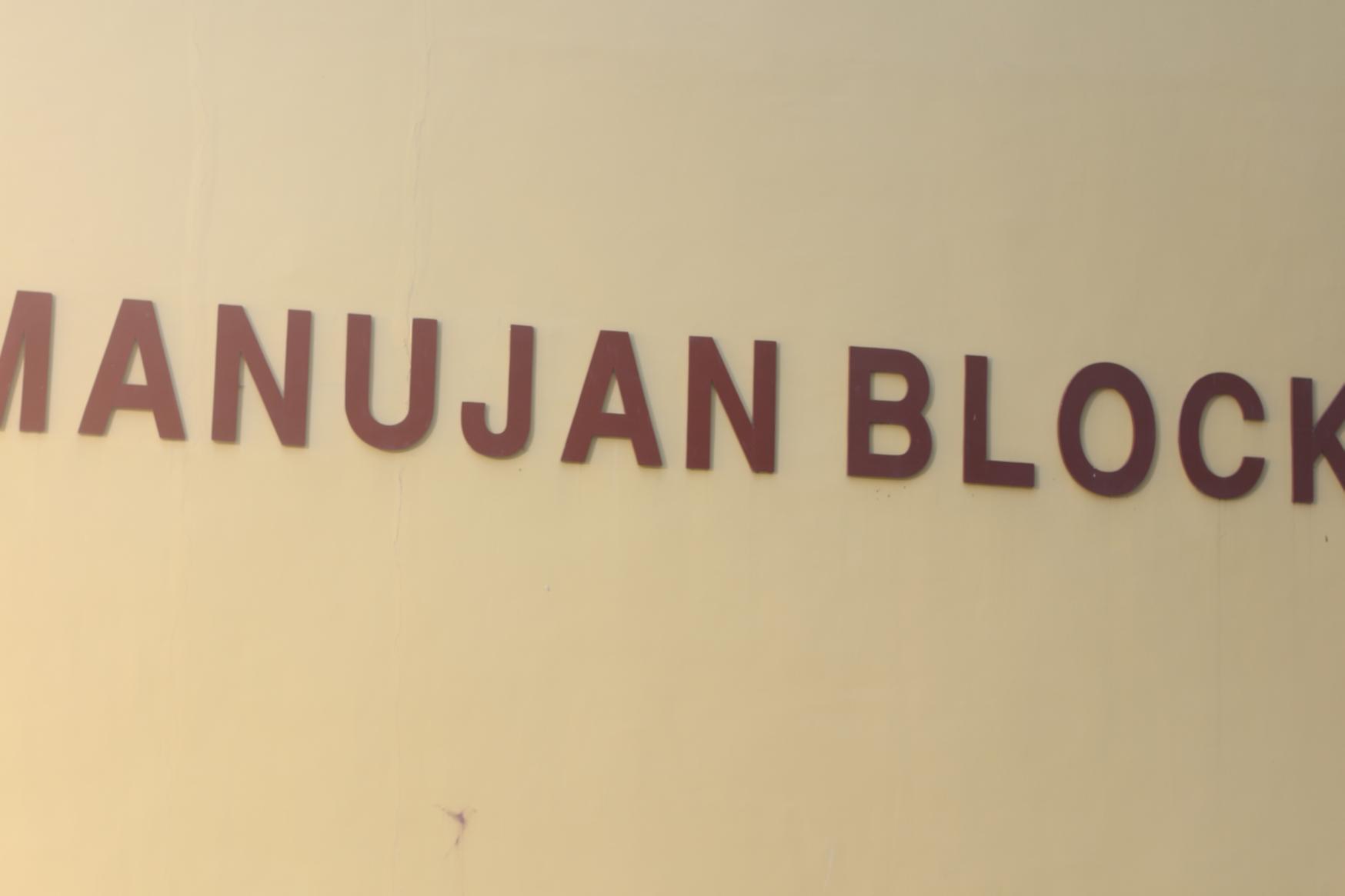} \hspace{-0.1cm} &
				\includegraphics[width=0.24\linewidth]{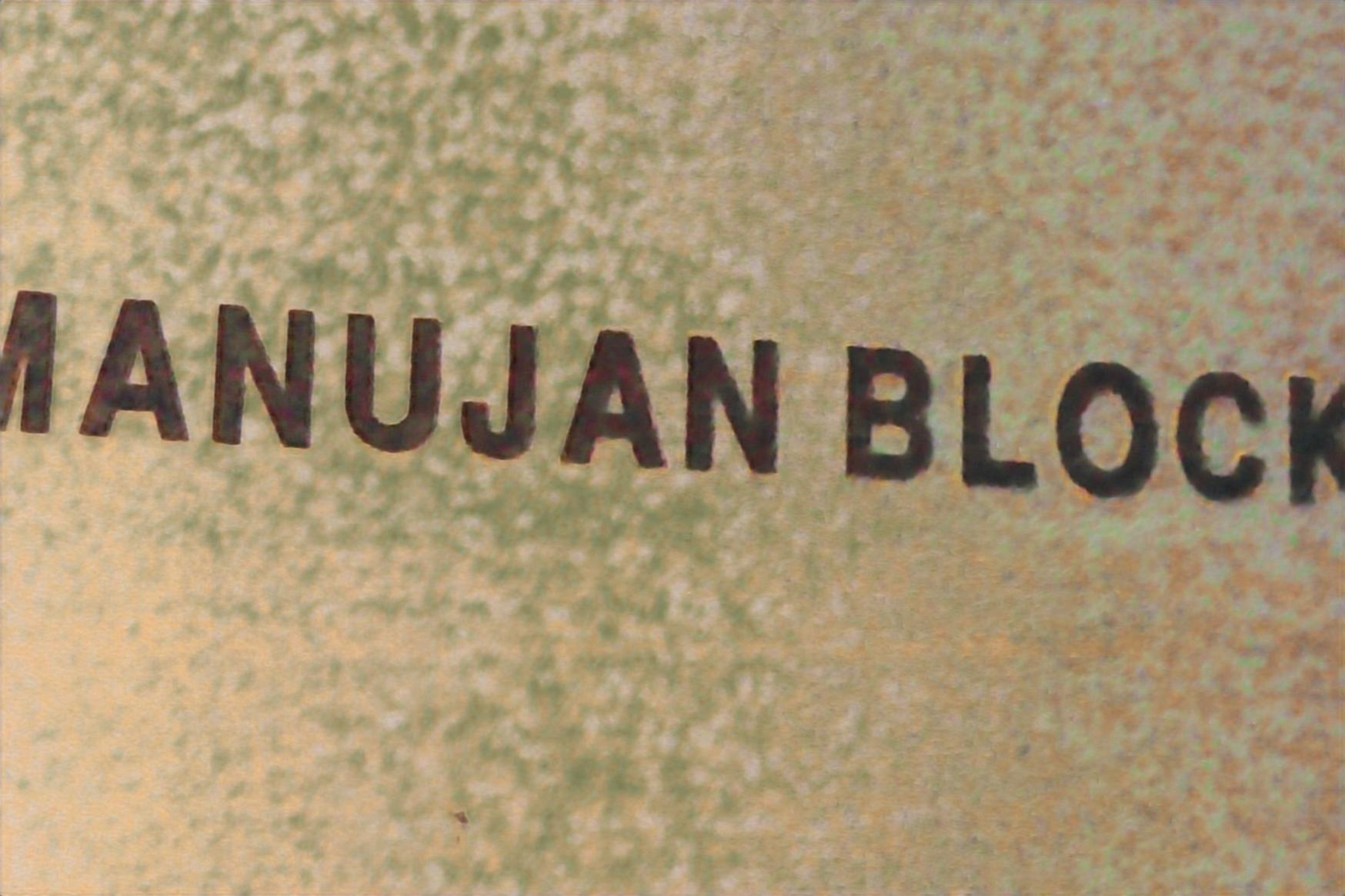} \hspace{-0.1cm} &
				\includegraphics[width=0.24\linewidth]{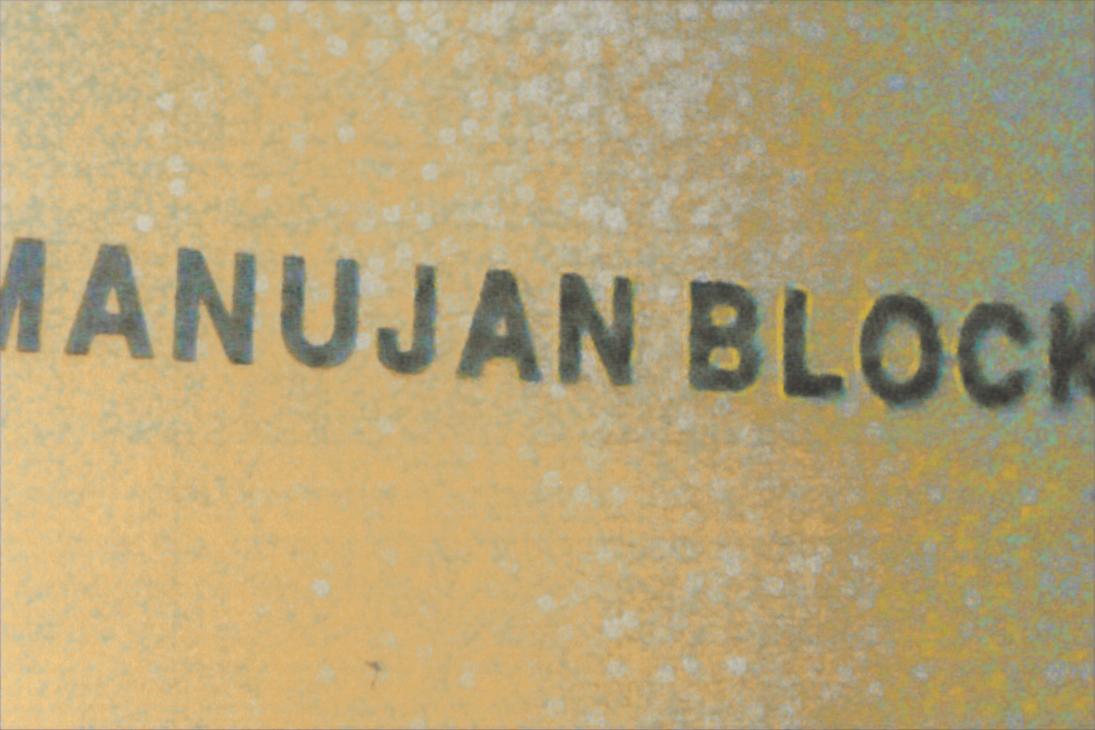} \hspace{-0.1cm} &
				\includegraphics[width=0.24\linewidth]{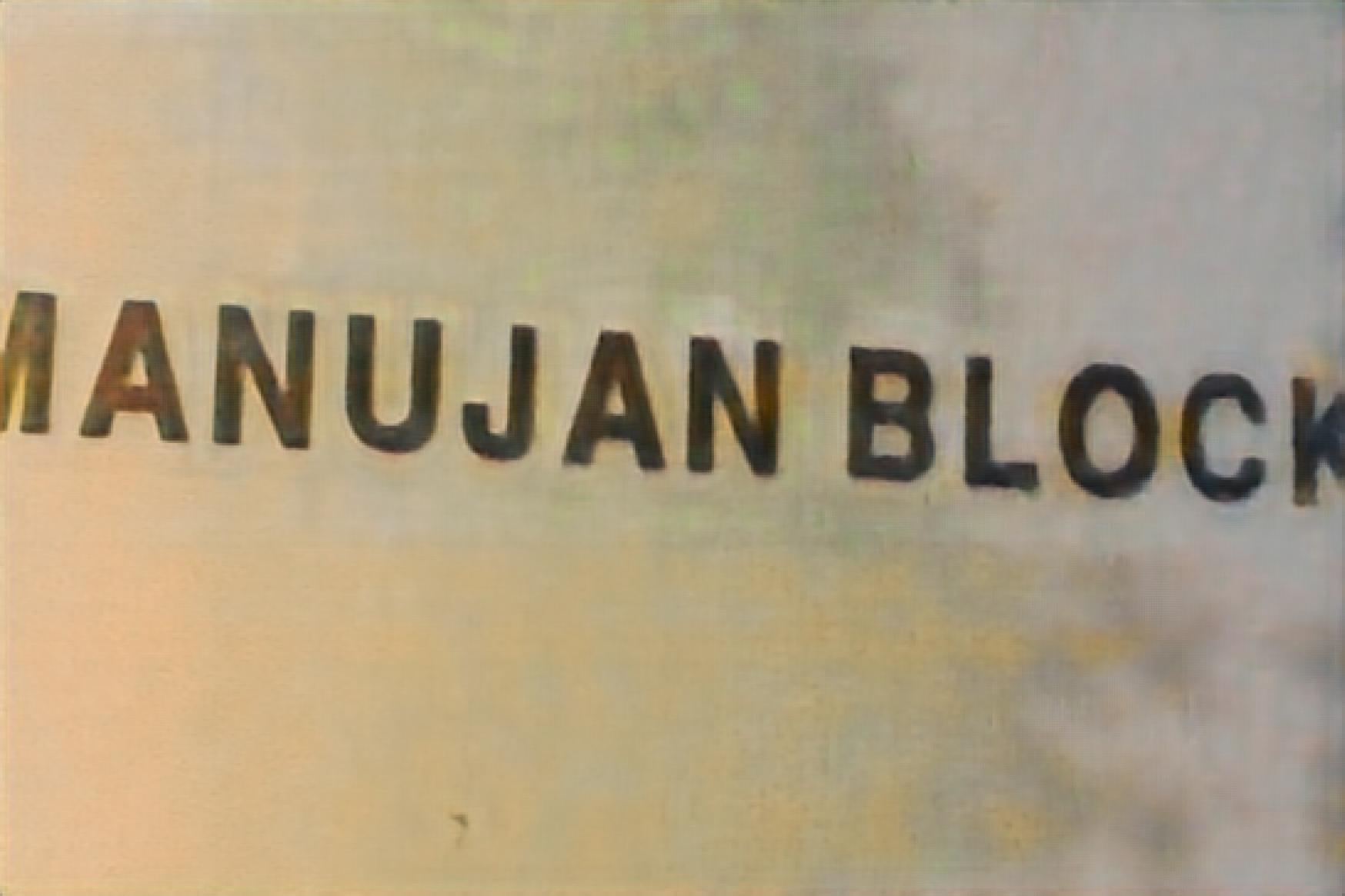} \hspace{-0.1cm} 			
				\vspace{-0.05cm}\\
				BRISQUE, NIQE& $47.68, 4.97$ & $44.71, 5.10$ & $\mathbf{38.35}, \mathbf{4.84}$ \\
				PSNR / SSIM& $17.75/\mathbf{0.92}$ & $20.48/0.85$ & $\mathbf{20.62}/0.83$ \vspace{0.1cm} \\ 
				\includegraphics[width=0.24 \linewidth]{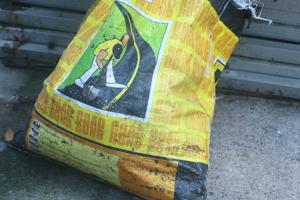} \hspace{-0.1cm} &
				\includegraphics[width=0.24\linewidth]{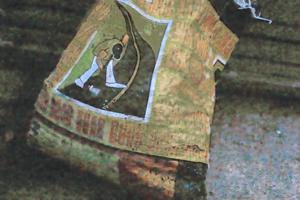} \hspace{-0.1cm} &
				\includegraphics[width=0.24\linewidth]{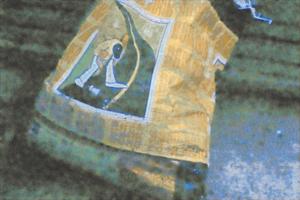} \hspace{-0.1cm} &
				\includegraphics[width=0.24\linewidth]{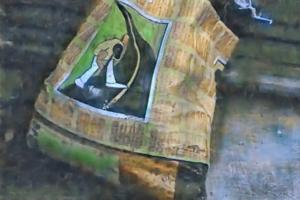} \hspace{-0.1cm} 			\vspace{-0.05cm}\\
				& $45.59, 5.07$ & $48.84, 5.11$ & $\mathbf{41.24}, \mathbf{3.92}$ \\
				& $17.92/0.42$ & $\mathbf{19.03}/0.47$ & $18.03/\mathbf{0.60}$ \vspace{0.1cm} \\
				
				\includegraphics[width=0.24 \linewidth]{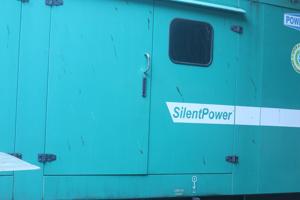} \hspace{-0.1cm} &
				\includegraphics[width=0.24\linewidth]{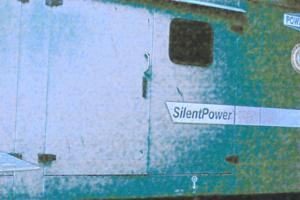} \hspace{-0.1cm} &
				\includegraphics[width=0.24\linewidth]{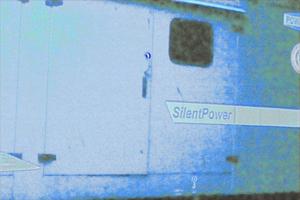} \hspace{-0.1cm} &
				\includegraphics[width=0.24\linewidth]{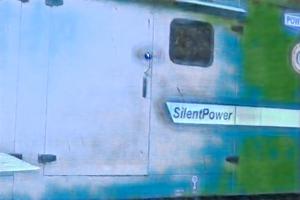} \hspace{-0.1cm} 			\vspace{-0.05cm}\\
				& $45.65, 4.99$ & $47.85, 5.26$ & $\mathbf{36.47}, \mathbf{4.36}$ \\
				& $\mathbf{12.76}/0.56$ & $12.16/0.58$ & $12.45/\mathbf{0.64}$ \vspace{0.1cm} \\
				\includegraphics[width=0.24 \linewidth]{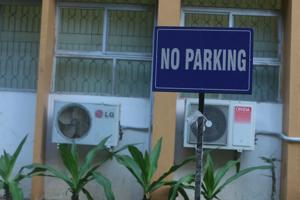} \hspace{-0.1cm} &
				\includegraphics[width=0.24\linewidth]{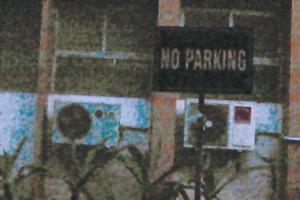} \hspace{-0.1cm} &
				\includegraphics[width=0.24\linewidth]{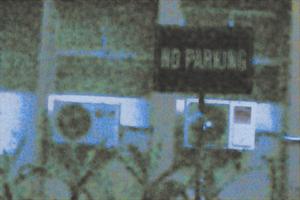} \hspace{-0.1cm} &
				\includegraphics[width=0.24\linewidth]{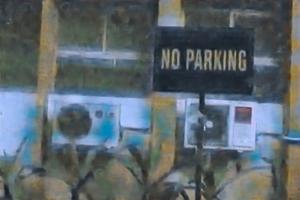} \hspace{-0.1cm} 			\vspace{-0.05cm}\\
				& $46.32, 5.22$ & $46.40, 5.26$ & $\mathbf{40.03}, \mathbf{3.53}$ \\
				& $19.68/0.36$ & $19.63/0.43$ & $\mathbf{20.77}/\mathbf{0.57}$

			\end{tabular}
		\end{adjustbox}			
	\end{tabular}
	\caption{Single-frame low-light DSLR restoration using the proposed pseudo-L3Fnet pipeline. BRISQUE and NIQE are recent perceptual metrics for single-frame DSLR images. Lower are these metrics, better is the perception.}
	\label{fig:pseudil3fnetResults}
	\vspace{-0.5cm}
\end{figure}

Some of the restoration results are shown in Fig. \ref{fig:pseudil3fnetResults}. The baseline and SID do good at denoising but very poorly on preserving the spatial smoothness, which is visible by the presence of color blobs spread all over the image. On the contrary, pseudo-L3Fnet obtains much gradual transitions mitigating the color blobs problem. This is because, as described in Sec. \ref{sec:pseudolf}, the pseudo-LF transformation awards L3Fnet with a large receptive field. \textcolor{black}{To verify this, we experimentally computed the receptive field and found that L3Fnet has a large receptive field of 830$\times$830, while for SID and Baseline it is just 252$\times$252 and 83$\times$83, respectively.}
The large receptive field of L3Fnet helps gather more contextual information for better restoration. 
The average PSNR(dB)/SSIM values for SID~is $17.21/0.61$, for baseline is $17.08/0.56$ and for the proposed Pseudo-LF is $19.04/0.62$. 
Further, the average NIQE \cite{niqe} values for SID~is $5.24$, for baseline is $5.01$ and $4.29$ for pseudo-LF. The average BRISQUE \cite{brisque} values for SID~is $48.30$, for baseline is $46.48$ and $41.45$ for the proposed pseudo-LF. A lower value for these metrics is considered better.

\section{Conclusion and Discussion}
The primary objective of this work was to enhance LF captured in low-light conditions. We showed that the existing single-frame low-light enhancements methods find it hard to preserve the LF geometry because they reconstruct each LF view independently. To this end, we proposed a low-light L3F dataset and a two-stage L3Fnet architecture. The effectiveness of L3Fnet was shown on LFs for varying levels of low-light by conducting experiments on L3F-20, L3F-50 and L3F-100 datasets. \textcolor{black}{Additionally, we used the Histogram Module to automatically tune the amplification factor $\gamma$.} With this pre-processing module, L3Fnet could now automatically adapt to different light levels, which was substantiated by showing results on the L3F-wild dataset.

We additionally showed that while single-frame methods are not conceptually suited for LF related tasks, our L3Fnet can be used for decent enhancement of single-frame low-light images also. This was achieved by converting the DSLR images into pseudo-LF and vice-versa. Of course, L3Fnet is better optimized for LF and may be modified in the future to equally suit both LF and DSLR images simultaneously.

As a future work, another interesting direction would be to explore which camera is more suited for low-light reconstruction: a single-frame DSLR camera or a Light Field camera? While DSLR cameras have high resolution, LF cameras may be helpful because of complementary information present in various LF views. Besides, depth estimation is a clear advantage for LF cameras.

\section{Supplementary}

\begin{figure}[h!]
	\centering
	\setlength{\tabcolsep}{1pt}	
	\begin{tabular}{cc}	
			\begin{tabular}{cccc}
			
				\includegraphics[width=0.23 \linewidth]{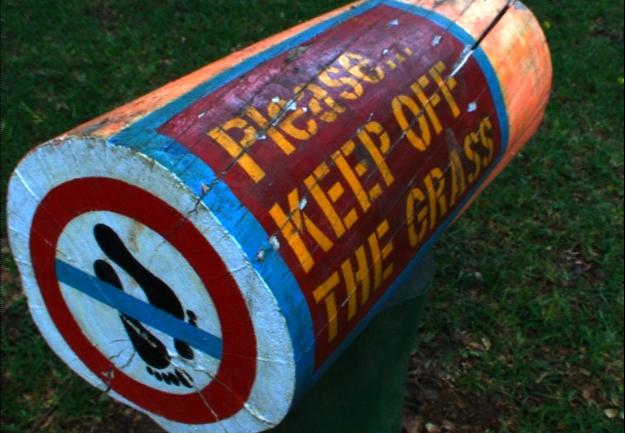}  & \includegraphics[width=0.23 \linewidth]{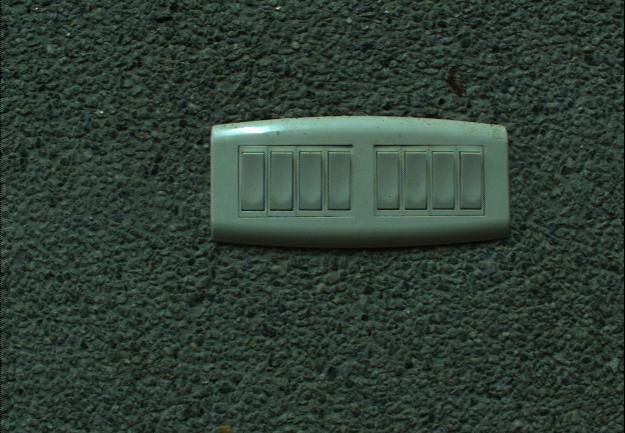}&\includegraphics[width=0.23 \linewidth]{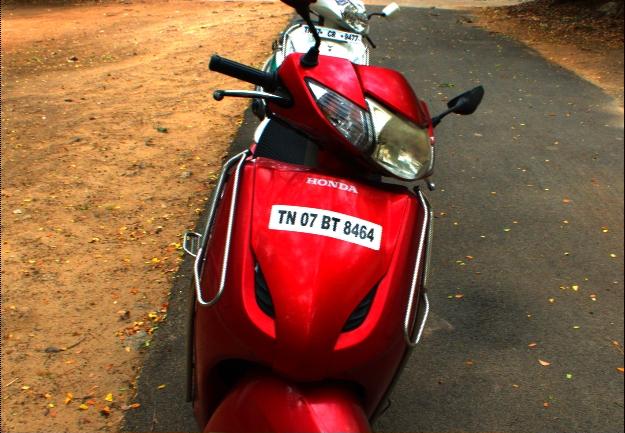}&\includegraphics[width=0.23 \linewidth]{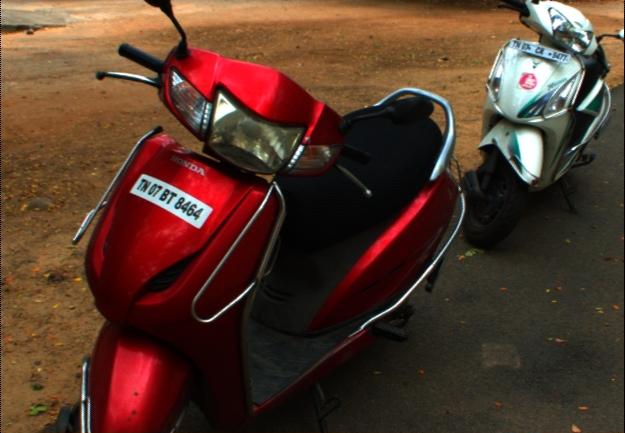}\\
				\includegraphics[width=0.23 \linewidth]{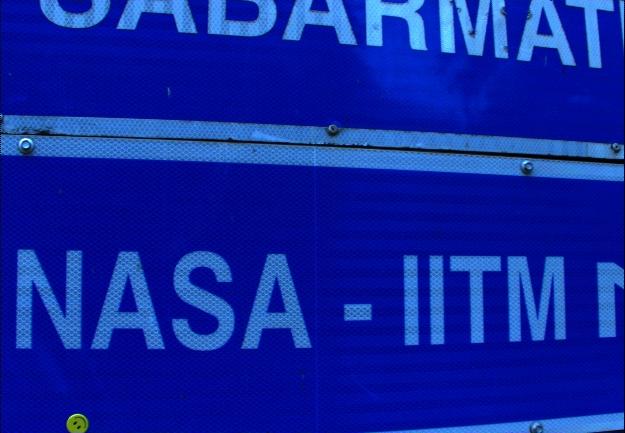}  & \includegraphics[width=0.23 \linewidth]{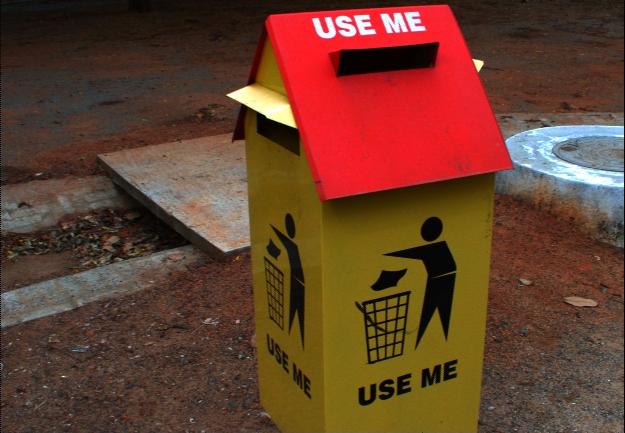}&\includegraphics[width=0.23 \linewidth]{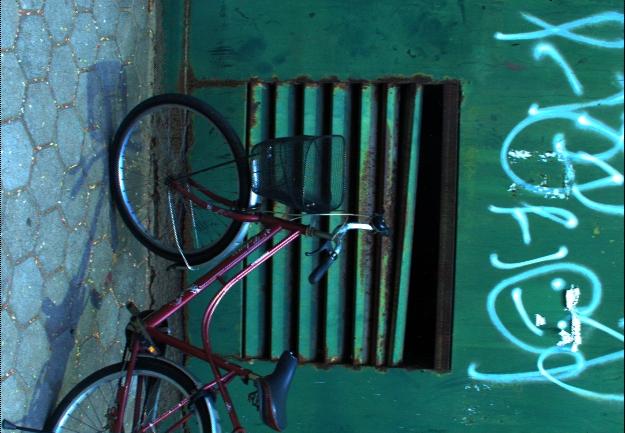}&\includegraphics[width=0.23 \linewidth]{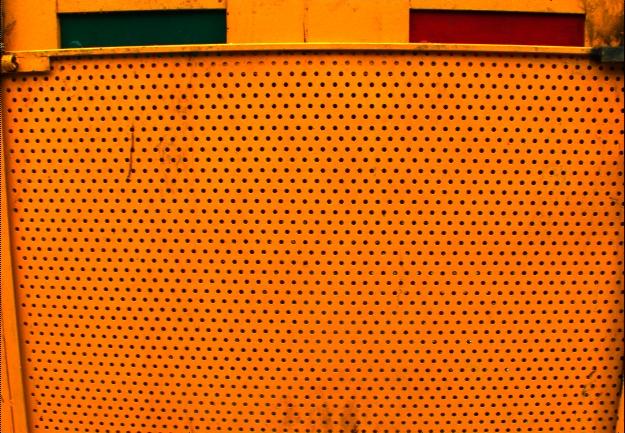}\\
				\includegraphics[width=0.23 \linewidth]{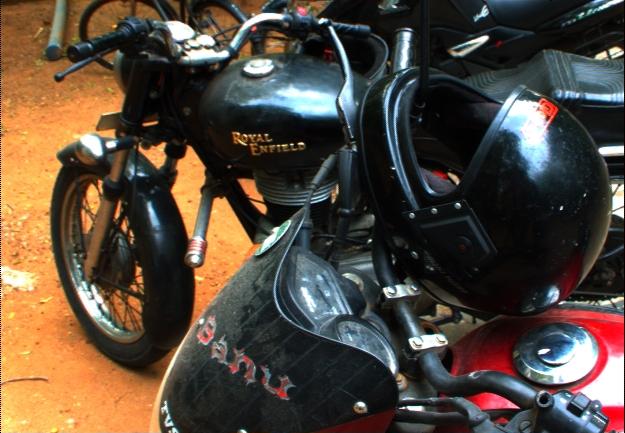}  & \includegraphics[width=0.23 \linewidth]{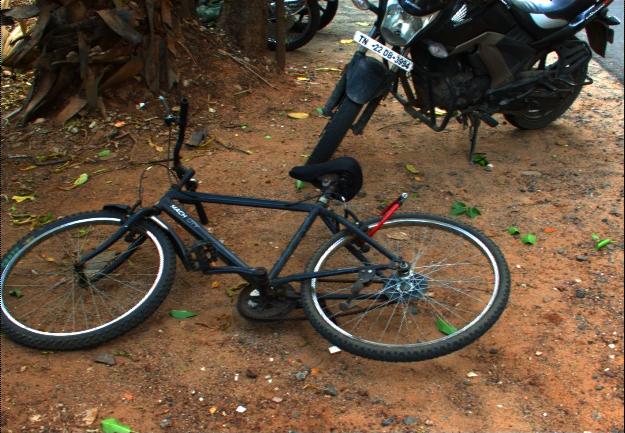}&\includegraphics[width=0.23 \linewidth]{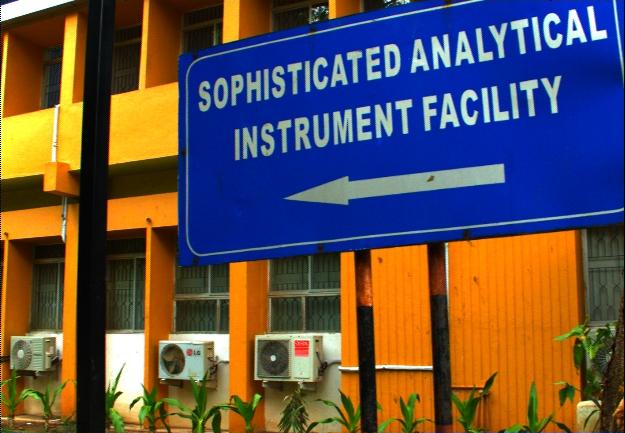}&\includegraphics[width=0.23 \linewidth]{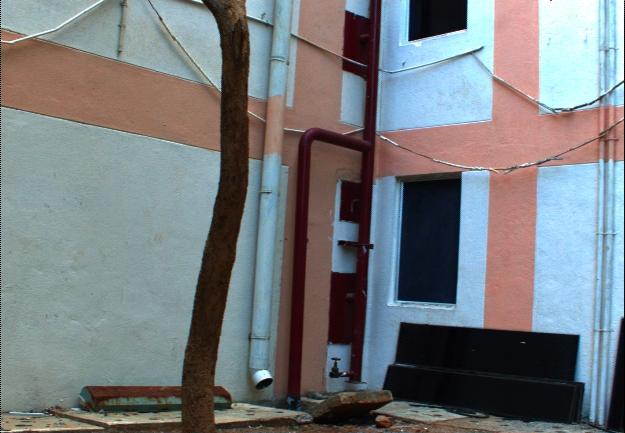}\\
				\includegraphics[width=0.23 \linewidth]{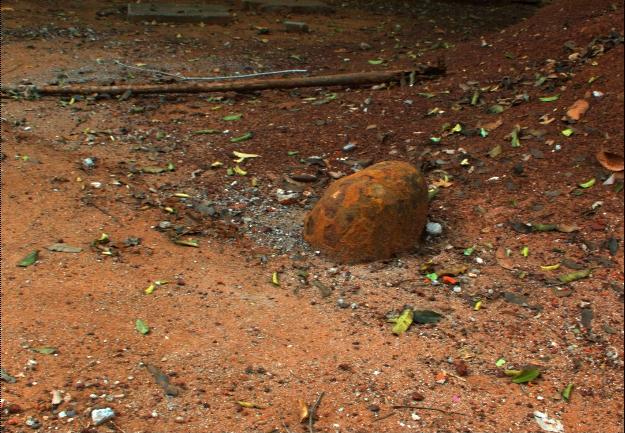}  & \includegraphics[width=0.23 \linewidth]{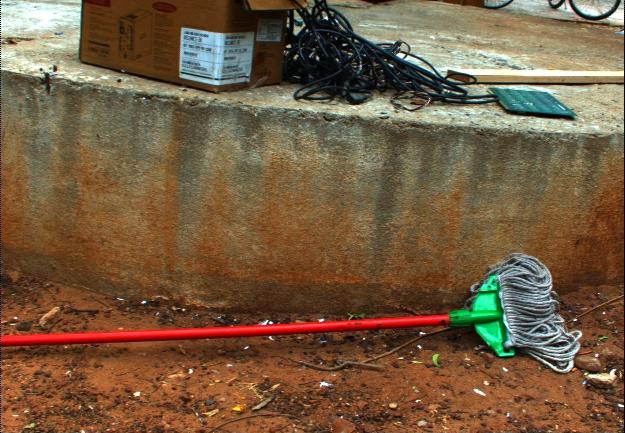}&\includegraphics[width=0.23 \linewidth]{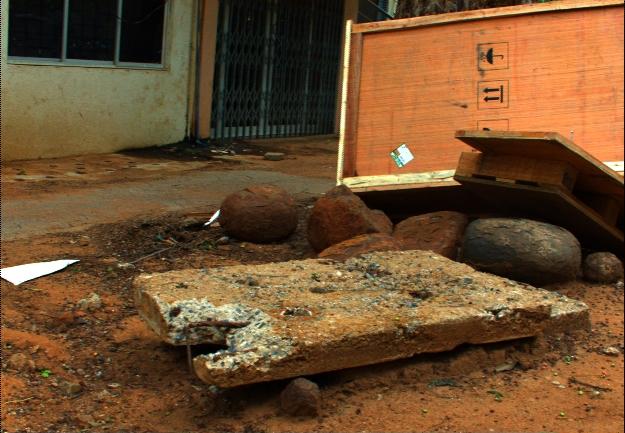}&\includegraphics[width=0.23 \linewidth]{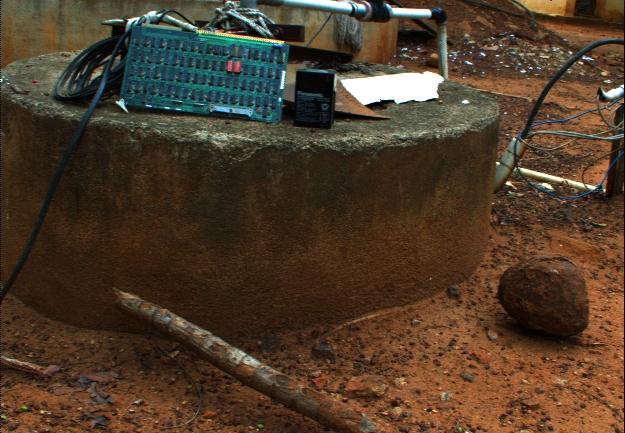}\\
				\includegraphics[width=0.23 \linewidth]{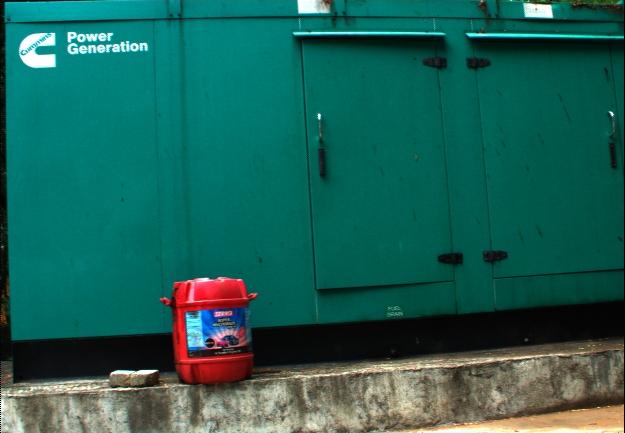}  & \includegraphics[width=0.23 \linewidth]{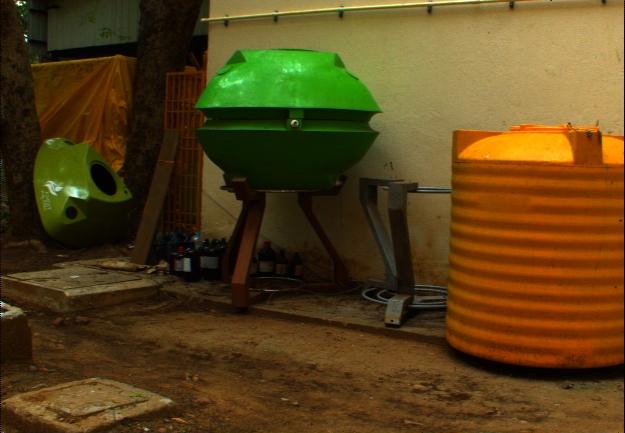}&&
				
				\end{tabular}
	\end{tabular}
	\caption{Central SAI of the ground truth LFs present in the training set of the proposed L3F dataset. This does not include the L3F-Wild dataset.}
\label{fig:train_set}

\end{figure}

\begin{figure}[h!]
	\centering
	\setlength{\tabcolsep}{1pt}	
	\begin{tabular}{c}	
			\begin{tabular}{ccc}
			
				\includegraphics[width=0.30 \linewidth]{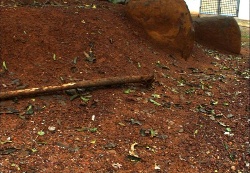}  & \includegraphics[width=0.30 \linewidth]{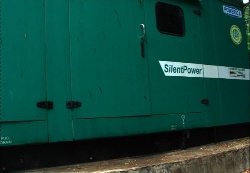}&\includegraphics[width=0.30 \linewidth]{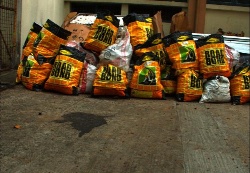}\\
				Forest & Generator & Cement\\
				
				\includegraphics[width=0.30 \linewidth]{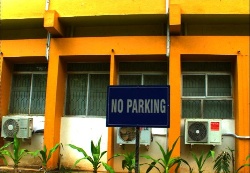}  & \includegraphics[width=0.30 \linewidth]{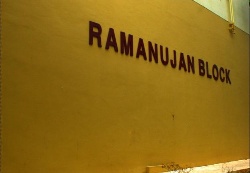}&\includegraphics[width=0.30 \linewidth]{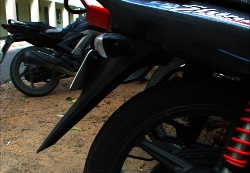}\\
				SignBoard & Ramanujam & Bikes\\
				
				\includegraphics[width=0.30 \linewidth]{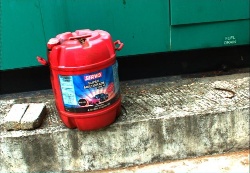}  & \includegraphics[width=0.30 \linewidth]{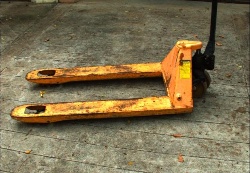}&
				\includegraphics[width=0.30 \linewidth]{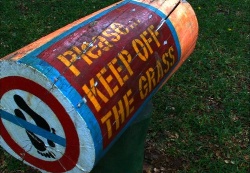}\\
				PaintBox & Crane & Grass
				
				\end{tabular}
	\end{tabular}
	\caption{Central SAI of the ground truth LFs present in the test set of the proposed L3F dataset. This does not include the L3F-Wild dataset. }
\label{fig:test_set}

\end{figure}

\begin{figure}[h!]
	\centering
	\setlength{\tabcolsep}{4pt}	

			\begin{tabular}{cc||cc}
			
			 SID & Ours &  SID & Ours  \\ 
				\includegraphics[trim=0 0 416 0,clip,width=0.23 \linewidth]{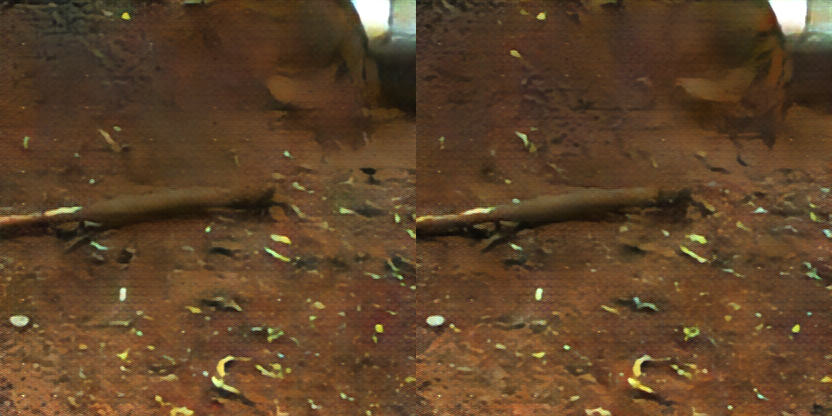} & \includegraphics[trim=0 0 416 0,clip,width=0.23 \linewidth]{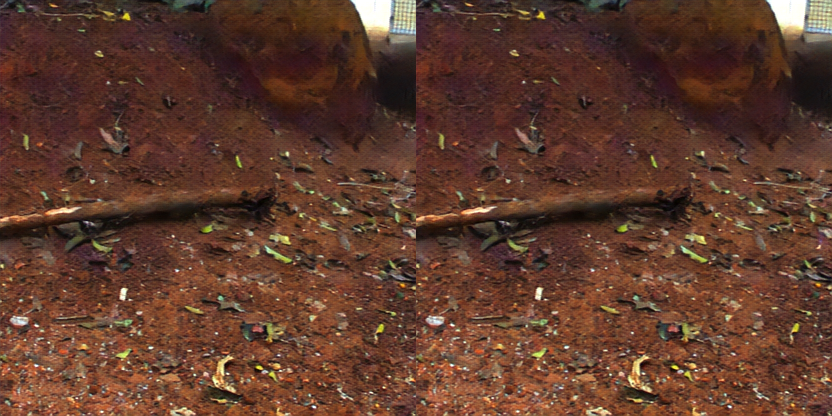}  &
				
				\includegraphics[trim=0 0 416 0,clip,width=0.23 \linewidth]{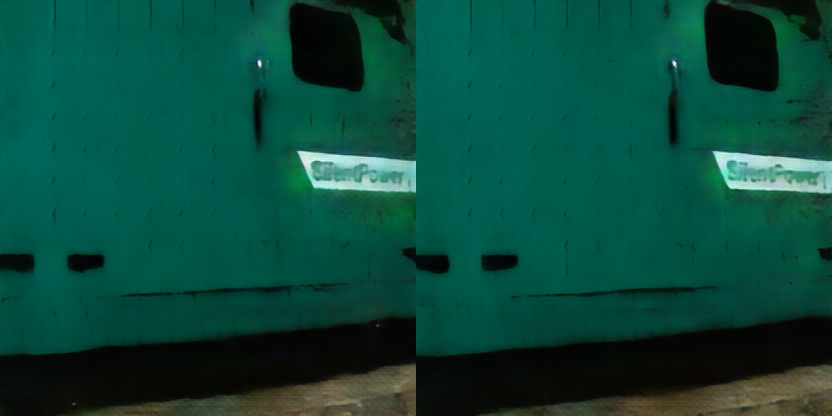}&\includegraphics[trim=0 0 416 0,clip,width=0.23 \linewidth]{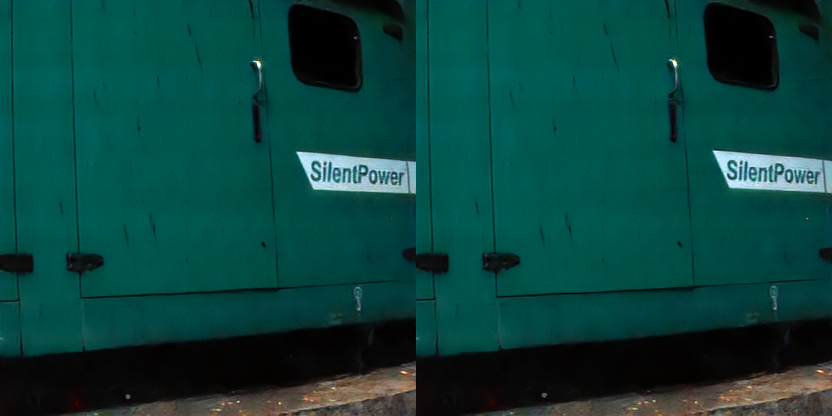}\\
				
				\includegraphics[trim=0 0 416 0,clip,width=0.23 \linewidth]{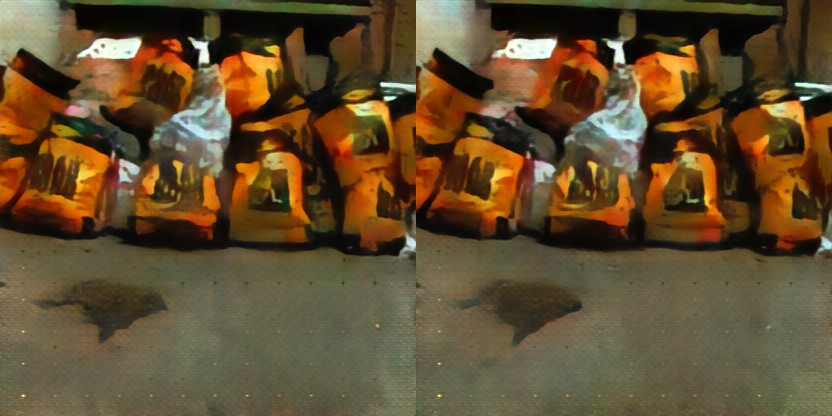}&\includegraphics[trim=0 0 416 0,clip,width=0.23 \linewidth]{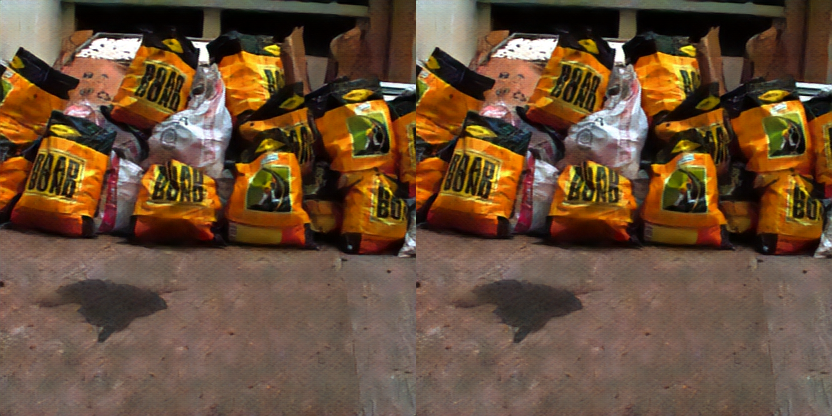}  &

				\includegraphics[trim=0 0 416 0,clip,width=0.23 \linewidth]{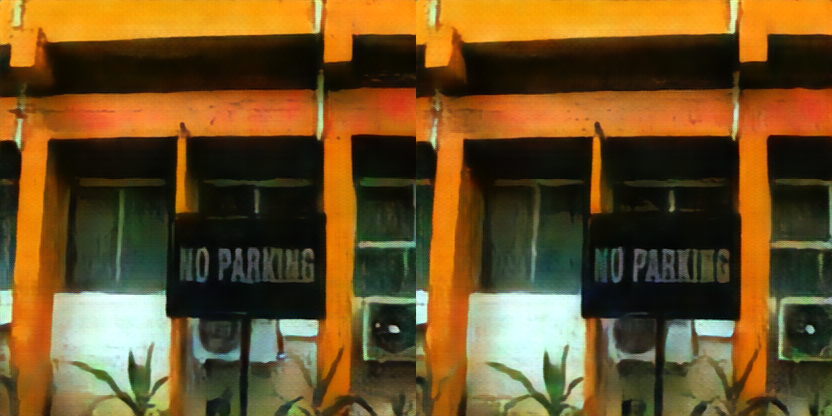}&\includegraphics[trim=0 0 416 0,clip,width=0.23 \linewidth]{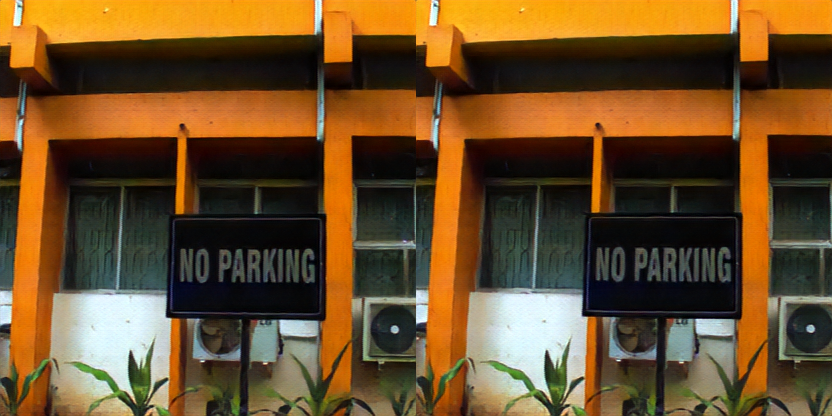}\\
				
				\includegraphics[trim=0 0 416 0,clip,width=0.23 \linewidth]{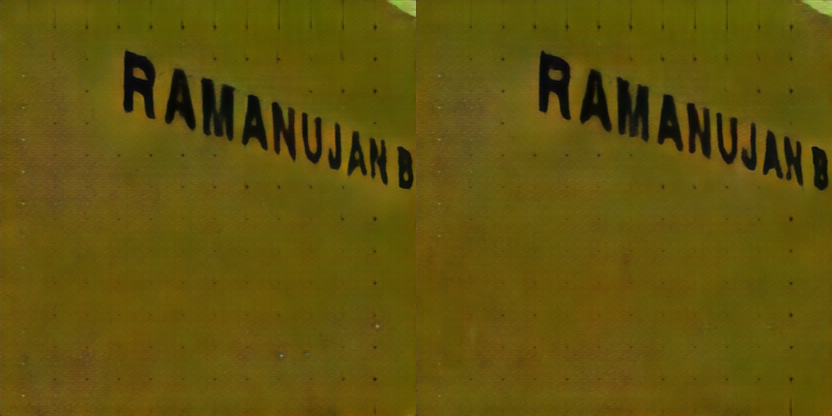}&\includegraphics[trim=0 0 416 0,clip,width=0.23 \linewidth]{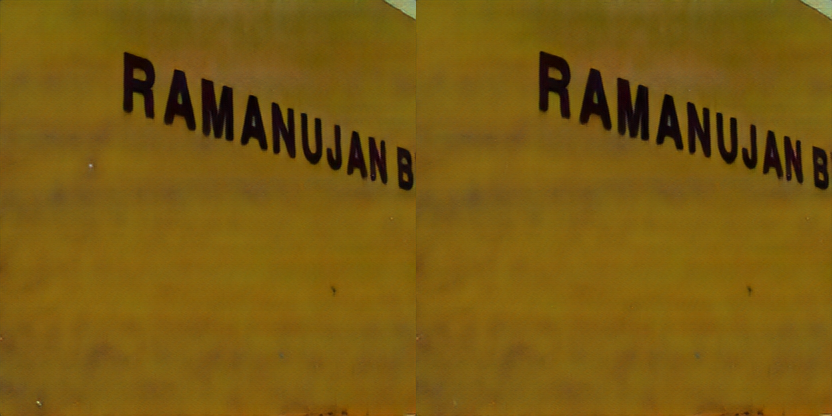}  &

				\includegraphics[trim=0 0 416 0,clip,width=0.23 \linewidth]{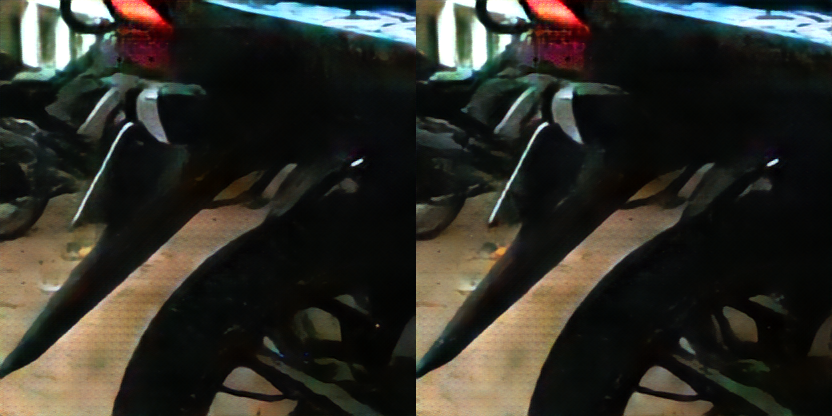}&\includegraphics[trim=0 0 416 0,clip,width=0.23 \linewidth]{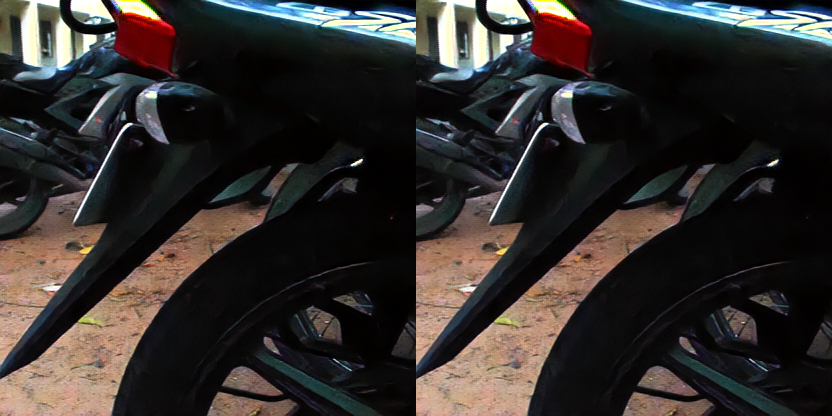}\\
				
				\includegraphics[trim=0 0 416 0,clip,width=0.23 \linewidth]{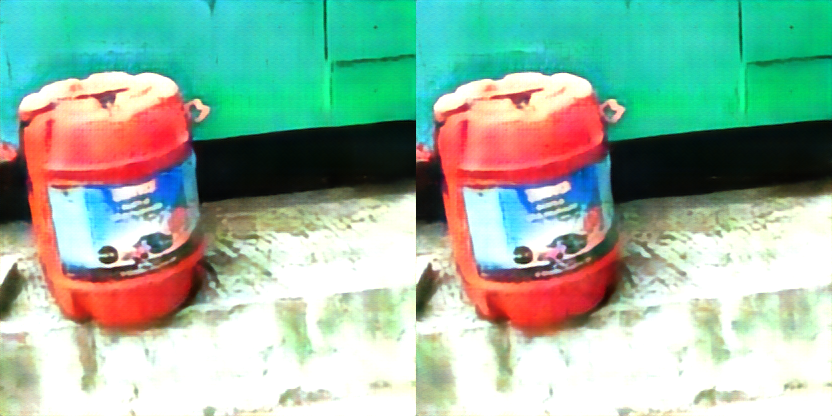}&\includegraphics[trim=0 0 416 0,clip,width=0.23 \linewidth]{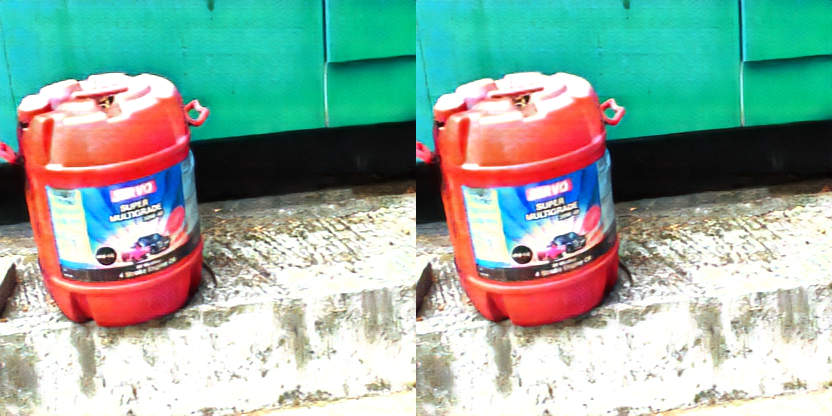}  &

				\includegraphics[trim=0 0 416 0,clip,width=0.23 \linewidth]{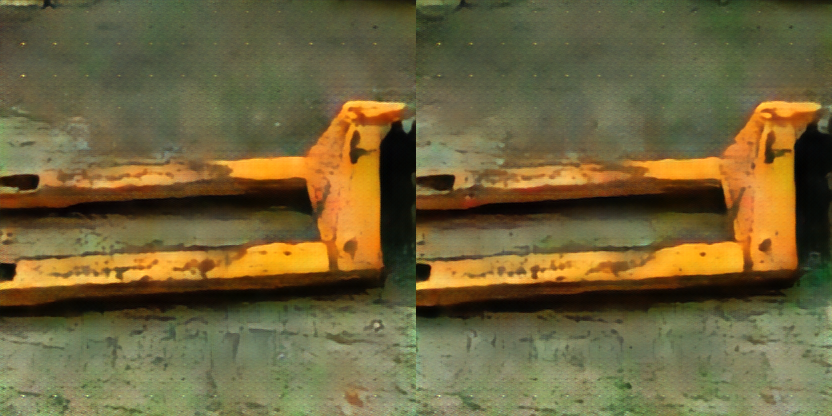}&\includegraphics[trim=0 0 416 0,clip,width=0.23 \linewidth]{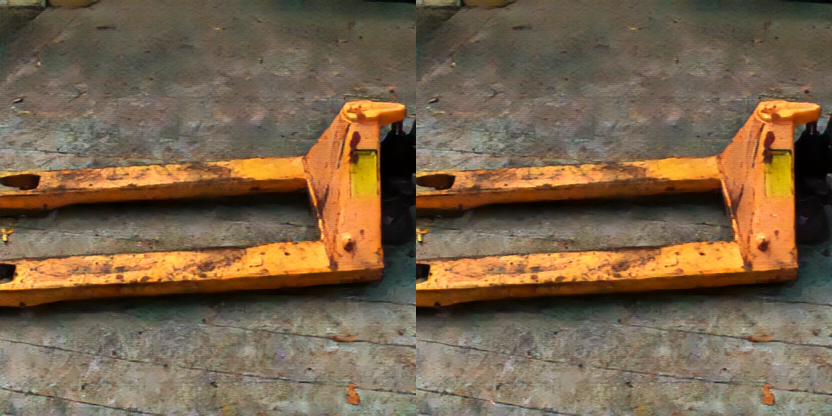}\\
				
				\includegraphics[trim=0 0 416 0,clip,width=0.23 \linewidth]{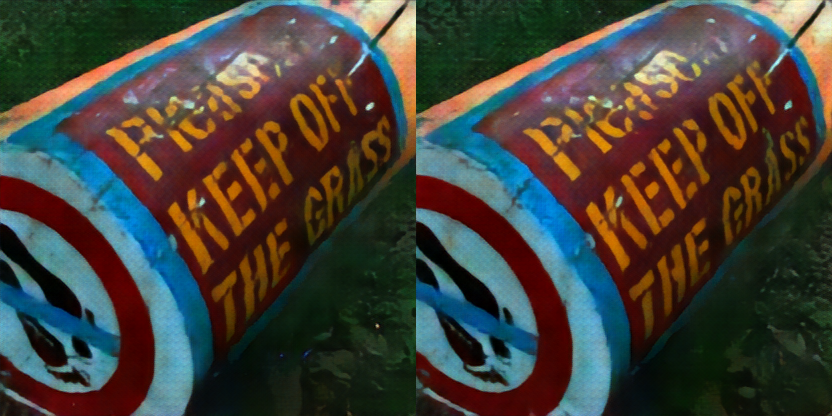}&\includegraphics[trim=0 0 416 0,clip,width=0.23 \linewidth]{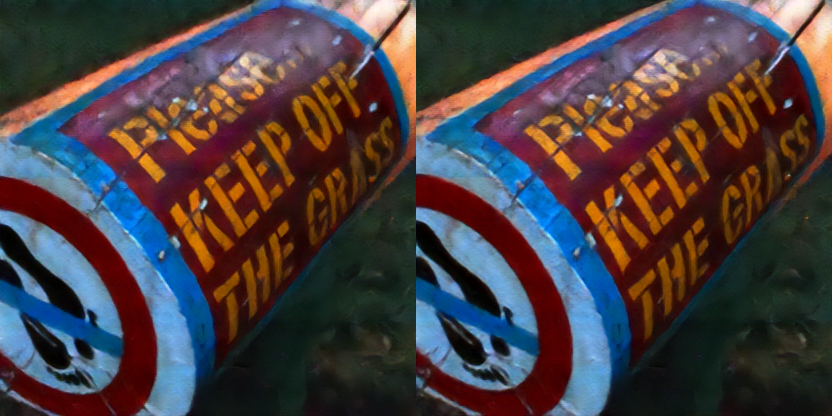} 
				\end{tabular}

\caption{L3F-100 Restoration results for the test data shown in Fig. \ref{fig:test_set}.}
\label{fig:test_results}
\end{figure}

\begin{table}[h!]
\centering
\caption{PSNR (dB)/ SSIM comparsion for the L3F test images shown in Fig. \ref{fig:test_set}}
\bgroup
\def\arraystretch{1.7}
\begin{tabular}{|c||c|c|}
\hline
\textbf{Image}       & \multicolumn{2}{c|}{\textbf{L3F-100 dataset}} \\ \hline
\textbf{}           & \textbf{SID \cite{chen2018learning2seeindark}} & \textbf{Our L3Fnet} \\ \hline \hline
\textbf{Forest}       & 22.03/0.38           & \textbf{24.79/0.67}    \\ \hline
\textbf{Generator}    & 28.01/0.80           & \textbf{31.30/0.85}    \\ \hline
\textbf{Cement}        & 21.22/0.50           & \textbf{21.44/0.62}    \\ \hline
\textbf{SignBoard}     & 18.44/0.59           & \textbf{19.41/0.68}    \\ \hline
\textbf{Ramanujam} & 19.89/0.72           & \textbf{22.86/0.79}    \\ \hline
\textbf{Bikes}  & 22.63/0.68           & \textbf{23.16/0.80}    \\ \hline
\textbf{PaintBox}    & 11.84/0.42           & \textbf{12.79/0.53}    \\ \hline
\textbf{Crane}    & 20.40/0.42           & \textbf{23.47/0.61}    \\ \hline
\textbf{Grass}    & 22.27/0.65           & \textbf{24.22/0.70}    \\ \hline
\end{tabular}
\egroup
\end{table}

\begin{figure}[h!]
	\centering
	\setlength{\tabcolsep}{1pt}	
			\begin{tabular}{c|c}
			\hline 
			Captured LF & Our Reconstructed\\ \hline~\\
				\includegraphics[width=0.40 \linewidth]{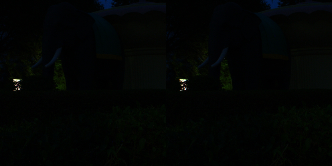}  & \includegraphics[width=0.40 \linewidth]{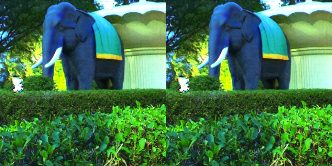}\\
				
				\includegraphics[width=0.40 \linewidth]{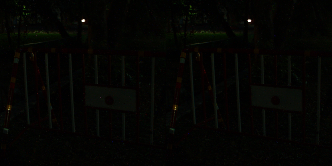}  & \includegraphics[width=0.40 \linewidth]{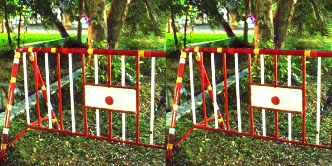}\\
				
				\includegraphics[width=0.40 \linewidth]{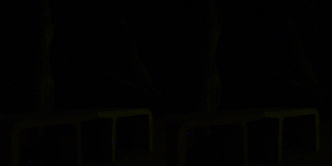}  & \includegraphics[width=0.40 \linewidth]{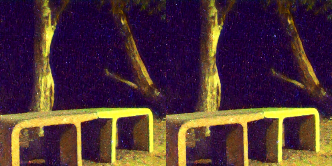}\\
				
				\includegraphics[width=0.40 \linewidth]{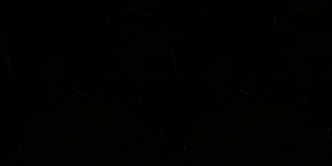}  & \includegraphics[width=0.40 \linewidth]{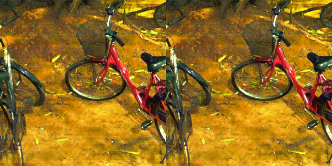}\\
				
				\includegraphics[width=0.40 \linewidth]{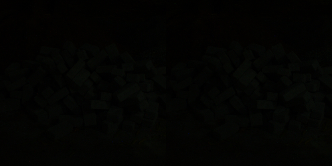}  & \includegraphics[width=0.40 \linewidth]{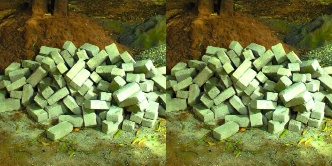}\\
				
				\includegraphics[width=0.40 \linewidth]{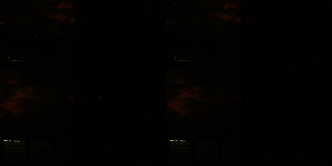}  & \includegraphics[width=0.40 \linewidth]{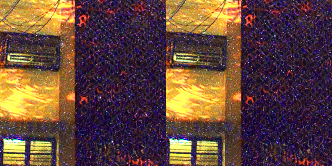}\\ \hline

				\end{tabular}
	\caption{Some L3F-Wild Reconstruction Results [Trained on L3F-20+L3F-50+L3F-100, see Fig. \ref{fig:train_set}] with Histogram Module Pre-processing. The figure shows two consecutive SAIs.  }
\label{fig:wild}

\end{figure}

\clearpage

\subsection{Effect of parameter count on performance}
\label{sec:Effect of parameter count on the performance}

The proposed L3Fnet with 3.7 Million parameters, outperforms the best performing baseline in our experiments SID \cite{chen2018learning2seeindark} which has more than 7 Million parameters. To further understand the effect of parameter count on model performance, we conducted additional experiments. In these experiments, the parameter count in L3Fnet were varied and then re-trained on the L3F-100 dataset (see Table \ref{table:parameter_count}). The details are given below:

\begin{table}[t!]
\caption{Variation in performance by changing the parameter count in the proposed L3Fnet. These networks have been re-training on the L3F-100 dataset. Even with much lower parameter count, we outperform SID \cite{chen2018learning2seeindark} which has more than 7 Million parameters. Notation: \textit{S1 = }Number of residual blocks in Stage-I, \textit{S2 = }Number of residual blocks in Stage-II, \textit{C =} Number of channels in each residual block, \textit{CT =} Number of Channels in the final Transposed convolution layer.}
\small
    \centering
    \renewcommand{\arraystretch}{1.5}
\begingroup
\setlength{\tabcolsep}{1pt}    \begin{tabular}{p{0.4 \linewidth}|c|c}
    \hline
    \textbf{Model description} & \textbf{Parameter count} & \textbf{PSNR(dB) / SSIM} \\ \hline \hline
    Proposed. \textit{S1=4, S2=6, C=128, CT=128} & 3.7 Million & 22.61/0.70 \\ \hline
    Reduce number of Residual Blocks. \textit{S1=2, S2=2, C=128, CT=128} & 2.2 Million & $22.60/0.69$ \\
    Increase the number of Residual Blocks \textit{S1=6, S2=9, C=128, CT=128} & 5.2 Million & 22.57/0.69 \\ \hline
    Reduce the channels in Residual Blocks. \textit{S1=4, S2=6, C=64, CT=128} & 1.4 Million & 21.87/0.68 \\
    Increase the channels in Residual Blocks. \textit{S1=4, S2=4, C=256, CT=128} & 13.4 Million & 22.50/0.69 \\ \hline
    Increase the richness of final activation layer. \textit{S1=4, S2=6, C=128, CT=1024} & 4.2 Million & 22.60/0.70 \\

      \hline
    \end{tabular}
    \endgroup

    \label{table:parameter_count}
\end{table}

\begin{itemize}
    \item \textbf{Reduce the number of residual blocks:} Currently, L3Fnet has 4 residual blocks for Stage-I and 6 for Stage-II. We reduce this to half and thus Stage-I has 2 residual blocks while Stage-II has 3. This reduces the parameter count from 3.7 Million to 2.2 Million. The performance is almost the same as the original network.   
    
    \item \textbf{Increase the number of residual blocks:} Currently, L3Fnet has 4 residual blocks for Stage-I and 6 for Stage-II. We increase this to 6 and 9 respectively. This increases the parameter count from 3.7 Million to 5.2 Million. The performance is almost the same as the original network. 
    
    \item \textbf{Reduce the channels in Residual Block: }Currently L3Fnet has 10 Residual Blocks each using 128 channels for each convolution. We reduced the number of channels to 64, which in turn reduced the parameter count from 3.7 Million to 1.4 Million. We observe a drop in performance. 
    
    \item \textbf{Increase the channels in Residual Block: }We increased the number of channels in the Residual Blocks from 128 to 256. This increased the parameter count from 3.7 Million to 13.4 Million. With this huge increase in the number of model parameters the performance decreases a bit possibly due to over-fitting.
    
    \item \textbf{Increase the richness of final Transposed Convolution: }To obtain the final 3 channel RGB image it is always good to average out a large feature map with richer activations. We thus increase the number of output channels in the Transpose-2D convolution form 128 to 1024. This increased the parameter count from 3.7 Million to 4.2 Million but with no change in performance. 
\end{itemize}
The results are summarized in Table \ref{table:parameter_count}. \textcolor{black}{Overall the performance of the proposed L3Fnet remains stable with the change in parameter count. Save and except for the case of $1.4~Million$ parameters, change in parameter count causes a variation of maximum 0.1 dB in the PSNR and 0.01 in the SSIM metric. A closer inspection of Table \ref{table:parameter_count} reveals that our results corroborate the `U-shaped' curve of the famous bias-variance trade-off. Initially when model parameters are less (close to 1 Million) the performance increases with parameter count. However, after some \textit{sweet spot} it starts to decline marginally with increase in model parameters.}

\subsection{Given a fixed number of Residual Blocks, how to distribute them among Stage-I \& Stage-II ?}
\label{sec:Changing the number of Residual Blocks in Stage-I and Stage-II}

\begin{table}[t!]
    \centering
    \caption{PSNR (dB) / SSIM metrics by swapping the number of Residual Blocks in Satge-I and Stage-II. In both cases L3Fnet was re-trained and tested on the L3F-100 dataset.}
    \renewcommand{\arraystretch}{1.5}
\begingroup
\setlength{\tabcolsep}{6pt}    \begin{tabular}{p{0.28 \linewidth}|p{0.28 \linewidth}|p{0.28 \linewidth}}
    \hline
      \textbf{Number of Residual Blocks in Stage-I}& \textbf{Number of Residual Blocks in Stage-II}& \textbf{PSNR(dB)/SSIM} \\ \hline \hline
       \centering 4 & \centering 6 & 22.61/0.70 (Proposed) \\
       \centering 6 & \centering 4 &  22.30/0.69\\
       \centering 3 & \centering 7 &  22.28/0.69\\
       \centering 7 & \centering 3 &  22.27/0.68\\
      \hline
    \end{tabular}
    \endgroup

    \label{tab:swap}
\end{table}

For the proposed L3Fnet we have used 4 residual blocks for Stage-I and 6 for Stage-II. We chose more residual blocks for Stage-II because our main goal is LF enhancement, which is being explicitly performed by Stage-II. Stage-I looks at all the 64 SAIs and computes a Global Representation.
In the restoration process however, it should not happen that to restore a pixel from a particular SAI $\phi$, the network is looking at all the SAIs without any discrimination. \textit{Rather to restore a pixel from SAI $\phi$ the network should gather more contextual information from $\phi$ and its immediate neighbors}. To allow this, we designed Stage-II which looks at the Global Representation of Stage-I but at the same time gives exclusive attention to the particular SAI (and its 4-nearest neighbors) which is being restored. Given that we want to give more weightage to the immediate neighbors, we allocated more residual blocks to Stage-II.

We conducted more experiments by varying the number of residual blocks in Stage I and II, see Table \ref{tab:swap}. For all these experiments, training and testing was carried out on L3F-100 dataset. We kept the total number of residual blocks to 10 so that the parameter count of L3Fnet is low (it is less than half of what is found in SID \cite{chen2018learning2seeindark} and yet outperform it). We observe that flipping the number of residual blocks between Stage I and II, results in a small drop in performance. We also tried $7$ residual blocks for stage-I and $3$ for stage-II and the other way around, see Table \ref{tab:swap}. For both the cases, the results are lower than the proposed L3Fnet.

\subsection{Trying different combinations of L1, Contextual and SSIM loss}

\begin{table}[t!]
    \centering
    \caption{PSNR/SSIM metrics of the proposed L3Fnet on the L3F-100 dataset for various loss functions.}
    \renewcommand{\arraystretch}{1.5}
\begingroup
\setlength{\tabcolsep}{6pt}
    \begin{tabular}{ccc|c}
    \hline
      \textbf{L1 loss}& \textbf{Contextual loss}& \textbf{SSIM loss}& \textbf{PSNR(dB)~/~SSIM}\\
      \hline \hline
       \cmark&\xmark&\xmark & 21.71/0.63\\
      \cmark&\cmark&\xmark & 22.61/0.701\\
      \cmark&\xmark&\cmark & 22.40/0.711\\
      \cmark&\cmark&\cmark & 22.45/0.710\\
      \hline
    \end{tabular}

    \endgroup
    
    \label{table:ssim_ablation}
\end{table}



The main reason why we chose Contextual Loss was to handle any minor camera shake that would have occurred during the dataset collection.
But we also tried using the SSIM loss. We do not find much difference in performance in either using SSIM or Contextual Loss, see Table \ref{table:ssim_ablation}. Using the SSIM loss, SSIM increases marginally, but the decrease in PSNR is slightly more. \textcolor{black}{And since recent methods such as Gu et al. \cite{gu2019self} and recently popular NTIRE challenges (\url{https://data.vision.ee.ethz.ch/cvl/ntire20/}, \url{https://data.vision.ee.ethz.ch/cvl/ntire19/}, \url{https://data.vision.ee.ethz.ch/cvl/ntire18/}) prefer PSNR over SSIM for deciding the best performance, we do not include the SSIM loss.}

{\color{black}
\subsection{On the use of Linear amplification vs Gamma Correction}
\textbf{Range of $\gamma$: }Table \ref{tab:gamma_values} shows the range of $ \gamma $. The learned $\gamma$ values follow our intuition that it should be higher for darker images.

\textbf{Can $\gamma$ cause over-saturation? : }\textcolor{black}{No, it is very unlikely that the Histogram Module can cause oversaturation. The Histogram Module is applied at the beginning of L3Fnet and not on the final restored image. As everything is learned end-to-end, the supervision provided by the GT images would modify the model weights such that the restored image is within the desired range. 
Besides this, the pixel intensities of the captured low-light LF are very small and the learned $\gamma$ values are less than 2 (see Table \ref{tab:gamma_values}). Thus multiplying the input low-light LF with $\gamma$ does not cause oversaturation and the pixel intensities still remain in the range [0,1] \footnote{In our experiments all images are normalized in the range [0,1]. Thus the maximum pixel intensity in our experiments is 1. }. Some restoration results are shown in Fig. \ref{fig:wild} .}

\textbf{Ablation study on linear amplification vs. Gamma correction: } In an ablation study we re-trained the Histogram Module with the proposed L3Fnet in an end-to-end fashion with the linear multiplication replaced with gamma correction. Specifically, $\mathcal{L}^{norm} = \gamma \times \mathcal{L}^{low}$ was changed to $\mathcal{L}^{norm} =  (\mathcal{L}^{low})^\gamma$.
In Fig. \ref{fig:gamma_correction} we find that using gamma correction for normalizing the images, oversaturates the bright pixels in order to boost the dark pixels and exhibits some artifacts in the vicinity of bright pixels. 
A possible reason for this could be the fact that Gamma correction introduces higher order noise terms which will be more difficult to remove by the later CNN stages. We elaborate on this in the next paragraph. 

Whenever we capture images in low light, it is affected by a large amount of noise. For simplicity, we can assume $y = x + n$, where $y$ is the captured image in low light, $x$ is the actual signal we intended to capture and $n$ is the large amount of noise present due to low-light conditions. Now if we use a linear amplification we obtain, $\gamma y = \gamma x + \gamma n$. If however, we use the power law (gamma correction), we obtain,
\begin{eqnarray}
y^\gamma &=& x^\gamma \left(1+\frac{n}{x}\right)^\gamma \nonumber \\
&=& x^\gamma + \frac{\gamma n  x^{\gamma-1}}{1!} + \frac{(\gamma)(\gamma-1) n^2  x^{\gamma-2}}{2!} \nonumber \\
&& \quad + \frac{(\gamma)(\gamma-1)(\gamma-2) n^3  x^{\gamma-3}}{3!} + ...
\end{eqnarray}

\begin{table}[t!]
    \centering
    \caption{\textcolor{black}{Range of $\gamma$ values learned for different datasets. We also compute the mean and maximum pixel intensity for each test image present in the dataset, and their range is shown in the table. The learned $\gamma$ values follow our intuition that $\gamma$ should be higher for darker images.}}
    \renewcommand{\arraystretch}{1.5}

    \begingroup
\setlength{\tabcolsep}{3pt}
    \begin{tabular}{c|c|c|c}
      \hline \textbf{Dataset} & \textbf{Mean pixel intensity} & \textbf{Max pixel intensity} & \textbf{$\gamma$ range} \\ \hline \hline
        L3F-20 & $0.0097-0.0257$& $0.06-0.77$ & $0.3-0.4$\\ 
        L3F-50 & $0.0043-0.0107$& $0.04-0.41$ & $0.8-0.9$ \\
        L3F-100 & $0.0026-0.0084$& $0.04-0.16$ & $1.4-1.7$\\ \hline
    \end{tabular}
    \endgroup
    
    \label{tab:gamma_values}
\end{table}

As for low-light $n$ is very large, the CNN will have a tough time to remove all the higher order terms.

\begin{figure}[t!]
	\centering
	\setlength{\tabcolsep}{1pt}
			\begin{tabular}{ccc}
			\includegraphics[width=0.25 \linewidth, ]{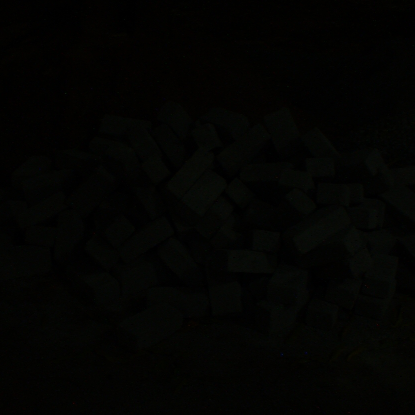} & \includegraphics[width=0.25 \linewidth ]{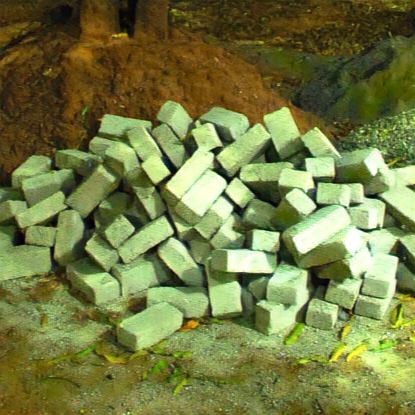} & \includegraphics[width=0.25 \linewidth ]{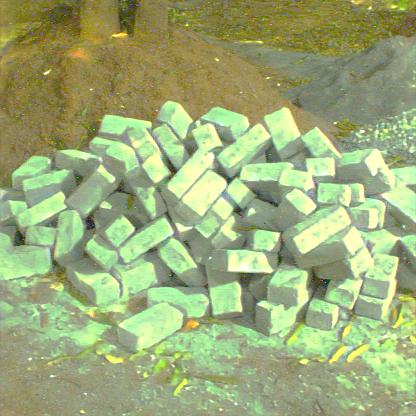} \\
			\includegraphics[width=0.25 \linewidth, ]{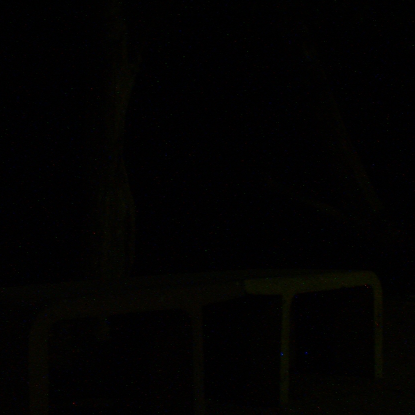} & \includegraphics[width=0.25 \linewidth ]{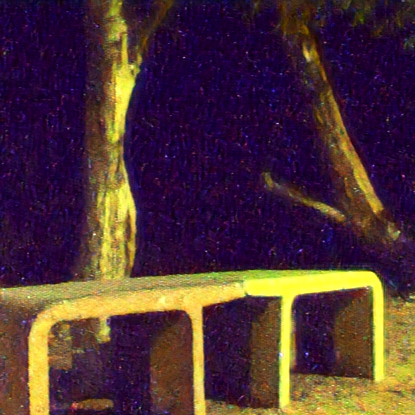} & \includegraphics[width=0.25 \linewidth ]{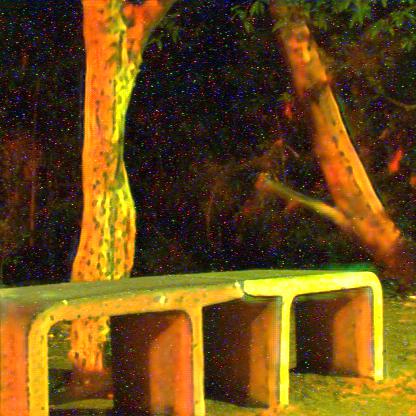}
			  \\
			 Input & Linear & Gamma \\
			 & Amplification & Correction 
\end{tabular}
\caption{Restoration results using L3Fnet by first normalizng the input LF using the proposed linear amplification or by using gamma correction. Using gamma correction for normalizing the images, oversaturates the bright pixels in order to boost the dark pixels and exhibits artifacts in the vicinity of bright pixels.}
\label{fig:gamma_correction}
\end{figure}

\subsection{Quantifying the misalignment in the proposed Dataset}

\begin{figure*}
    \centering
    \begin{tabular}{cc}
         \includegraphics[width=0.48\linewidth]{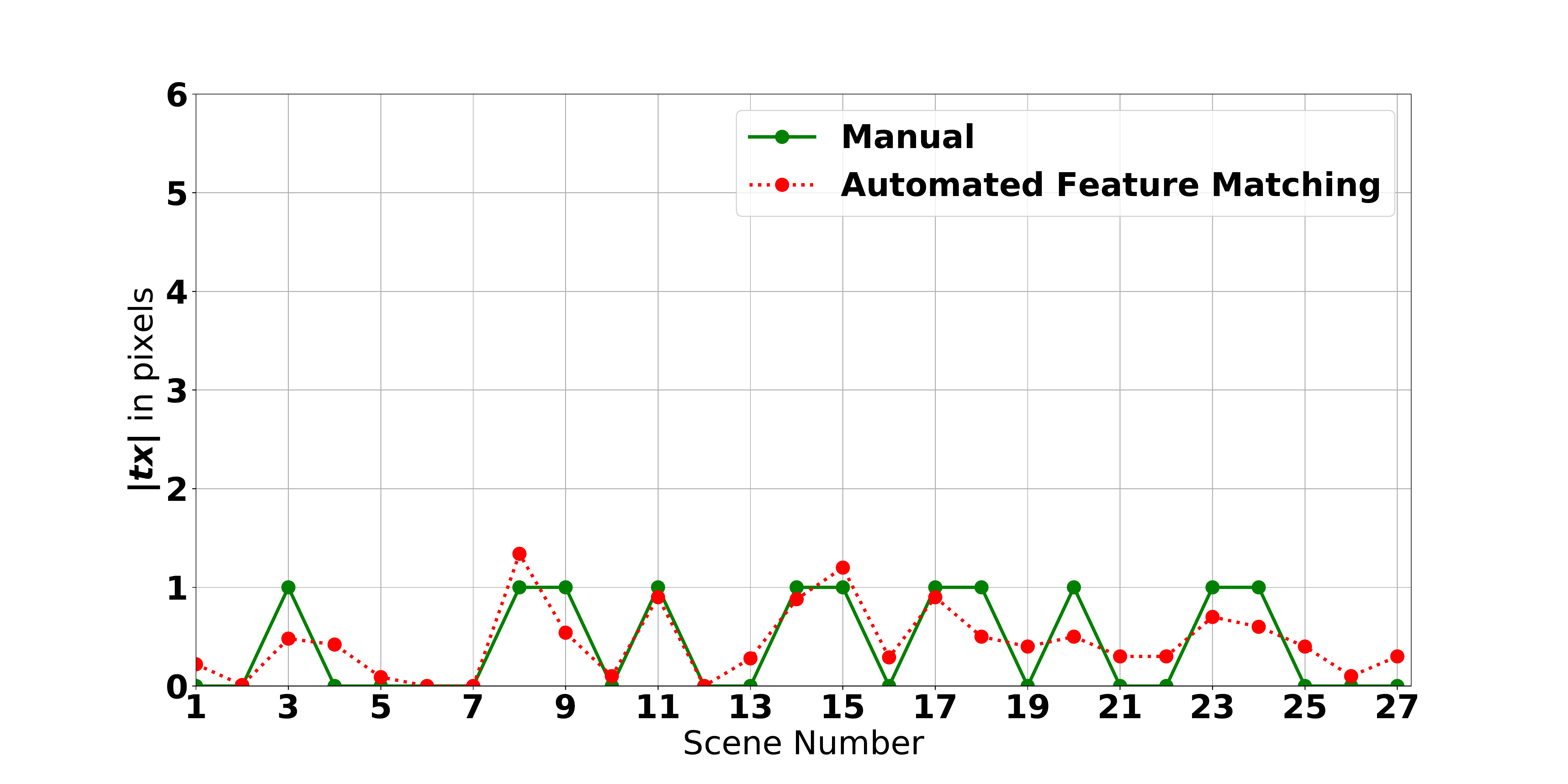} &
         \includegraphics[width=0.48\linewidth]{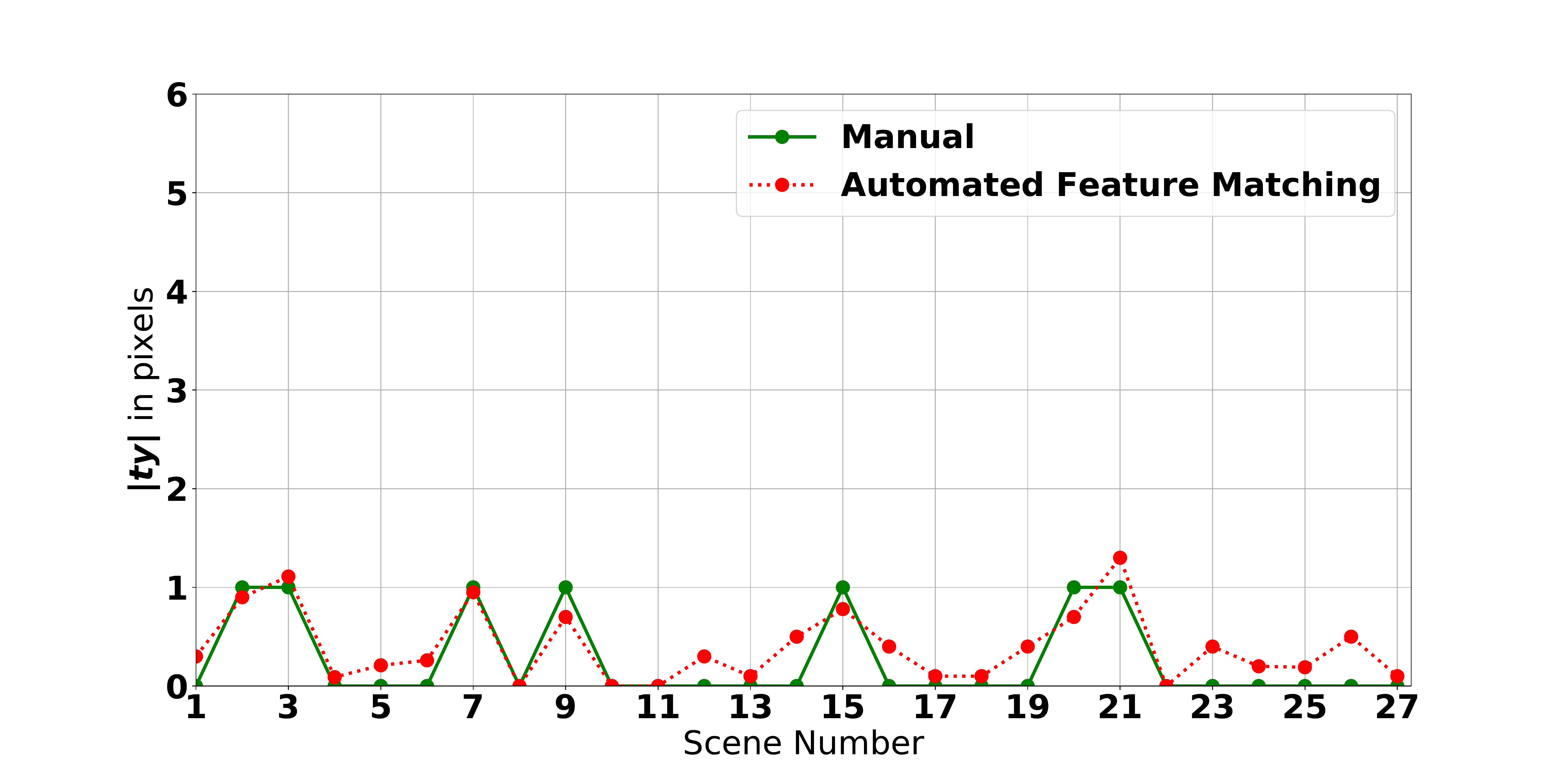} \\
    \multicolumn{2}{c}{(a) L3F-20 dataset} \vspace{10pt} \\ 
    \includegraphics[width=0.48\linewidth]{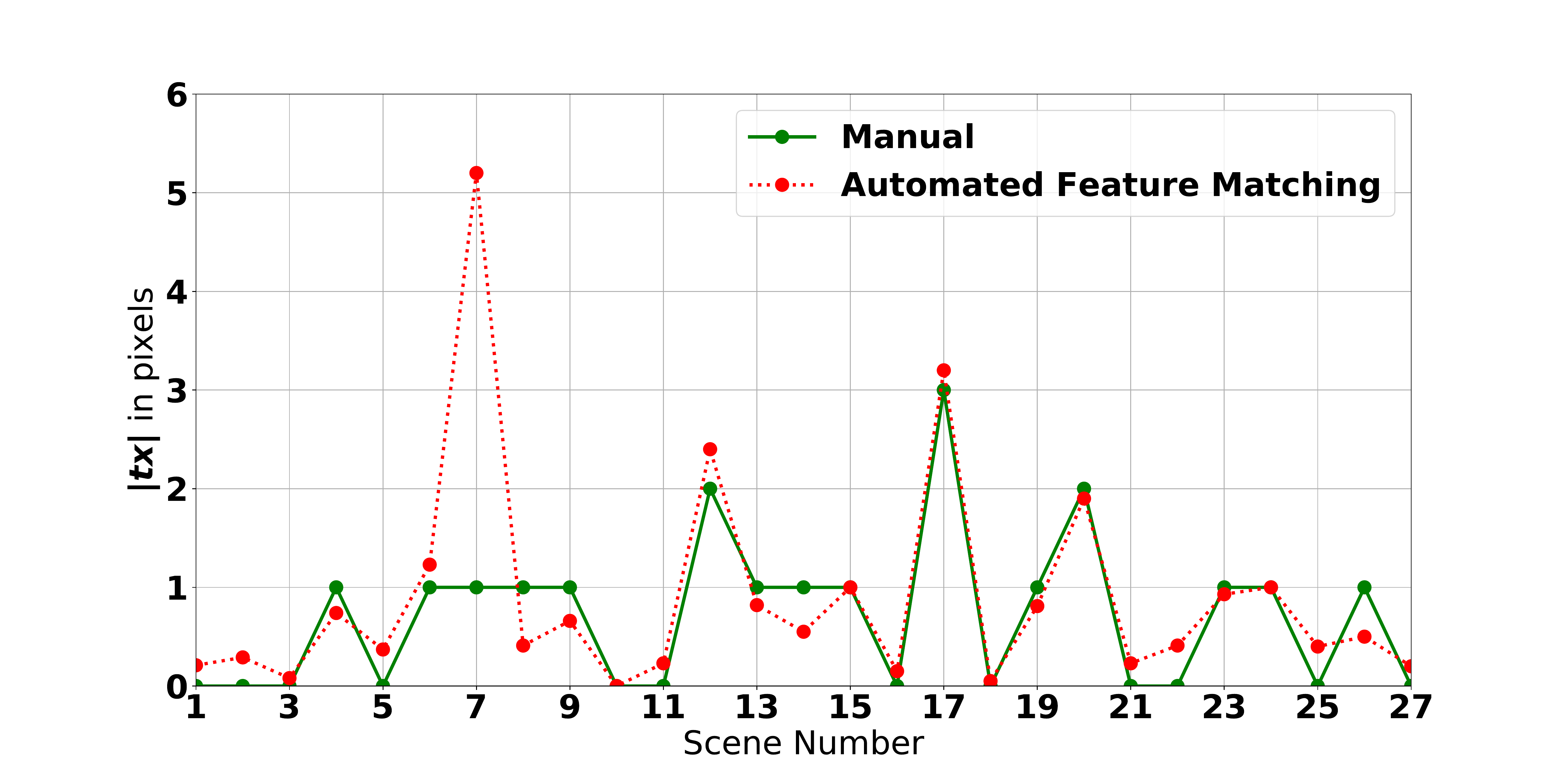} &
         \includegraphics[width=0.48\linewidth]{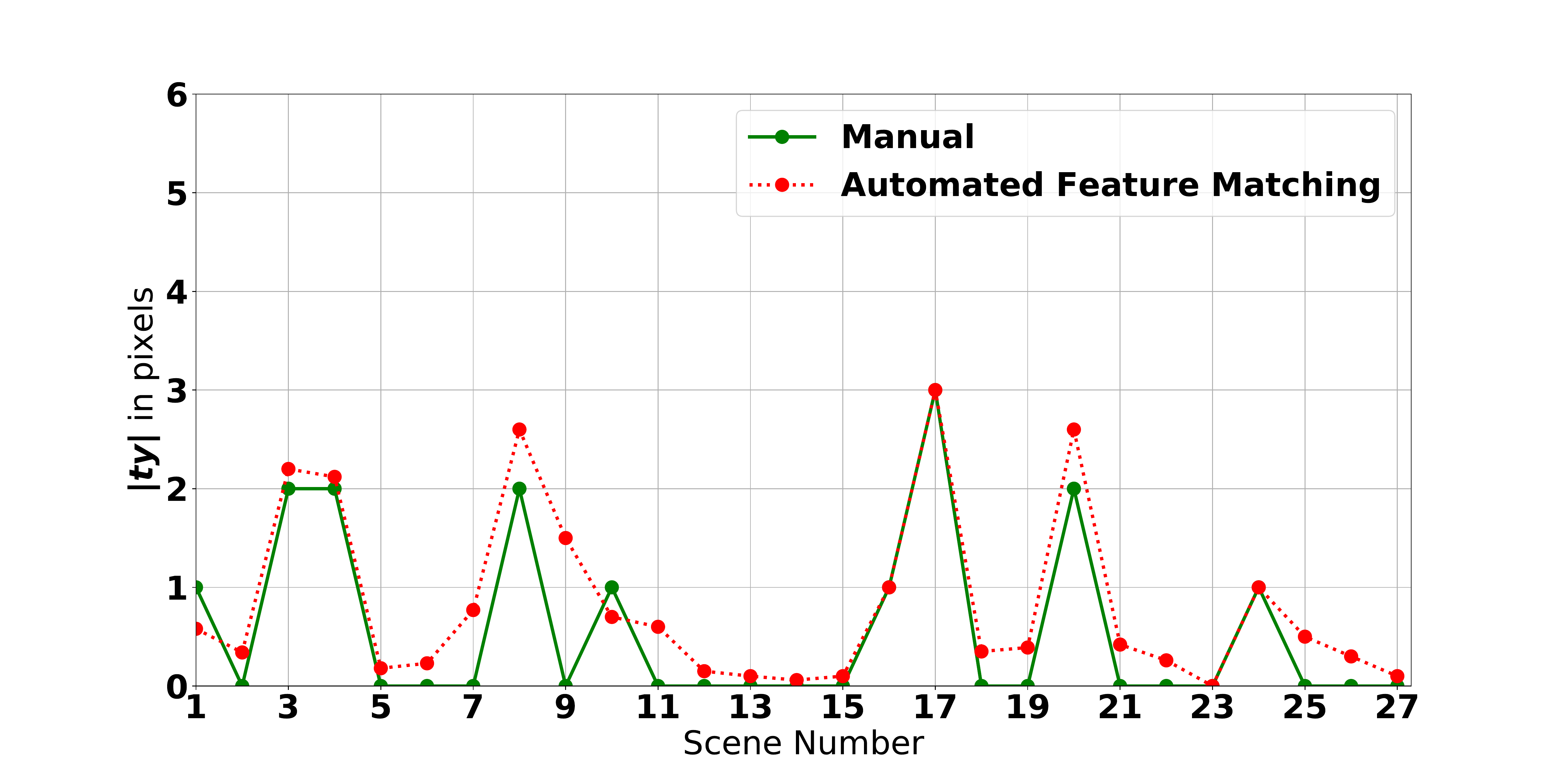} \\
    \multicolumn{2}{c}{(b) L3F-50 dataset} \vspace{10pt} \\
    \includegraphics[width=0.48\linewidth]{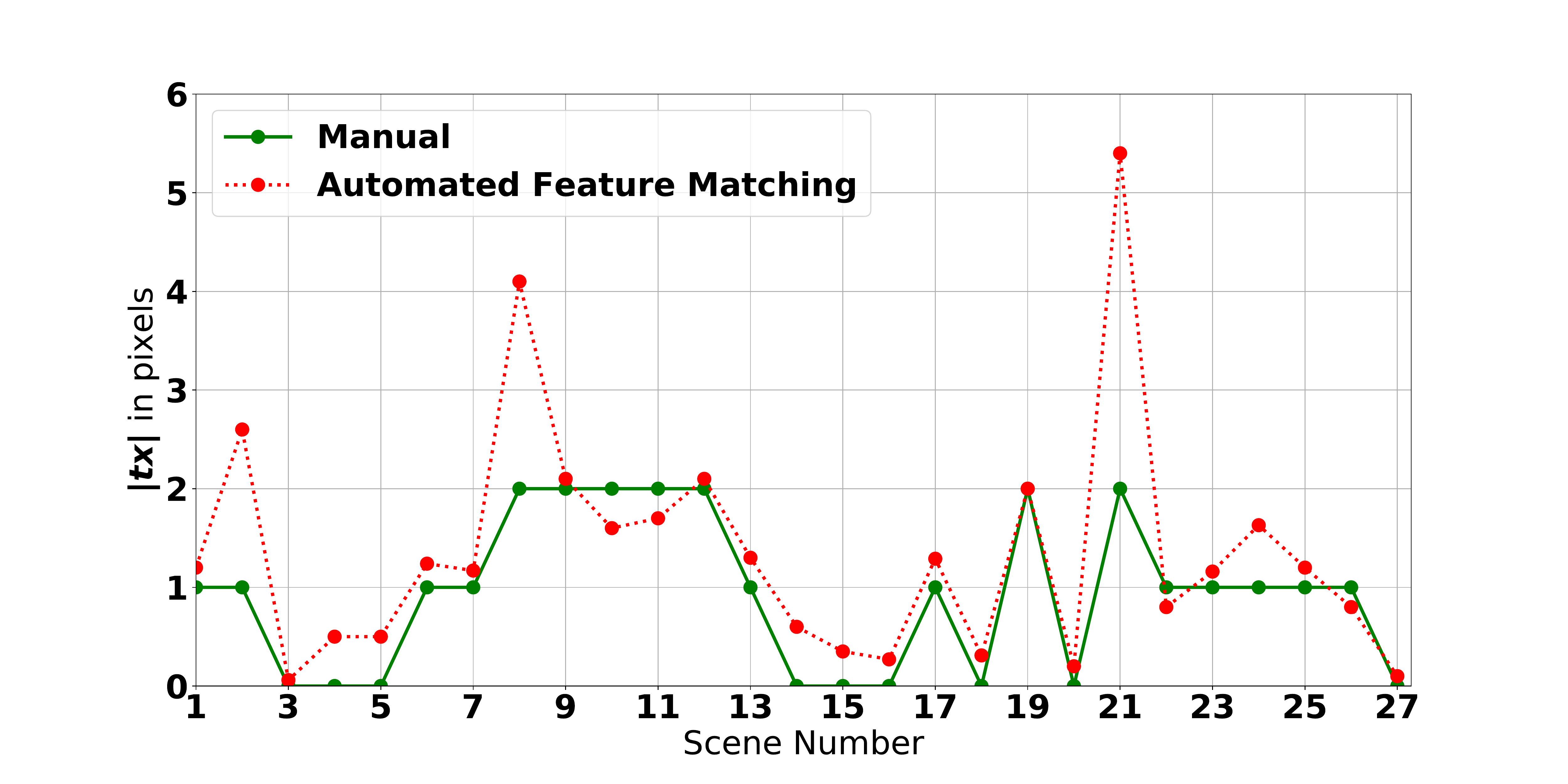} &
         \includegraphics[width=0.48\linewidth]{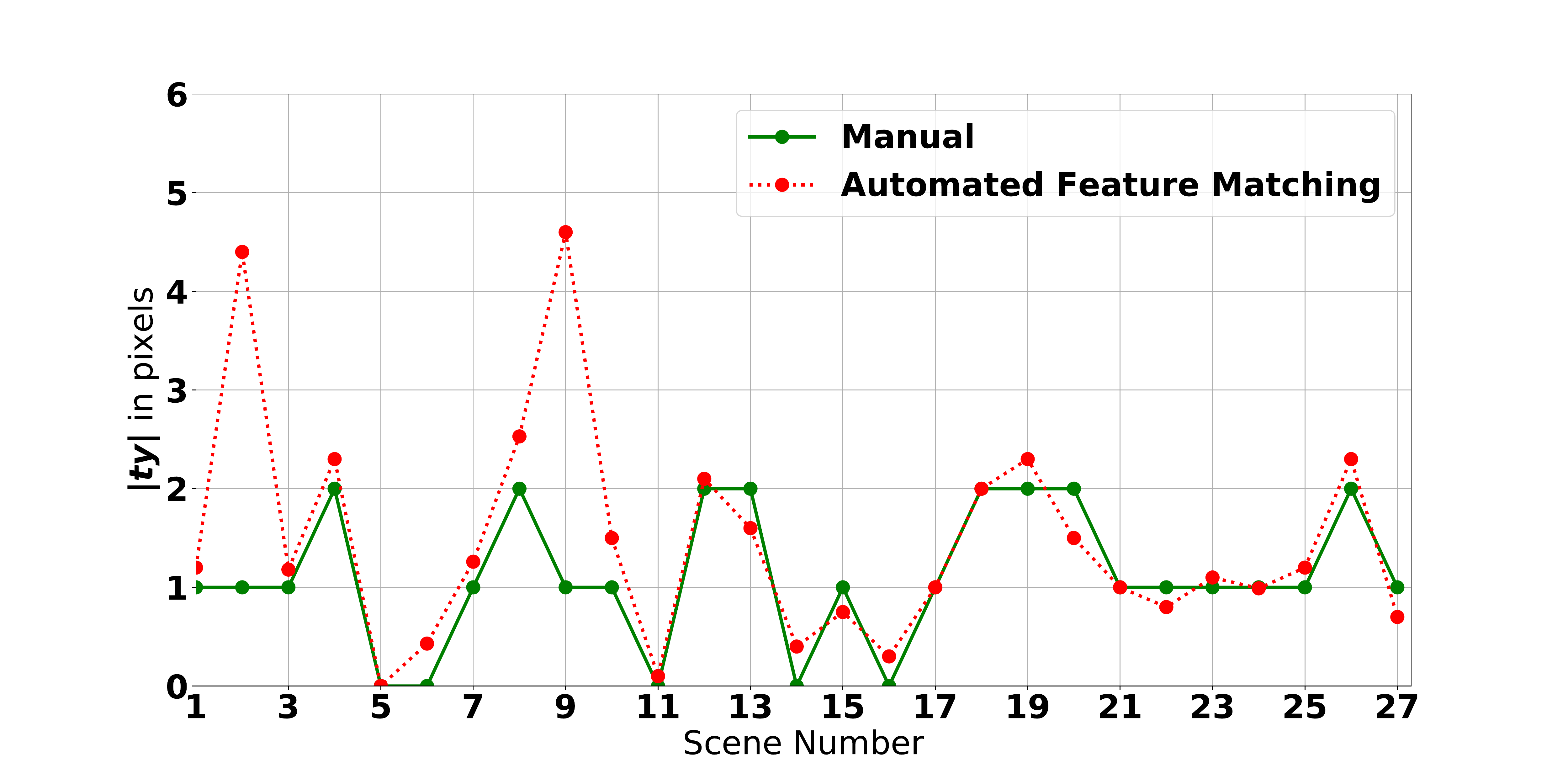} \\
    \multicolumn{2}{c}{(b) L3F-100 dataset}
    \end{tabular}
    \caption{\small The figure shows the misalignment in the horizontal (tx) and vertical (ty) direction. Data for rotation is not plotted as both manual and automated feature matching revealed almost zero rotation. The first nine images on the x-axis are the test images and the remaining are the images used for training.}
    \label{fig:translation}
\end{figure*}

To quantitatively examine the small misalignment in our data, we have adopted two approaches: (a) Use automated feature correspondences to estimate the transformation required to warp the low-light image onto the GT image. The estimated transformation thereby describes the misalignment quantitatively. (b) Manually inspecting the images to detect and quantify misalignment. {\color{black}Before describing these methods in detail, we summarise the results in Fig. \ref{fig:translation}. The figure shows the misalignment in the horizontal and vertical directions for all the scenes in our dataset. The average misalignment is $<1$ pixel. We now describe the two approaches in detail and then report the results in Sec. \ref{sec:results}.}

\subsection{Using automated feature matching}
Employing matching algorithms for extreme low-light images often results in wrong correspondences. This is especially true for the proposed L3F-100 dataset, where we had a tough time finding reliable matched pairs. This is because the input extreme low-light LF has very poor visibility and SNR, marked by an arbitrarily high amount of noise. A more reliable alternative is to manually inspect the image pairs for misalignment. But as manual inspection for all the image pairs across the entire dataset is virtually impossible, we use a good number of precautionary measures to find a reliable set of feature correspondences using the SIFT descriptor. They are described below:
\begin{itemize}
    \item The GT image is well-illuminated, and we faced no trouble in finding good descriptors in them. However, we tried several descriptors such as SIFT, SURF, ORB, etc, but they could not detect features in our extreme low-light images. We thus linearly amplified them. The L3F-20 dataset was amplified by $10 \times$, L3F-50 by $30 \times$ and L3F-100 by $50 \times$. Although a larger amplification gives better overall brightness level, but also increases the noise --- causing inconsistent pairs. We thus tried using the minimum amplification factor required for detecting features in the low-light image. These amplification factors were chosen after empirically experimenting with different amplification factors. The grayscale version of the color images are used for feature matching.
    
    \item {\color{black}We then matched the features from the GT image and the low-light image by computing the L1 distance between the feature descriptors.} To now remove the inconsistent pairs we employed a technique called `\textit{Cross-Check}'. Let  $\mathcal G = \{g_1, g_2, ..., g_n\}$ and $\mathcal D = \{d_1, d_2, ..., d_m\}$ be the set of detected features in the GT and the input dark images, respectively. Now for each descriptor $g_i~\forall~i \in [1,n]$, we find the descriptor $d_j \in \mathcal{D}$ such that,
    {\color{black}
    \begin{equation}
      ||g_i - d_j|| < ||g_i - d_k||~\forall ~d_k\in\mathcal{D}\text{ but }~d_k\neq d_j.  
    \end{equation}
    This pair $(g_i,d_j)$ is called a  consistent feature pair, if additionally,
    \begin{equation}
      ||g_i - d_j|| < ||g_k - d_j||~\forall ~g_k\in\mathcal{G}\text{ but }~g_k\neq g_i,  
    \end{equation}}
 otherwise the pair is called inconsistent and is not used for subsequent calculations.
 
 \item To further ensure the matched pair's consistency, they were subjected to the `\textit{Ratio Test}' mentioned by David Lowe in his seminal paper on SIFT~\cite{sift}. {\color{black}The threshold ratio was set to $70\%$.}
 
 \item After these pre-eliminary checks were made, the RANSAC algorithm was used to remove the outliers with a confidence of $0.99$.
 
 \item Despite all these precautions, we also randomly checked the matches returned after employing these automatic subroutines and once a while had to manually delete the incorrect correspondences as illustrated in Fig.~\ref{fig:manual_remove}.
\end{itemize}

\begin{figure}
    \centering
    \includegraphics[width = 0.8\linewidth]{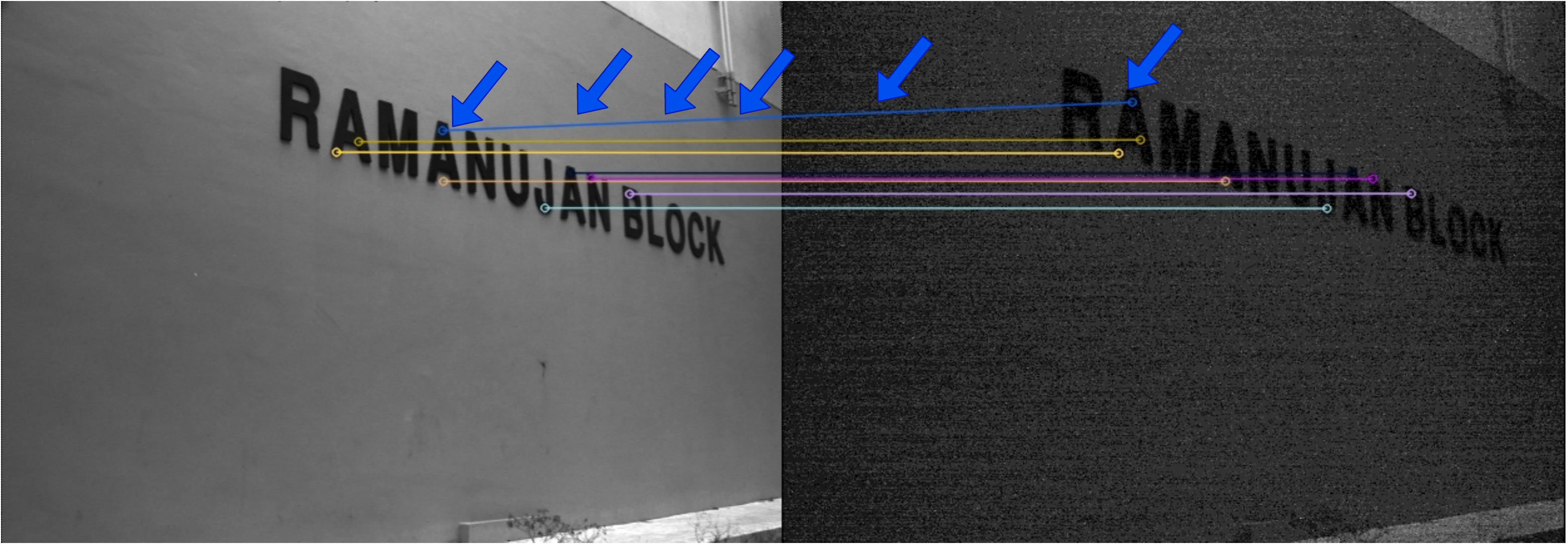}
    \caption{\small {\color{black}Here we show the matched feature pairs between a well-lit GT scene (on the left) and extreme low-light scene taken from the L3F-50 dataset. The low-light image was linearly amplified by $30 \times$.} These pairs were obtained after the Cross-Check, Ratio Test and RANSAC, checks were made. But despite all these precautions, we had to also manually check the matched pairs for inconsistencies. The blue arrows point to an inconsistent matched pair that was deleted after manual inspection. }
    \label{fig:manual_remove}
\end{figure}

Some examples of reliable matched pairs between the low-light and GT SAI are shown in Fig. \ref{fig:consistent_pairs}. Here the visualisation is shown for the (7,7) SAIs lying in a $15 \times 15$ grid.
Once we had the matched pairs we estimated the {\color{black}translation and rotation matrix} required to warp the low-light SAI onto the GT SAI. Our dataset's misalignment is not because of casual photography but because of a very small but unavoidable camera shake due to pressing of camera button and movement of the camera's internal mirrors. Thus it is safe to assume that the captured dataset is free from scaling, skew and projective distortions.
This is also evident from Fig. \ref{fig:animation} in which we show the GIF animation of toggling between the GT and low-light SAI. One can hardly see any scaling or skew distortions in this animation.
We thus estimate the following transformation,

\begin{equation}
    \begin{bmatrix}
x_g \\ y_g \\ 1
\end{bmatrix} =
    \left[
         \begin{array}{ccc}
         cos \theta & -sin \theta & tx          \\
         sin \theta & cos \theta & ty \\
         0&0&1
        \end{array}
    \right]~ \begin{bmatrix}
x_d \\ y_d \\ 1
\end{bmatrix}.
\end{equation}
Here $x$ and $y$ denote the spatial coordinates and the subscripts $g$ and $d$ denote GT and dark input image, respectively. $\theta$ is the rotation angle and $tx, ty$ are the translation along the horizontal and vertical direction.

\begin{figure*}
    \centering
    \begin{tabular}{cc}
    \includegraphics[width=0.48\linewidth]{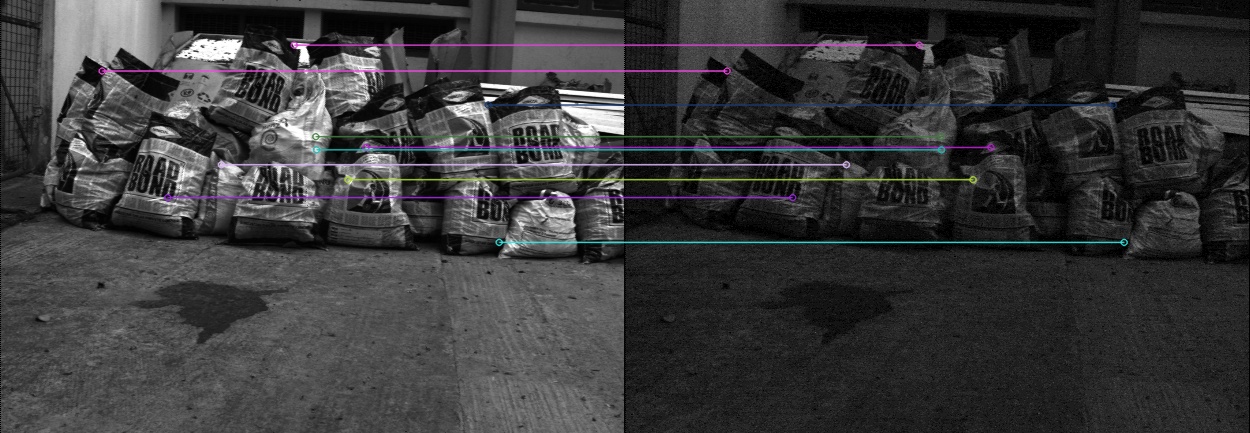} & \includegraphics[width=0.48\linewidth]{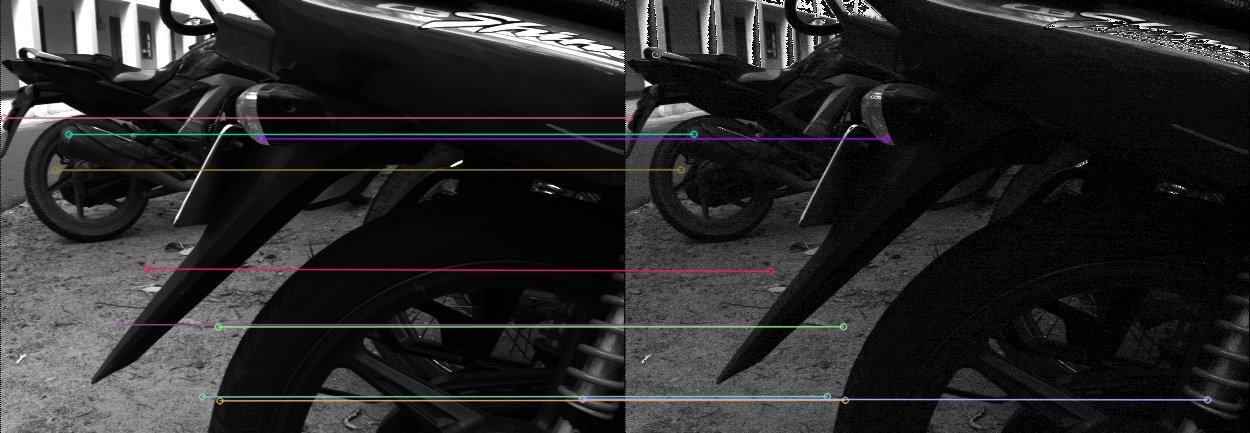}\\
    \multicolumn{2}{c}{(a) L3F-20 dataset, amplified $10 \times$} \vspace{10pt} \\ 
    
    \includegraphics[width=0.48\linewidth]{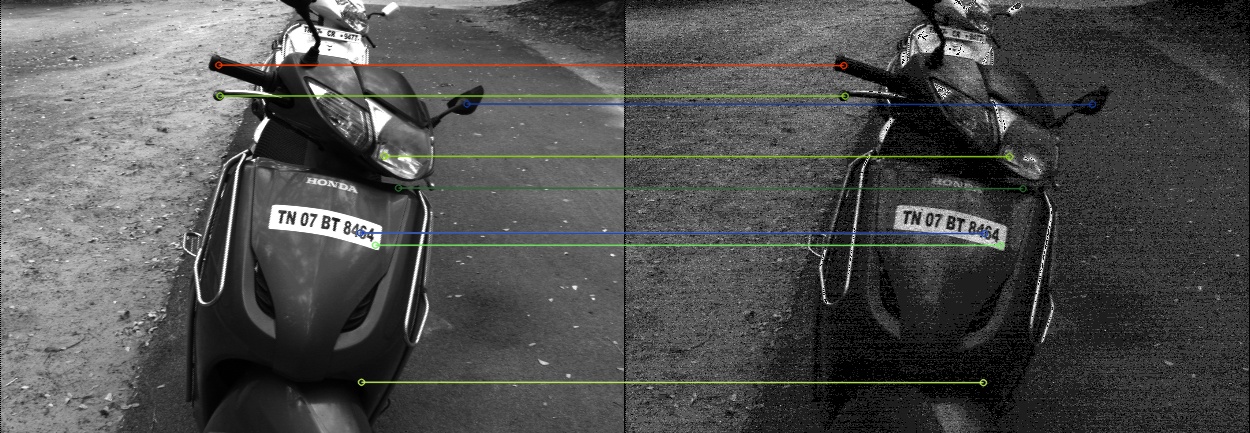} & \includegraphics[width=0.48\linewidth]{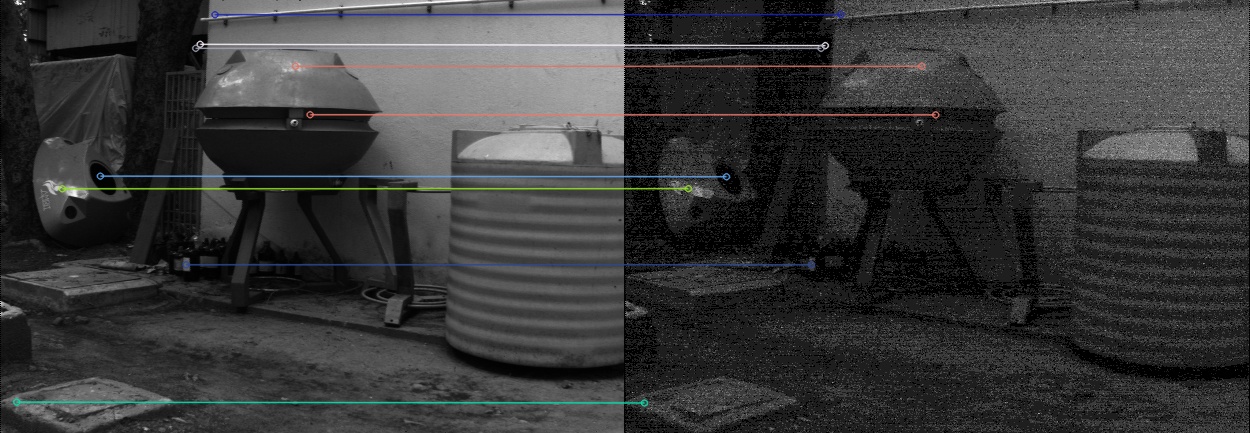}\\
    \multicolumn{2}{c}{(b) L3F-50 dataset, amplified $30 \times$} \vspace{10pt} \\ 
    
    \includegraphics[width=0.48\linewidth]{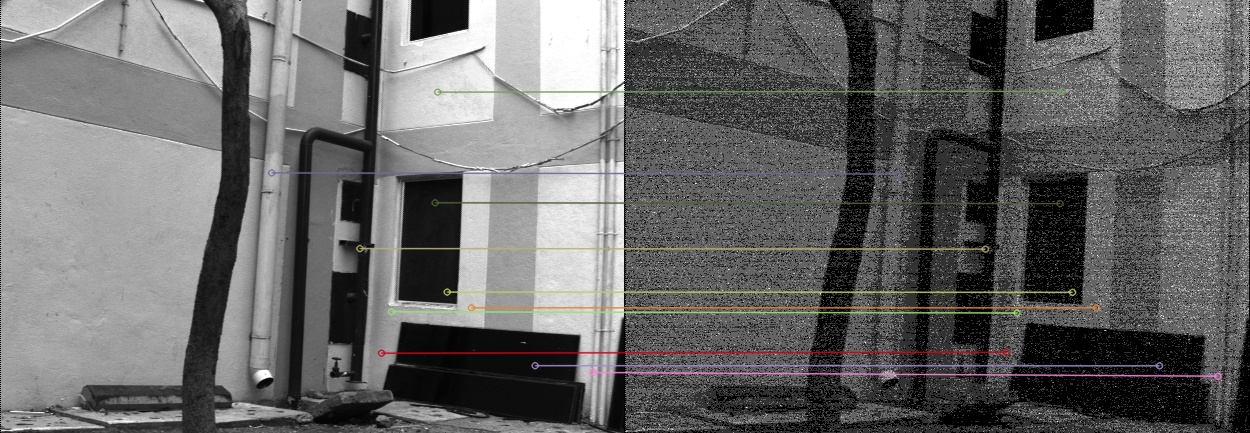} & \includegraphics[width=0.48\linewidth]{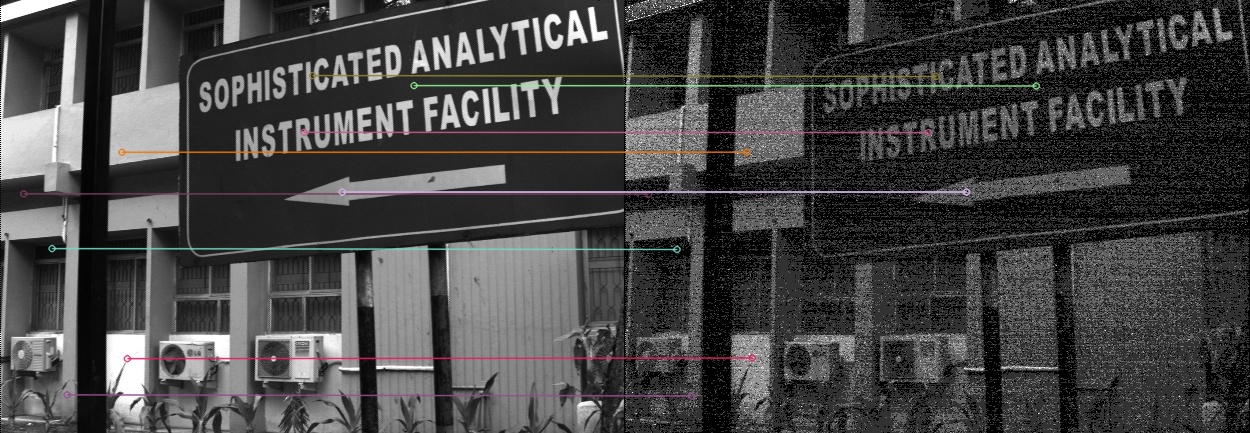}\\
    \multicolumn{2}{c}{(c) L3F-100 dataset, amplified $50 \times$} \\ 
    \end{tabular}
    \caption{\small Some examples of reliable matched pairs between the low-light and GT SAI. Here the visualisation is shown for the (7,7) SAIs lying in a $15 \times 15$ grid.}
    \label{fig:consistent_pairs}
\end{figure*}

\begin{figure*}[t!]
    \centering
    \scriptsize
    \setlength{\tabcolsep}{1pt}
    \begin{tabular}{ccc}
     \animategraphics[loop,autoplay,width=0.25\linewidth]{1}{second_rebuttal/20matched/animation/one/}{2}{1}
    
      &
       \animategraphics[loop,autoplay,width=0.25\linewidth]{1}{second_rebuttal/50matched/animation/one/}{2}{1}
       
       &
       \animategraphics[loop,autoplay,width=0.25\linewidth]{1}{second_rebuttal/100matched/animation/one/}{2}{1} \\
       
       \animategraphics[loop,autoplay,width=0.25\linewidth]{1}{second_rebuttal/20matched/animation/two/}{2}{1}
    
      &
       \animategraphics[loop,autoplay,width=0.25\linewidth]{1}{second_rebuttal/50matched/animation/two/}{2}{1}
       
       &
       \animategraphics[loop,autoplay,width=0.25\linewidth]{1}{second_rebuttal/100matched/animation/two/}{2}{1} \\
       
       (a) L3F-20, $10\times$ amplified & (b) L3F-50, $30\times$ amplified & (c) L3F-100, $50\times$ amplified

    \end{tabular}
    \caption{\small GIF animation of the toggling operation between the GT and the input extreme low-light SAI. \textit{View in Adobe PDF Reader or any other PDF viewer with JavaScript engine enabled.}}
    \label{fig:animation}
\end{figure*}


\subsection{Manual Inspection}

Automated feature matching is a quick way of estimating the misalignment. Still, occasionally we had to see failure cases because, in our low-light images, the noise significantly dominates the scene visibility. This is especially true for the L3F-100 dataset, where localizing the features is very hard. We thus manually inspected the image pairs to estimate the misalignment. {\color{black}For this, we overlapped the SAIs with $50\%$ transparency and quickly toggled between the GT and dark SAIs to detect the regions showing some displacement. Then with a good amount of zoom-in, we noted the pixels' spatial coordinates in those regions, which we could uniquely identify in both images.} This is a very slow and tedious task but returned a couple of matched pairs with a very high probability of correct correspondence.

\subsection{Results and conclusion}
\label{sec:results}
As each SAI present in a LF image captured using Lytro Illum has pixels from all parts of the LF sensor, we focus on analyzing the {\color{black}(7,7) SAIs. The results are shown in Fig. \ref{fig:translation}. In this figure, we plot $|tx|$ and $|ty|$ for the (7,7) SAIs obtained from each scene present in our dataset.} The figure shows the estimates obtained using both manual inspection (denoted as `\textit{Manual}' in the legend) and using automated feature matching (denoted as `\textit{Automated Feature Matching}' in the legend). The first nine scenes denoted on the x-axis are the scenes reserved for testing and the remaining scenes were used for training. We do not plot $\theta$ because the rotation angle was $<0.01~degress$.

We observe that for the L3F-20 dataset the estimation via both methods is in almost perfect harmony. But as the noise increases, especially for the L3F-100 dataset, we sometimes see sporadic peaks for the automated feature matching method. This is because in the L3F-100 dataset the noise dominates the true signal by a large margin. This makes precise localization very difficult, resulting in wrong estimates. Thus, manual inspection of misalignment in such cases is very essential.
{\color{black}Please note that, while manually inspecting the L3F-100 dataset, it was quite challenging for us also to do precise localization. At several instants, we found misalignment to be $<1$ pixel. But given the huge dominance of noise over the true signal, we decided not to report a conservative estimate and thus, for some scenes upto $2$ pixel misalignment is reported.}
Still, overall both the manual and automated methods estimate nearly the same value for most scenes. On an average, the misalignment in both directions is $<1$ pixel. We also repeated this experiment for all the SAIs present in the central $9\times9$ grid and the automated feature matching estimated roughly the same misalignment of 1 pixel on an average. 
}


%


\ifCLASSOPTIONcaptionsoff
  \newpage
\fi



%

%

{\small
\bibliographystyle{IEEEtran}
\bibliography{ref}
}
\end{document}